\documentclass[longauth]{aa}
\usepackage[varg]{txfonts}

\usepackage{graphicx}
\usepackage{natbib}
\usepackage{comment}
\usepackage{amsmath,yhmath,upgreek}
\usepackage{textcomp}
\usepackage{color}
\usepackage{dcolumn}

\bibpunct{(}{)}{;}{a}{}{,} 
\newcolumntype{d}[1]{D{.}{.}{#1}}

\begin{document}

\title{A LOFAR census of millisecond pulsars}

\author{V.~I.~Kondratiev\inst{\ref{astron},\ref{asc}} 
        \and 
        J.~P.~W.~Verbiest\inst{\ref{bielefeld},\ref{mpifr}} 
        \and
        J.~W.~T.~Hessels\inst{\ref{astron},\ref{uva}}
        \and
        A.~V.~Bilous\inst{\ref{nijmegen}}
        \and
        B.~W.~Stappers\inst{\ref{jb}}
        \and
        M.~Kramer\inst{\ref{jb},\ref{mpifr}}
        \and
        E.~F.~Keane\inst{\ref{swinburne},\ref{caastro},\ref{skahq}}
        \and
        A.~Noutsos\inst{\ref{mpifr}}
        \and
        S.~Os\l{}owski\inst{\ref{bielefeld},\ref{mpifr}}
        \and
        R.~P.~Breton\inst{\ref{southampton}}
        \and
        T.~E.~Hassall\inst{\ref{southampton}}
        \and
        A.~Alexov\inst{\ref{stsi}}
        \and
        S.~Cooper\inst{\ref{jb}}
        \and
        H.~Falcke\inst{\ref{nijmegen},\ref{astron}}
        \and
        J.-M.~Grie{\ss}meier\inst{\ref{lpc2e},\ref{nancay}}
        \and
        A.~Karastergiou\inst{\ref{oxford},\ref{western_cape},\ref{rhodes}}
        \and
        M.~Kuniyoshi\inst{\ref{naoj}}
        \and
        M.~Pilia\inst{\ref{inaf}}
        \and
        C.~Sobey\inst{\ref{astron}}
        \and
        S.~ter~Veen\inst{\ref{astron}}
        \and
        J.~van~Leeuwen\inst{\ref{astron},\ref{uva}}
        \and
        P.~Weltevrede\inst{\ref{jb}}
        \and
        M.~E.~Bell\inst{\ref{csiro},\ref{caastro}}
        \and
        J.~W.~Broderick\inst{\ref{oxford},\ref{southampton}}
        \and
        S.~Corbel\inst{\ref{diderot},\ref{nancay}}
        \and
        J.~Eisl\"offel\inst{\ref{tautenburg}}
        \and
        S.~Markoff\inst{\ref{uva}}
        \and
        A.~Rowlinson\inst{\ref{csiro}}
        \and
        J.~D.~Swinbank\inst{\ref{princeton}}
        \and
        R.~A.~M.~J.~Wijers\inst{\ref{uva}}
        \and
        R.~Wijnands\inst{\ref{uva}}
        \and
        P.~Zarka\inst{\ref{lesia},\ref{nancay}}
        }

\institute{ASTRON, the Netherlands Institute for Radio Astronomy, Postbus
2, 7990 AA Dwingeloo, The Netherlands \\ \email{kondratiev@astron.nl}\label{astron} 
\and
Astro Space Centre, Lebedev Physical Institute, Russian Academy of Sciences, Profsoyuznaya Str. 84/32, Moscow 117997, Russia\label{asc} 
\and
Fakult\"at f\"ur Physik, Universit\"at Bielefeld, Postfach 100131, 33501 Bielefeld, Germany\label{bielefeld} 
\and
Max-Planck-Institut f\"ur Radioastronomie, Auf dem H\"ugel 69, 53121 Bonn, Germany\label{mpifr} 
\and
Anton Pannekoek Institute for Astronomy, University of Amsterdam, Science Park 904, 1098 XH Amsterdam, The Netherlands\label{uva}
\and
Department of Astrophysics/IMAPP, Radboud University Nijmegen, P.O. Box 9010, 6500 GL Nijmegen, The Netherlands\label{nijmegen}
\and
Jodrell Bank Centre for Astrophysics, School of Physics and Astronomy, University of Manchester, Manchester M13 9PL, UK\label{jb}
\and
Centre for Astrophysics and Supercomputing, Swinburne University of Technology, Mail H30, PO Box 218, VIC 3122, Australia\label{swinburne}
\and
ARC Centre of Excellence for All-sky Astrophysics (CAASTRO), The University of Sydney, 44 Rosehill Street, Redfern, NSW 2016, Australia\label{caastro}
\and
SKA Organisation, Jodrell Bank Observatory, Lower Withington, Macclesfield, Cheshire, SK11 9DL, UK\label{skahq}
\and
School of Physics  and  Astronomy,  University  of  Southampton, SO17 1BJ, UK\label{southampton}
\and
Space Telescope Science Institute, 3700 San Martin Drive, Baltimore, MD 21218, USA\label{stsi}
\and
LPC2E - Universit\'{e} d'Orl\'{e}ans / CNRS, 45071 Orl\'{e}ans cedex 2, France\label{lpc2e}
\and
Station de Radioastronomie de Nan\c{c}ay, Observatoire de Paris, PSL Research University, CNRS, Univ. Orl\'{e}ans, OSUC, 18330 Nan\c{c}ay, France\label{nancay}
\and
Oxford Astrophysics, Denys Wilkinson Building, Keble Road, Oxford OX1 3RH, UK\label{oxford}
\and
Department of Physics \& Astronomy, University of the Western Cape, Private Bag X17, Bellville 7535, South Africa\label{western_cape}
\and
Department of Physics and Electronics, Rhodes University, PO Box 94, Grahamstown 6140, South Africa\label{rhodes}
\and
NAOJ Chile Observatory, National Astronomical Observatory of Japan, 2-21-1 Osawa, Mitaka, Tokyo 181-8588, Japan\label{naoj}
\and
INAF - Osservatorio Astronomico di Cagliari, via della Scienza 5, 09047 Selargius (Cagliari), Italy\label{inaf}
\and
CSIRO Astronomy and Space Science, PO Box 76, Epping, NSW 1710, Australia\label{csiro}
\and
Laboratoire AIM (CEA/IRFU - CNRS/INSU - Universit\'e Paris Diderot), CEA DSM/IRFU/SAp, F-91191 Gif-sur-Yvette, France\label{diderot}
\and
Th\"uringer Landessternwarte, Sternwarte 5, D-07778 Tautenburg, Germany\label{tautenburg}
\and
Department of Astrophysical Sciences, Princeton University, Princeton, NJ 08544, USA\label{princeton}
\and
LESIA, Observatoire de Paris, CNRS, UPMC, Universit\'e Paris-Diderot, 5 place Jules Janssen, 92195 Meudon, France\label{lesia}
}

\date{Received August 2015 / Accepted 2015}

\abstract{
We report the detection of 48 millisecond pulsars (MSPs) out of 75 observed thus far using the LOw-Frequency ARray (LOFAR) 
in the frequency range 110--188\,MHz.  
We have also detected three MSPs out of nine observed in the frequency range 38--77\,MHz.  
This is the largest sample of MSPs ever observed at these 
low frequencies, and half of the detected MSPs were observed for the first time at frequencies below 200\,MHz. 
We present the average pulse profiles of the detected MSPs, their effective pulse widths, and flux densities and compare 
these with higher observing frequencies.
The flux-calibrated, multifrequency LOFAR pulse profiles are publicly available via the European Pulsar 
Network Database of Pulsar Profiles.
We also present average values of dispersion measures (DM) and discuss DM and profile variations.
About 35\% of the MSPs show strong narrow profiles, another 25\% exhibit scattered profiles, and
the rest are only weakly detected.
A qualitative comparison of the LOFAR MSP profiles with those at higher
radio frequencies shows constant separation between profile components.
Similarly, the profile widths are consistent with those observed at higher frequencies,
unless scattering dominates at the lowest frequencies.
This is very different from what is observed for normal pulsars and suggests a compact
emission region in the MSP magnetosphere.
The amplitude ratio of the profile components, on the other hand, can dramatically change towards low frequencies, often with 
the trailing component becoming dominant. As previously demonstrated this can be caused by aberration and retardation.
This data set enables high-precision studies of pulse profile evolution with frequency, dispersion, 
Faraday rotation, and scattering in the interstellar medium. Characterising and correcting these systematic effects 
may improve pulsar-timing precision at higher observing frequencies, where pulsar timing array projects
aim to directly detect gravitational waves.
}

\titlerunning{A LOFAR Census of MSPs}
\authorrunning{Kondratiev et al.}

\maketitle

\section{Introduction}\label{intro}

Radio millisecond pulsars (MSPs) are commonly held to have been spun up by
mass transfer from a binary companion \citep{acrs82,rs81} and are 
 evolutionarily linked with X-ray binary systems
\citep{bh91,asr+09,pfbr+13,sah+14}. Their observed properties are in many respects quite 
different from those
of the long-period, 'normal' pulsars. 
Both the spin periods, $P$, and period derivatives, $\dot{P}$, of MSPs ($P\sim 1$--10\,ms, $\dot{P}\sim 10^{-19}$\,s\,s$^{-1}$) 
are a few orders of magnitude lower than for normal pulsars
($P\sim 1$\,s, $\dot{P}\sim 10^{-15}$\,s\,s$^{-1}$),
and thus the inferred values of surface magnetic fields for MSPs are three to four orders of magnitude lower
than the typical value of $10^{12}$\,G for normal pulsars. Despite these differences, MSP studies at 1.4, 2.7, and
4.9\,GHz by \citet{kramer1998,kramer1999} and \citet{xilouris1998} suggest that the radio emission mechanisms for normal 
pulsars and MSPs are essentially the same -- even though their magnetospheres
differ in size by two to three orders of magnitude.

This conclusion is mainly based on the similarity of their spectra and
profile complexity. The same result was also obtained by \citet{tbms98} for a sample of MSPs in the
southern hemisphere at frequencies between 0.4 and 1.6\,GHz. This is also confirmed by single-pulse studies at 1.4\,GHz 
of PSR~J0437$-$4715 by \citet{jak+98}, who find that its individual pulses are very similar to those of slowly rotating pulsars.
Finally, \citet{es03} came to the same conclusion in their study of individual pulses from several pulsars with the 
Westerbork Synthesis Radio Telescope (WSRT) at  several frequencies between 328 and 2240\,MHz.

Even though the radio emission mechanism of both MSPs and normal pulsars
seems to be the same, there are a number of differences in the observed emission properties.  Multifrequency
studies above 400\,MHz generally\footnote{There are notable exceptions, such as PSRs J2145$-$0750 and J0751+1807, that show large frequency evolution 
in the amplitude ratio of their profile components \citep{kramer1999,kl96}, and a few other MSPs have new profile components appearing 
at higher/lower frequencies.} 
show remarkably little frequency evolution in the average profiles of MSPs;
in particular, the width of components and their separation typically remain almost constant \citep{kramer1998,kramer1999}.
This is different from what is generally observed in normal pulsars \citep[see e.g.][]{thorsett1991,xkjwt96,pilia2015}, and
may result from a small radio emission region in the very compact magnetospheres of MSPs. 
\citet{kramer1998} also show that MSPs are somewhat less luminous and less efficient radio emitters than normal pulsars, 
with isolated MSPs being even less luminous than their binary counterparts. These findings are in accordance with
results from recent works by \citet{lbb+13} and \citet{bbb+13}.
All these results suggest that MSP
magnetospheres might not simply represent scaled versions of the magnetospheres of normal pulsars \citep{xilouris1998},
and the differences could be caused by the different evolutionary histories of normal pulsars and MSPs.

MSP studies at low radio frequencies, i.e. below $\lesssim 300$\,MHz, are valuable, as they 
expand our ability to compare flux-density spectra, polarisation and profile evolution to those of normal pulsars and 
probe the compactness
of MSP emission regions \citep{c78}.  Normal pulsars often show a spectral turnover between 
100 and 250\,MHz \citep[e.g.][]{sieber73,kms+78,ikms81,malofeev2000,mms00}, raising the question of whether MSPs behave similarly.  A 
study of 30 MSPs with narrow bandwidths at frequencies of 102 and 111\,MHz by \citet{kl01} concluded that the spectra of the 
MSPs they studied mostly do not show any low-frequency turnover.  Only in the case of \object{PSR J1012+5307} did they find a possible turnover near 100\,MHz. 
\citet{kl99} analysed the frequency dependence of profile widths of 12 MSPs between
102 and 1400\,MHz. They found that profile widths remained nearly constant within this frequency range, and in the case 
of PSR J2145$-$0750, the component separation became even smaller at 102\,MHz than at
higher frequencies \citep{kl96}. Both results provide further evidence for the compactness of the
emission region of MSPs.

\begin{table*}[phtb]
\centering
\caption{Standard setup for the LOFAR observations of MSPs.
}\label{obssetup}
\begin{tabular}{ccccccccc}
\hline\hline
Array & Core     & Bits & Frequency & Sub-bands\tablefootmark{a} & Sub-band\tablefootmark{a} & Sampling & Observing & Data \\
      & stations &  per & range     &          & width   & time     & time      & products \\ 
      &          & sample & (MHz)   &          & (kHz)   & ($\upmu$s) & (min)     & \\
\hline
HBA Dual & 23\tablefootmark{b} & 8 & 110--188 & 51--450 & 195.312 & 5.12 & 20 & XXYY\tablefootmark{c} \\
\hline
\end{tabular}
\tablefoot{
\tablefoottext{a}{The second polyphase filter \citep{haarlem2013} was skipped in the online processing, thus the number of sub-bands and channels
was the same, and we will refer to the original sub-bands as channels in the rest of the paper.}
\tablefoottext{b}{The core station CS013 has an incorrect orientation of dipoles with respect to the rest of the array and was always excluded in
our observations. Typically two--three other core stations were not available for different reasons, and we used only 20--21 HBA stations
out of 24 total. On a few occasions we only had 18--19 HBA stations available.}
\tablefoottext{c}{Raw complex voltages of two linear polarisations.}
}
\end{table*}

Although \citet[and references therein]{kl01} provided a first
insight into low-frequency MSP profiles and flux densities, their data
suffered from poor time resolution (only 0.128--0.64\,ms).  This was mainly due to the necessity to form many narrow frequency channels 
to minimise 
dispersion smearing. 
Observations of MSPs at such low frequencies indeed
present a challenge because of the large pulse broadening caused by both scattering and dispersion in the interstellar medium (ISM). 
Dispersion in the cold ionised plasma of the ISM scales as $f^{-2}$.
Though computationally intensive, it can be fully accounted for by coherent dedispersion 
\citep{hankins1971,hankins_rickett1975}, unlike scattering, which also
has an even stronger frequency dependence (scales as $f^{-4.4}$ for a Kolmogorov
spectrum of electron density inhomogeneities in the ISM).
Because the pulses of the MSPs have much shorter durations than those of normal pulsars, they can be 
completely scattered out, preventing detection. 
 
On average, pulsars have a steep negative spectral index 
of $-1.4$ \citep{blv13}, and MSP spectra are found to be consistent with the normal pulsar population \citep{kramer1998}.
However, the background sky temperature of the Galactic synchrotron radiation is also
frequency dependent, and this dependence is steeper \citep[$f^{-2.55}$;][]{lmop87}.
Thus, even if one assumes no turnover in the spectra of MSPs above 100\,MHz, the majority of the MSPs, especially those
located along the Galactic plane, will be more difficult to detect 
below 250\,MHz than at higher frequencies.

In the past decade, higher-quality, low-frequency MSP observations have been enabled by a new generation
 of pulsar backends that provide wide-bandwidth, coherent dedispersion (sometimes in real-time).  For example, \citet{stappers2008} used WSRT and the 
 PuMaII pulsar backend \citep{puma2} in the frequency range of 115--175\,MHz to coherently dedisperse the data to a final 
 time resolution of 25.6--51.2\,$\upmu$s.  They detected eight out of 14 MSPs observed and did not find any clear relationship between the detectability
of a source and any combination of its period, DM or flux density at higher frequencies. 
Nevertheless, the majority of the non-detections were pulsars with lower flux densities.
\citet{stappers2008} did not detect two MSPs, PSRs J1024$-$0719 and J1713+0747, that were detected 
by \citet{kl01}\footnote{Pulsar J0218+4232 was also not detected by \citet{stappers2008}, though
\citet{kl01} reported it to be detected and provided its flux density estimates, but did not present its profile.}.
They did not find any obvious reason for these non-detections.
The authors suggested scattering to be the main reason
for all of their six non-detections. For PSR J2145$-$0750 \citet{stappers2008} also found
the reduced separation between the peaks of the components with regard to the high frequencies and even found this
separation to be smaller than in \citet{kl96}, which can be attributed to residual dispersive
smearing. However, \citet{stappers2008} 
cast doubt on whether these components are the same as at high frequencies, because the leading
component at 150\,MHz more resembles the trailing component in the high-frequency profile.  Recently, \citet{drt+13}
also found the dominant component at low frequencies to be narrower than reported by \citet{kl96} in
their observations with the Long Wavelength Array (LWA) at 37--81\,MHz.

Since the work of \citet{stappers2008} and until the recent study of PSR J2145$-$0750 by \citet{drt+13}, 
there have been no other MSP studies at frequencies below 200\,MHz that have provided
high-quality profiles 
at these low frequencies. 
At the same time, the number of known MSPs in the Galactic field (many at distances $< 2$\,kpc, and at high Galactic latitudes) 
has substantially grown, owing to many 
detections in {\it Fermi/LAT} unassociated sources \citep[see e.g.][]{kjr+11,rrc+11,kcj+12,bgc+13}.
About 75 MSPs were discovered in the Galactic field between the discovery of the first MSP 
\citep[PSR B1937+21;][]{backer1982} and the {\it Fermi} launch in 2008.
In the last five years we have experienced an MSP renaissance as this number has almost tripled to 218
\citep[ATNF catalogue\footnote{\tt http://www.atnf.csiro.au/people/pulsar/psrcat/};][and ongoing pulsar surveys]{atnf}.
This is mainly a result of {\it Fermi} discoveries, but also includes a number of discoveries by ongoing pulsar surveys. 
These include the Green Bank Telescope (GBT) 350-MHz Drift-scan \citep{boyles2013,lynch2013} 
and Green-Bank Northern Celestial Cap \citep[GBNCC;][]{slr+14} surveys,
the Pulsar Arecibo L-band Feed Array (PALFA) survey \citep{nab+13,csl+12,dfc+12}, 
and the Parkes/Effelsberg High Time Resolution Universe Survey \citep[HTRU; e.g.][]{kjv+10,kjb+12,bck+13}.
Therefore, there is also a significantly larger sample of MSPs that can be observed for the first time below 200\,MHz.

The LOw-Frequency ARray (LOFAR) is a digital aperture array radio telescope operating from 10--240\,MHz -- i.e. 
the lowest four octaves of the `radio window' observable from the Earth's surface (see \citealt{haarlem2013} for a description of LOFAR).
It is an interferometric array of dipole antenna stations distributed over Europe,
with the Core of 24 stations located within 
an area approximately 2~\!km across near the Dutch village of Exloo.
Compared to previous low-frequency radio telescopes, LOFAR offers many advantages in its observing 
capabilities \citep{stappers2011}. Among them are the large fractional 
bandwidth (80\,MHz of bandwidth in the 10--90 or 110--250\,MHz range), sensitivity that is at least 20 times
better than the low-frequency frontends at WSRT \citep{stappers2008}, and an unrestricted ability to track sources from rise to set, thus accumulating
a large number of pulses in a single observing session. The latter is very valuable as many of the old low-frequency arrays,
such as the Large Phased Array in Pushchino (Russia), are transit instruments.

In this paper we present the results of the first exploratory observations of MSPs with LOFAR.
In Sect.~\ref{obs} we describe the LOFAR observing setup and data analysis. 
In Sect.~\ref{results} we present the results of our MSP observations, give lists of 
detected and non-detected pulsars, show folded pulse profiles and provide measurements of flux densities at 110--188\,MHz.
In Sect.~\ref{discussion} we discuss the detectability of MSPs, provide new measurements of their DMs and compare these with 
previously published values.
We also discuss DM and profile variations of MSPs and qualitatively compare profiles for
several pulsars with profiles at higher frequencies.
We give a summary in Sect.~\ref{summary}.

\section{Observations and data reduction}\label{obs}

The observations presented in this paper were taken between 19 December 2012 and 3 November 2014 with the LOFAR Core stations
using the High-Band Antennas (HBAs) in the frequency range 110--188\,MHz and the Low-Band Antennas (LBAs) from 10--88\,MHz
(see \citealt{haarlem2013}; \citealt{stappers2011} for more information on the LOFAR observing system and set-up).
A single clock system allows all Core stations to be combined into one or multiple coherent tied-array beams, and an eight-bit sampling mode 
allows 80\,MHz of instantaneous bandwidth (limited by filters that avoid the FM frequency range from 90--110\,MHz).  
This provides a factor $\sim 5$ increase in sensitivity compared with early LOFAR pulsar observations that were restricted 
to using the six stations on the `Superterp' and 48\,MHz of bandwidth \citep[e.g.][]{kondr+2013}.

\subsection{Data acquisition}

After coherent addition of the data streams from all available Core stations (typically 20--23 stations), 
the raw complex-voltage (CV)
data were recorded at 5.12\,$\upmu$s.
The 78-MHz bandwidth is split into 400 channels of 195.312\,kHz each.
The typical length of our observations was 20\,min, although subsequent observations for
monitoring and timing purposes were adjusted depending on the pulsar's signal-to-noise ratio (S/N) in the LOFAR band.

The standard observing setup is summarised in Table~\ref{obssetup}. During the observations the data from the 
stations were streamed for further online processing (beamforming, etc.)
to the Blue Gene/P (BG/P) supercomputer\footnote{After April 18, 2014 (MJD 56765) the new LOFAR GPU correlator/beam-former \emph{Cobalt}
was used for MSP observations.} in Groningen. The output data products (in our case the raw CV data)
were written in HDF5\footnote{\mbox{Hierarchical Data Format 5}, {\tt http://www.hdfgroup.org/HDF5/}} format
to the CEP2 cluster\footnote{LOFAR's 2nd CEntral Processing (CEP2) computer cluster. It will be replaced with the next generation CEP4 cluster
by the start of 2016.} for further offline processing.

\subsection{Pulsar pipeline}

After each observation, we ran the LOFAR PULsar Pipeline (PULP), a Python-based suite of scripts
that provides basic offline pulsar processing (dedispersion and folding) for standard observing modes including CV data. 
PULP collects information about the observation itself and location of the data on the CEP2 nodes and then executes data reduction tools from the
{\tt PRESTO}\footnote{\tt http://www.cv.nrao.edu/\symbol{126}sransom/presto/} \citep{presto},
{\tt dspsr}\footnote{\tt http://dspsr.sourceforge.net} \citep{dspsr} 
and {\tt PSRCHIVE}\footnote{\tt http://psrchive.sourceforge.net} \citep{psrchive} software suites, and 
creates a number of summary diagnostic plots and {\tt PSRFITS}\footnote{\tt http://www.atnf.csiro.au/research/pulsar/index.html? n=Main.Psrfits} 
\citep{psrchive} data files.

To process CV data we used {\tt dspsr} to read raw LOFAR data 
via the LOFAR Direct Access Library (DAL) interface. 
Each channel was
coherently dedispersed and folded with a pulsar ephemeris. 
Where possible, we used the most recent published timing models for folding, but in some cases we used updated 
ephemerides from ongoing timing campaigns that are not yet published.
We set the length of the folding sub-integrations
to be 5\,s and the number of pulse profile bins equal to the closest and lowest power of two for $P/\Delta t$ 
(where $P$ is the MSP
period, and $\Delta t=5.12\,\upmu$s is the sampling interval), but not higher than 1024 bins per period.
The data were combined in frequency using {\tt psradd}. In early observations, we used {\tt paz -r} to perform
zapping of a radio frequency interference (RFI) with a median smoothed automatic
channel zapping algorithm.  In later observations, we used the {\tt clean.py} tool 
from {\tt CoastGuard}\footnote{\tt https://github.com/plazar/coast\_guard} (Lazarus et al., in prep).
This RFI excision routine with the {\tt -F surgical} option
provided much better excision, and for this publication all pipeline products were reprocessed using this tool.
The data were then dedispersed, i.e. appropriate time delays between frequency channels (note that each channel 
was already coherently dedispersed with {\tt dspsr}) were applied, using {\tt pam}.
In the last step we ran
the {\tt pdmp} tool that enables the determination of a best DM and $P$ by optimising the S/N of the profile. 
The output data files were further updated with these best DM and $P$ values.
Diagnostic plots were then produced to enable quick inspection of the data quality.

\section{Results}\label{results}

At present we have observed 75 Galactic field MSPs with the LOFAR HBAs, and detected 48 of them. This is the 
largest sample of MSPs ever observed/detected at these low frequencies.
Of these sources, 25 are recent MSP discoveries and have never before been detected at 110--188\,MHz.
We discuss the detectability of MSPs in more detail in Sect.~\ref{detectability}. The lists of detected and non-detected MSPs are given
in Tables~\ref{msps_summary_detected} and \ref{msps_summary_nondetected}, respectively. For the detected MSPs we list 
the profile significance
of the best individual observation and its S/N, observing epoch/ID, and integration length, $T_\mathrm{obs}$.
For both detected and non-detected
MSPs we also list whether they were previously observed with the WSRT (\citealt{stappers2008}; \citealt{asr+09} for \object{PSR J1023+0038})
or by the Large Phased Array (BSA) in Pushchino at 102/111\,MHz \citep{kl01}. 
From the 27 MSPs that we did not detect, only one, PSR~J0613$-$0200, was observed by \citet{stappers2008} 
and they also did not detect it.
Three out of these 27 have flux density measurements by \citet{kl01} (including PSR J0613$-$0200).  
However, they did not publish the profiles
of these MSPs and their flux density estimates have very large uncertainties.

\subsection{Pulse profiles}\label{profs}

\begin{figure*}[phtb]
\centering
 \includegraphics[scale=0.9]{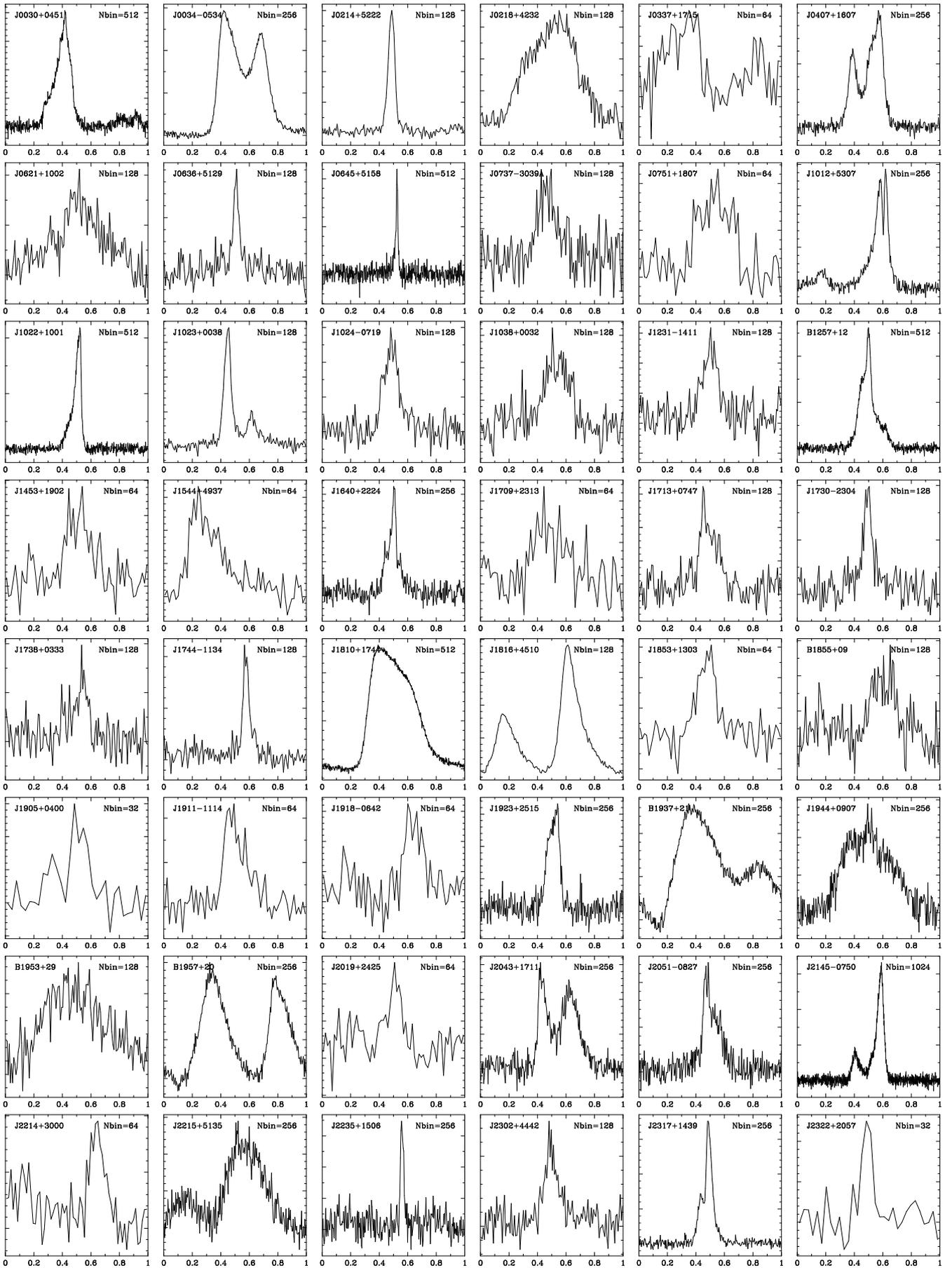}
 \caption{LOFAR total-intensity pulse profiles for all detected MSPs from the best (highest S/N) individual observations.
 The horizontal axes show the pulse phase, and the vertical axes are the flux density in arbitrary units.
 }
 \label{mspprof}
\end{figure*}

We present the best individual profiles for all 48 detected MSPs in Fig.~\ref{mspprof}. 
They are the sum of $\sim 3\times10^5$ spin periods, ranging from about $3\times10^4$ for the 
59.8-ms PSR~J2235+1506 to about $1.3\times10^6$ periods for the 1-h observation of \object{PSR J0337+1715}
\citep[the pulsar triple system;][]{ransom2014}.

Many of the MSPs are detected with high S/N. 
For example, the black widow pulsar J1810+1744, discovered in the 350-MHz GBT Survey of faint {\it Fermi} $\gamma$-ray 
sources \citep{hrm+11}, was detected with S/N $=65$ using a 20-min LOFAR observation. 
Furthermore, the eclipsing pulsar J1816+4510, discovered in the GBNCC pulsar survey \citep{slr+14}, 
was detected with S/N $=159$ using a sum of four 5-min LOFAR observations.
LOFAR profiles for four PSRs J0214+5222, J0636+5129, J0645+5158, and J1816+4510 were also presented in \citet{slr+14}
together with the GBT profiles at 350, 820, and 1500\,MHz.
About 35\% of the detected MSPs show strong narrow profiles, while only 25\% of MSPs have scattered profiles with a clear
exponential tail. The profiles of the remaining MSPs ($\sim40$\%) are weak, which could be due to the
intrinsic spectrum, and/or due to scattering, where the exponential scattering tail may approach or exceed a pulse period.
Later in this Section we discuss the HBA detections of a few individual MSPs, and in Sect.~\ref{comparison} 
we discuss the profiles in general and make a comparison to high frequencies. 
Flux-calibrated, multifrequency LOFAR profiles from this paper are available via the European Pulsar Network (EPN) Database of
Pulsar Profiles\footnote{\tt http://www.epta.eu.org/epndb/} \citep{ljs+98}.

\begin{sidewaystable*}
\caption{List of the 48 detected MSPs.
}\label{msps_summary_detected}
\centering
\begin{tabular}{ld{2.3}d{2.3}ccclccccccc}
\hline\hline
PSR & \multicolumn{1}{c}{$P$} & \multicolumn{1}{c}{DM} & Binary? & WSRT & BSA & LOFAR & Epoch & $T_\mathrm{obs}$\hphantom{\tablefootmark{$\dagger$}} & Prof.\hphantom{\tablefootmark{$\dagger$}} & S/N\tablefootmark{$\ast$}\hphantom{\tablefootmark{$\dagger$}} & LBA & Ref \\
    & \multicolumn{1}{c}{(ms)}   & \multicolumn{1}{c}{(pc\,cm${}^{-3}$)} &  & detected? & detected? & ObsID & (MJD) & (min)\hphantom{\tablefootmark{$\dagger$}} & Sign.\hphantom{\tablefootmark{$\dagger$}} &  & detected? & \\
\hline
J0030+0451 & 4.865 & 4.333 & Isolated & $\ldots$ & y & L83021 & 56304.694 & 20\hphantom{\tablefootmark{$\dagger$}} & 251\hphantom{\tablefootmark{$\dagger$}} & 31\hphantom{\tablefootmark{$\dagger$}} & y & \hphantom{0}1 \\ 
J0034$-$0534 & 1.877 & 13.765 & He WD & y & y & L81272 & 56286.738 & 20\hphantom{\tablefootmark{$\dagger$}} & 543\hphantom{\tablefootmark{$\dagger$}} & 63\hphantom{\tablefootmark{$\dagger$}} & y & \hphantom{0}1 \\ 
J0214+5222 & 24.575 & 22.037 & He WD/sdB? & $\ldots$ & $\ldots$ & L196378 & 56646.791 & 20\tablefootmark{$\dagger$} & 116\tablefootmark{$\dagger$} & 41\tablefootmark{$\dagger$} & $\ldots$ & \hphantom{0}4 \\
J0218+4232 & 2.323 & 61.252 & He WD & n & y & L155442 & 56473.304 & 20\hphantom{\tablefootmark{$\dagger$}} & 126\hphantom{\tablefootmark{$\dagger$}} & 19\hphantom{\tablefootmark{$\dagger$}} & $\ldots$ & \hphantom{0}1 \\ 
\smallskip
J0337+1715 & 2.733 & 21.316 & He WD$+$He WD & $\ldots$ & $\ldots$ & L167133 & 56512.229 & 60\hphantom{\tablefootmark{$\dagger$}} & \hphantom{0}50\hphantom{\tablefootmark{$\dagger$}} & 11\hphantom{\tablefootmark{$\dagger$}} & $\ldots$ & \hphantom{0}2 \\
J0407+1607 & 25.702 & 35.65 & He WD & $\dots$ & $\ldots$ & L227494 & 56790.515 & 20\hphantom{\tablefootmark{$\dagger$}} & 194\hphantom{\tablefootmark{$\dagger$}} & 30\hphantom{\tablefootmark{$\dagger$}} & $\ldots$ & 20 \\
J0621+1002 & 28.854 & 36.601 & CO WD & n & y & L81270 & 56289.023 & 20\hphantom{\tablefootmark{$\dagger$}} & \hphantom{0}47\hphantom{\tablefootmark{$\dagger$}} & \hphantom{0}9\hphantom{\tablefootmark{$\dagger$}} & $\ldots$ & \hphantom{0}3 \\ 
J0636+5129 & 2.869 & 11.107 & BW? & $\ldots$ & $\ldots$ & L196371 & 56648.046 & 10\tablefootmark{$\dagger$} & \hphantom{0}30\tablefootmark{$\dagger$} & 10\tablefootmark{$\dagger$} & $\ldots$ & \hphantom{0}4 \\
J0645+5158 & 8.853 & 18.247 & Isolated & $\ldots$ & $\ldots$ & L85909 & 56322.009 & 20\hphantom{\tablefootmark{$\dagger$}} & \hphantom{0}59\hphantom{\tablefootmark{$\dagger$}} & 19\hphantom{\tablefootmark{$\dagger$}} & $\ldots$ & \hphantom{0}4 \\ 
\smallskip
J0737$-$3039A & 22.699 & 48.920 & PSR & $\ldots$ & $\ldots$ & L85911 & 56321.940 & 60\hphantom{\tablefootmark{$\dagger$}} & \hphantom{0}25\hphantom{\tablefootmark{$\dagger$}} & \hphantom{0}6\hphantom{\tablefootmark{$\dagger$}} & $\ldots$ & \hphantom{0}5 \\ 
J0751+1807 & 3.479 & 30.249 & He WD & $\ldots$ & y & L81051 & 56280.049 & 20\hphantom{\tablefootmark{$\dagger$}} & \hphantom{0}24\hphantom{\tablefootmark{$\dagger$}} & \hphantom{0}7\hphantom{\tablefootmark{$\dagger$}} & $\ldots$ & \hphantom{0}1 \\ 
J1012+5307 & 5.256 & 9.023 & He WD & y & y & L81268 & 56289.149 & 20\hphantom{\tablefootmark{$\dagger$}} & 183\hphantom{\tablefootmark{$\dagger$}} & 32\hphantom{\tablefootmark{$\dagger$}} & n & \hphantom{0}1 \\ 
J1022+1001 & 16.453 & 10.252 & CO WD & y & y & L81254 & 56296.126 & 20\hphantom{\tablefootmark{$\dagger$}} & 274\hphantom{\tablefootmark{$\dagger$}} & 49\hphantom{\tablefootmark{$\dagger$}} & n & \hphantom{0}6 \\ 
J1023+0038 & 1.688 & 14.325 & MS/Redback & y & $\ldots$ & L85233 & 56315.163 & 20\hphantom{\tablefootmark{$\dagger$}} & 105\hphantom{\tablefootmark{$\dagger$}} & 30\hphantom{\tablefootmark{$\dagger$}} & $\ldots$ & \hphantom{0}7 \\ 
\smallskip
J1024$-$0719 & 5.162 & 6.485 & Isolated & n & y & L81049 & 56280.168 & 20\hphantom{\tablefootmark{$\dagger$}} & \hphantom{0}40\hphantom{\tablefootmark{$\dagger$}} & 10\hphantom{\tablefootmark{$\dagger$}} & n & \hphantom{0}1 \\ 
J1038+0032 & 28.852 & 26.59 & Isolated & $\ldots$ & $\ldots$ & L227490 & 56782.804 & 20\hphantom{\tablefootmark{$\dagger$}} & \hphantom{0}39\hphantom{\tablefootmark{$\dagger$}} & \hphantom{0}8\hphantom{\tablefootmark{$\dagger$}} & $\ldots$ & 21 \\
J1231$-$1411 & 3.684 & 8.090 & He WD & $\ldots$ & $\ldots$ & L227492 & 56789.862 & 20\hphantom{\tablefootmark{$\dagger$}} & \hphantom{0}27\hphantom{\tablefootmark{$\dagger$}} & \hphantom{0}7\hphantom{\tablefootmark{$\dagger$}} & $\ldots$ & 17 \\
B1257+12 & 6.219 & 10.166 & Planets & y & y & L81253 & 56296.230 & 20\hphantom{\tablefootmark{$\dagger$}} & 313\hphantom{\tablefootmark{$\dagger$}} & 44\hphantom{\tablefootmark{$\dagger$}} & n & \hphantom{0}8 \\ 
J1453+1902 & 5.792 & 14.049 & Isolated & $\ldots$ & $\ldots$ & L227293 & 56779.988 & 20\hphantom{\tablefootmark{$\dagger$}} & \hphantom{0}18\hphantom{\tablefootmark{$\dagger$}} & \hphantom{0}6\hphantom{\tablefootmark{$\dagger$}} & $\ldots$ & 22 \\
\smallskip
J1544+4937 & 2.159 & 23.226 & BW & $\ldots$ & $\ldots$ & L227294 & 56780.003 & 20\hphantom{\tablefootmark{$\dagger$}} & \hphantom{0}89\hphantom{\tablefootmark{$\dagger$}} & 23\hphantom{\tablefootmark{$\dagger$}} & $\ldots$ & 23 \\
J1640+2224 & 3.163 & 18.426 & He WD & $\ldots$ & y & L81266 & 56289.368 & 20\hphantom{\tablefootmark{$\dagger$}} & \hphantom{0}66\hphantom{\tablefootmark{$\dagger$}} & 15\hphantom{\tablefootmark{$\dagger$}} & $\ldots$ & \hphantom{0}1 \\ 
J1709+2313 & 4.631 & 25.347 & He WD & $\ldots$ & $\ldots$ & L249810 & 56964.533 & 20\hphantom{\tablefootmark{$\dagger$}} & \hphantom{0}14\hphantom{\tablefootmark{$\dagger$}} & \hphantom{0}5\hphantom{\tablefootmark{$\dagger$}} & $\ldots$ & 24 \\
J1713+0747 & 4.570 & 15.992 & He WD & n & y & L149156 & 56465.941 & 60\hphantom{\tablefootmark{$\dagger$}} & \hphantom{0}30\hphantom{\tablefootmark{$\dagger$}} & \hphantom{0}8\hphantom{\tablefootmark{$\dagger$}} & $\ldots$ & \hphantom{0}1 \\ 
J1730$-$2304 & 8.123 & 9.617 & Isolated & $\ldots$ & y & L164998 & 56490.889 & 30\hphantom{\tablefootmark{$\dagger$}} & \hphantom{0}32\hphantom{\tablefootmark{$\dagger$}} & \hphantom{0}9\hphantom{\tablefootmark{$\dagger$}} & $\ldots$ & \hphantom{0}1 \\ 
\hline
\end{tabular}
\tablefoot{
$P$ and DM are from the best ephemerides we used.
Col.~4 shows the binary companion or system type, where `BW' denotes `black widow' eclipsing systems with
ultra-light (UL) companion ($M<0.08M_\sun$), `Redback'~-- compact eclipsing systems with more massive
companion ($M\sim0.1$--$0.2M_\sun$), `MS'~-- main-sequence star companion, `WD'~-- white dwarf
companion, either helium (He), or CO/ONeMg (CO), `sdB'~-- subdwarf B star.
WSRT observations are from \citet{stappers2008} and from \citet{asr+09} in the case of PSR J1023+0038. 
BSA is the Large Phased Array in Pushchino and observations are those reported by \citet{kl01}. The corresponding flux density
measurements are given in Table~\ref{msps_fluxes} for those pulsars detected with LOFAR. 
The LOFAR ObsID, Epoch, $T_\mathrm{obs}$,
S/N and profile significance (Prof. Sign.) are for the best individual observation.
Profile significance is defined as in \citet[p.~167, Eq. 7.1]{handbook}.
The last column gives a reference to the ephemerides used.
\tablefoottext{$\dagger$}{Corresponding value is for the sum of several 5-min observations.}
\tablefoottext{$\ast$}{Given the number of profile bins as in Fig.~\ref{mspprof}.}
}
\tablebib{
(1)~\citet{desvignes2015}; (2)~\citet{ransom2014}; (3)~\citet{sna+02}; (4)~\citet{slr+14}; (5)~M.~Kramer (private communication);
(6)~\citet{vbc+09}; (7)~\citet{asr+09}; (8)~\citet{kw03}; (9)~J.~Hessels (private communication); (10)~\citet{gsf+11};
(11)~\citet{tsb+99}; (12)~\citet{lynch2013}; (13)~\citet{clm+05}; (14)~\citet{aft94}; (15)~\citet{gfc+12}; (16)~\citet{dlk+01};
(17)~\citet{rrc+11}; (18)~\citet{cnt96}; (19)~\citet{cgj+11}; (20)~\citet{lxf+05}; (21)~\citet{bjd+06}; (22)~\citet{lmcs07};
(23)~\citet{brr+13}; (24)~\citet{lwf+04}; (25)~\citet{kbv+13}.
}
\end{sidewaystable*}

\begin{sidewaystable*}
\setcounter{table}{1}
\caption{continued.
}\label{msps_summary_detected1}
\centering
\begin{tabular}{ld{2.3}d{2.3}ccclccccccc}
\hline\hline
PSR & \multicolumn{1}{c}{$P$} & \multicolumn{1}{c}{DM} & Binary? & WSRT & BSA & LOFAR & Epoch & $T_\mathrm{obs}$\hphantom{\tablefootmark{$\dagger$}} & Prof.\hphantom{\tablefootmark{$\dagger$}} & S/N\tablefootmark{$\ast$}\hphantom{\tablefootmark{$\dagger$}} & LBA & \hphantom{00}Ref \\
    & \multicolumn{1}{c}{(ms)}   & \multicolumn{1}{c}{(pc\,cm${}^{-3}$)} &  & detected? & detected? & ObsID & (MJD) & (min)\hphantom{\tablefootmark{$\dagger$}} & Sign.\hphantom{\tablefootmark{$\dagger$}} &  & detected? & \\
\hline
J1738+0333 & 5.850 & 33.778 & He WD & $\ldots$ & $\ldots$ & L124889 & 56399.183 & 20\hphantom{\tablefootmark{$\dagger$}} & \hphantom{0}21\hphantom{\tablefootmark{$\dagger$}} & \hphantom{00}6\hphantom{\tablefootmark{$\dagger$}} & $\ldots$ & \hphantom{4, 2}1 \\ 
J1744$-$1134 & 4.075 & 3.139 & Isolated & y & y & L81264 & 56293.440 & 20\hphantom{\tablefootmark{$\dagger$}} & \hphantom{0}37\hphantom{\tablefootmark{$\dagger$}} & \hphantom{0}15\hphantom{\tablefootmark{$\dagger$}} & $\ldots$ & \hphantom{4, 2}1 \\ 
J1810+1744 & 1.663 & 39.659 & BW & $\ldots$ & $\ldots$ & L81263 & 56293.456 & 20\hphantom{\tablefootmark{$\dagger$}} & 861\hphantom{\tablefootmark{$\dagger$}} & \hphantom{0}65\hphantom{\tablefootmark{$\dagger$}} & n & \hphantom{4, 2}9 \\ 
J1816+4510 & 3.193 & 38.887 & He? WD/sdB? & $\ldots$ & $\ldots$ & L212736 & 56737.245 & 20\tablefootmark{$\dagger$} & 888\tablefootmark{$\dagger$} & 159\tablefootmark{$\dagger$} & $\ldots$ & 4, 25 \\
\smallskip
J1853+1303 & 4.092 & 30.570 & He WD & $\ldots$ & $\ldots$ & L84523 & 56311.438 & 20\hphantom{\tablefootmark{$\dagger$}} & \hphantom{0}24\hphantom{\tablefootmark{$\dagger$}} & \hphantom{00}7\hphantom{\tablefootmark{$\dagger$}} & $\ldots$ & \hphantom{4, }10 \\ 
B1855+09 & 5.362 & 13.300 & He WD & $\ldots$ & y & L131365 & 56417.151 & 20\hphantom{\tablefootmark{$\dagger$}} & \hphantom{0}23\hphantom{\tablefootmark{$\dagger$}} & \hphantom{00}5\hphantom{\tablefootmark{$\dagger$}} & $\ldots$ & \hphantom{4, 2}1 \\ 
J1905+0400 & 3.784 & 25.692 & Isolated & $\ldots$ & $\ldots$ & L249826 & 56964.657 & 20\hphantom{\tablefootmark{$\dagger$}} & \hphantom{0}18\hphantom{\tablefootmark{$\dagger$}} & \hphantom{00}7\hphantom{\tablefootmark{$\dagger$}} & $\ldots$ & \hphantom{4, }10 \\
J1911$-$1114 & 3.626 & 30.975 & He WD & y & y & L81277 & 56286.527 & 20\hphantom{\tablefootmark{$\dagger$}} & \hphantom{0}25\hphantom{\tablefootmark{$\dagger$}} & \hphantom{00}8\hphantom{\tablefootmark{$\dagger$}} & $\ldots$ & \hphantom{4, }11 \\ 
J1918$-$0642 & 7.646 & 26.554 & He WD & $\ldots$ & $\ldots$ & L81276 & 56286.543 & 20\hphantom{\tablefootmark{$\dagger$}} & \hphantom{0}28\hphantom{\tablefootmark{$\dagger$}} & \hphantom{00}7\hphantom{\tablefootmark{$\dagger$}} & $\ldots$ & \hphantom{4, 2}1 \\ 
\smallskip
J1923+2515 & 3.788 & 18.858 & Isolated & $\ldots$ & $\ldots$ & L85594 & 56318.472 & 20\hphantom{\tablefootmark{$\dagger$}} & \hphantom{0}58\hphantom{\tablefootmark{$\dagger$}} & \hphantom{0}12\hphantom{\tablefootmark{$\dagger$}} & $\ldots$ & \hphantom{4, }12 \\ 
B1937+21 & 1.558 & 71.040 & Isolated & $\ldots$ & $\ldots$ & L138647 & 56434.134 & 30\hphantom{\tablefootmark{$\dagger$}} & 275\hphantom{\tablefootmark{$\dagger$}} & \hphantom{0}25\hphantom{\tablefootmark{$\dagger$}} & $\ldots$ & \hphantom{4, 2}1 \\ 
J1944+0907 & 5.185 & 24.34 & Isolated & $\ldots$ & $\ldots$ & L84521 & 56311.472 & 20\hphantom{\tablefootmark{$\dagger$}} & 103\hphantom{\tablefootmark{$\dagger$}} & \hphantom{0}12\hphantom{\tablefootmark{$\dagger$}} & $\ldots$ & \hphantom{4, }13 \\ 
B1953+29 & 6.133 & 104.501 & He WD & $\ldots$ & $\ldots$ & L84522 & 56311.456 & 20\hphantom{\tablefootmark{$\dagger$}} & \hphantom{0}44\hphantom{\tablefootmark{$\dagger$}} & \hphantom{00}6\hphantom{\tablefootmark{$\dagger$}} & $\ldots$ & \hphantom{4, }10 \\ 
B1957+20 & 1.607 & 29.117 & BW & y & $\ldots$ & L81275 & 56286.559 & 20\hphantom{\tablefootmark{$\dagger$}} & 230\hphantom{\tablefootmark{$\dagger$}} & 24\hphantom{\tablefootmark{$\dagger$}} & $\ldots$ & \hphantom{4, }14 \\ 
\smallskip
J2019+2425 & 3.935 & 17.203 & He WD & $\ldots$ & y & L146225 & 56457.103 & 20\hphantom{\tablefootmark{$\dagger$}} & \hphantom{0}18\hphantom{\tablefootmark{$\dagger$}} & \hphantom{00}6\hphantom{\tablefootmark{$\dagger$}} & $\ldots$ & \hphantom{4, 2}1 \\ 
J2043+1711 & 2.380 & 20.710 & He WD & $\ldots$ & $\ldots$ & L84518 & 56311.522 & 20\hphantom{\tablefootmark{$\dagger$}} & \hphantom{0}96\hphantom{\tablefootmark{$\dagger$}} & \hphantom{0}14\hphantom{\tablefootmark{$\dagger$}} & $\ldots$ & \hphantom{4, }15 \\ 
J2051$-$0827 & 4.509 & 20.745 & BW & n & y & L85592 & 56318.504 & 20\hphantom{\tablefootmark{$\dagger$}} & \hphantom{0}65\hphantom{\tablefootmark{$\dagger$}} & \hphantom{0}12\hphantom{\tablefootmark{$\dagger$}} & $\ldots$ & \hphantom{4, }16 \\ 
J2145$-$0750 & 16.052 & 8.998 & CO WD & y & y & L81259 & 56293.607 & 20\hphantom{\tablefootmark{$\dagger$}} & 324\hphantom{\tablefootmark{$\dagger$}} & \hphantom{0}37\hphantom{\tablefootmark{$\dagger$}} & y & \hphantom{4, 2}1 \\ 
J2214+3000 & 3.119 & 22.557 & BW & $\ldots$ & $\ldots$ & L146228 & 56457.159 & 20\hphantom{\tablefootmark{$\dagger$}} & \hphantom{0}27\hphantom{\tablefootmark{$\dagger$}} & \hphantom{00}7\hphantom{\tablefootmark{$\dagger$}} & $\ldots$ & \hphantom{4, }17 \\ 
\smallskip
J2215+5135 & 2.610 & 69.2 & MS/Redback & $\ldots$ & $\ldots$ & L85588 & 56318.567 & 20\hphantom{\tablefootmark{$\dagger$}} & 114\hphantom{\tablefootmark{$\dagger$}} & \hphantom{0}13\hphantom{\tablefootmark{$\dagger$}} & $\ldots$ & \hphantom{4, 2}9 \\ 
J2235+1506 & 59.767 & 18.09 & Isolated & $\ldots$ & y & L168068 & 56521.018 & 30\hphantom{\tablefootmark{$\dagger$}} & \hphantom{0}25\hphantom{\tablefootmark{$\dagger$}} & \hphantom{00}9\hphantom{\tablefootmark{$\dagger$}} & $\ldots$ & \hphantom{4, }18 \\
J2302+4442 & 5.192 & 13.762 & He(?) WD & $\ldots$ & $\ldots$ & L84516 & 56311.603 & 20\hphantom{\tablefootmark{$\dagger$}} & \hphantom{0}40\hphantom{\tablefootmark{$\dagger$}} & \hphantom{0}10\hphantom{\tablefootmark{$\dagger$}} & $\ldots$ & \hphantom{4, }19 \\ 
J2317+1439 & 3.445 & 21.907 & He WD & $\ldots$ & y & L83022 & 56304.635 & 20\hphantom{\tablefootmark{$\dagger$}} & 176\hphantom{\tablefootmark{$\dagger$}} & \hphantom{0}40\hphantom{\tablefootmark{$\dagger$}} & n & \hphantom{4, 2}1 \\ 
J2322+2057 & 4.808 & 13.372 & Isolated & $\ldots$ & y & L146234 & 56460.218 & 20\hphantom{\tablefootmark{$\dagger$}} & \hphantom{0}14\hphantom{\tablefootmark{$\dagger$}} & \hphantom{00}9\hphantom{\tablefootmark{$\dagger$}} & $\ldots$ & \hphantom{4, 2}1 \\ 
\hline
\end{tabular}
\tablefoot{
$P$ and DM are from the best ephemerides we used.
Col.~4 shows the binary companion or system type, where `BW' denotes `black widow' eclipsing systems with
ultra-light (UL) companion ($M<0.08M_\sun$), `Redback'~-- compact eclipsing systems with more massive
companion ($M\sim0.1$--$0.2M_\sun$), `MS'~-- main-sequence star companion, `WD'~-- white dwarf
companion, either helium (He), or CO/ONeMg (CO), `sdB'~-- subdwarf B star.
WSRT observations are from \citet{stappers2008} and from \citet{asr+09} in the case of PSR J1023+0038. 
BSA is the Large Phased Array in Pushchino and observations are those reported by \citet{kl01}. The corresponding flux density
measurements are given in Table~\ref{msps_fluxes} for those pulsars detected with LOFAR. 
The LOFAR ObsID, Epoch, $T_\mathrm{obs}$,
S/N and profile significance (Prof. Sign.) are for the best individual observation.
Profile significance is defined as in \citet[p.~167, Eq. 7.1]{handbook}.
The last column gives a reference to the ephemerides used.
\tablefoottext{$\dagger$}{Corresponding value is for the sum of several 5-min observations.}
\tablefoottext{$\ast$}{Given the number of profile bins as in Fig.~\ref{mspprof}.}
}
\tablebib{
(1)~\citet{desvignes2015}; (2)~\citet{ransom2014}; (3)~\citet{sna+02}; (4)~\citet{slr+14}; (5)~M.~Kramer (private communication);
(6)~\citet{vbc+09}; (7)~\citet{asr+09}; (8)~\citet{kw03}; (9)~J.~Hessels (private communication); (10)~\citet{gsf+11};
(11)~\citet{tsb+99}; (12)~\citet{lynch2013}; (13)~\citet{clm+05}; (14)~\citet{aft94}; (15)~\citet{gfc+12}; (16)~\citet{dlk+01};
(17)~\citet{rrc+11}; (18)~\citet{cnt96}; (19)~\citet{cgj+11}; (20)~\citet{lxf+05}; (21)~\citet{bjd+06}; (22)~\citet{lmcs07};
(23)~\citet{brr+13}; (24)~\citet{lwf+04}; (25)~\citet{kbv+13}.
}
\end{sidewaystable*}

\begin{sidewaystable*}
\vskip 1cm
\caption{List of the 27 non-detected MSPs.
}\label{msps_summary_nondetected}
\centering
\begin{tabular}{ld{2.3}d{2.3}ccclccccc}
\hline\hline
PSR & \multicolumn{1}{c}{$P$} & \multicolumn{1}{c}{DM} & Binary? & WSRT & BSA & LOFAR & Epoch & $T_\mathrm{obs}$ & $S{}_\mathrm{lim}$ & KL01 flux density & \hphantom{000}Ref \\
    & \multicolumn{1}{c}{(ms)} & \multicolumn{1}{c}{(pc\,cm${}^{-3}$)} &  & detected? & detected? & ObsID & (MJD) & (min) & (mJy) & (mJy @ 102\,MHz) & \\
\hline
J0023+0923 & 3.050 & 14.326 & BW & $\ldots$ & $\ldots$ & L84514  & 56311.679 & 20 & \hphantom{00}4 & $\ldots$ & \hphantom{38, 3}1      \\ 
           &       &        &    &          &          & L155443 & 56473.224 & 20 & & &        \\ 
J0340+4130 & 3.299 & 49.575 & Isolated & $\ldots$ & $\ldots$ & L85586  & 56318.785 & 20 & \hphantom{00}3 & $\ldots$ & \hphantom{38, 3}9 \\ 
           &       &        &          &          &          & L155441 & 56473.360 & 20 & & &   \\ 
\smallskip           
J0613$-$0200 & 3.062 & 38.781 & He WD & n & y & L81271  & 56289.007 & 20 & \hphantom{00}5 & $240\pm100$ & \hphantom{38, 3}1 \\ 
             &       &        &       &   &   & L155440 & 56473.467 & 20 &  & & \\ 
J1327$-$0755 & 2.678 & 27.912 & He WD & $\ldots$ & $\ldots$ & L243060 & 56907.580 & 20 & \hphantom{00}5 & $\ldots$ & \hphantom{38, }29 \\
J1400$-$1438\tablefootmark{$\ast$} & 3.084 & 4.928 & He WD & $\ldots$ & $\ldots$ & L173398 & 56533.626 & 20 & \hphantom{0}10 & $\ldots$ & \hphantom{38, }42 \\
J1455$-$3330 & 7.987 & 13.562 & He WD & $\ldots$ & $\ldots$ & L85908 & 56322.243 & 20 & \hphantom{0}60 & $\ldots$ & \hphantom{38, 3}1 \\ 
\smallskip
J1551$-$0658 & 7.09  & 21.6 & Binary? & $\ldots$ & $\ldots$ & L227493 & 56789.992 & 20 & \hphantom{00}7 & $\ldots$ & \hphantom{38, }30 \\
J1600$-$3053 & 3.598 & 52.317 & He WD & $\ldots$ & $\ldots$ & L146215 & 56456.919 & 20 & \hphantom{0}70 & $\ldots$ & \hphantom{38, 3}1 \\ 
J1614$-$2230 & 3.151 & 34.487 & CO WD & $\ldots$ & $\ldots$ & L85907 & 56322.269 & 20 & \hphantom{0}25 & $\ldots$ & \hphantom{38, }25 \\ 
B1620$-$26 & 11.076 & 62.865 & He WD$+$UL & $\ldots$ & y & L85906 & 56322.285 & 20 & \hphantom{0}40 & \hphantom{\tablefootmark{\S}}$450\pm200$\tablefootmark{\S} & \hphantom{38, 3}1 \\ 
J1643$-$1224 & 4.622 & 62.377 & He WD & $\ldots$ & $\ldots$ & L81278 & 56286.424 & 20 & \hphantom{0}15 & $\ldots$ & \hphantom{38, 3}1 \\ 
\smallskip
             &       &        &       &          &          & L85231 & 56315.324 & 20 &  & & \\ 
J1646$-$2142 & 5.85 & 29.8 & Binary & $\ldots$ & $\ldots$ & L227491 & 56787.043 & 20 & \hphantom{0}30 & $\ldots$ & \hphantom{38, }31 \\
J1719$-$1438 & 5.79 & 36.862 & UL/Carbon WD? & $\ldots$ & $\ldots$ & L249812 & 56964.547 & 20 & \hphantom{0}15 & $\ldots$ & 32, 33 \\
J1721$-$2457 & 3.497 & 47.758 & Isolated & $\ldots$ & $\ldots$ & L249814 & 56964.562 & 20 & \hphantom{0}45 & $\ldots$ & 34, 35 \\
J1741+1351 & 3.747 & 24.0 & He WD & $\ldots$ & $\ldots$ & L81281 & 56281.492 & 20 & \hphantom{00}4 & $\ldots$ & \hphantom{38, }26 \\ 
\smallskip
           &       &      &       &          &          & L83025 & 56304.456 & 20 &  & & \\

J1745+1017 & 2.652 & 23.970 & BW? & $\ldots$ & $\ldots$ & L249816 & 56964.576 & 20 & \hphantom{00}4 & $\ldots$ & \hphantom{38, }36 \\
J1751$-$2857 & 3.915 & 42.808 & He WD & $\ldots$ & $\ldots$ & L249818 & 56964.591 & 20 & 220 & $\ldots$ & \hphantom{38, }37 \\
J1804$-$2717 & 9.343 & 24.674 & He WD & $\ldots$ & $\ldots$ & L249820 & 56964.606 & 20 & 110 & $\ldots$ & 38, 39 \\
J1828+0625 & 3.63 & 22.4 & Binary & $\ldots$ & $\ldots$ & L249822 & 56964.620 & 20 & \hphantom{00}6 & $\ldots$ & \hphantom{38, }31 \\
\smallskip
J1832$-$0836 & 2.719 & 28.18 & Isolated & $\ldots$ & $\ldots$ & L249824 & 56964.635 & 20 & \hphantom{0}40 & $\ldots$ & \hphantom{38, }40 \\
J1903+0327 & 2.150 & 297.548 & MS & $\ldots$ & $\ldots$ & L85596 & 56318.440 & 20 & \hphantom{0}20 & $\ldots$ & \hphantom{38, 3}1 \\ 
J1910+1256 & 4.984 & 38.097 & He WD & $\ldots$ & $\ldots$ & L85595 & 56318.456 & 20 & \hphantom{0}10 & $\ldots$ & \hphantom{38, 3}1 \\ 
J1911+1347 & 4.626 & 30.99 & Isolated & $\ldots$ & $\ldots$ & L243072 & 56907.790 & 20 & \hphantom{00}5 & $\ldots$ & \hphantom{38, }41 \\
J1949+3106 & 13.138 & 164.126 & CO WD & $\ldots$ & $\ldots$ & L85593 & 56318.487 & 20 & \hphantom{00}4 & $\ldots$ & \hphantom{38, }27 \\ 
\smallskip
J2010$-$1323 & 5.223 & 22.181 & Isolated & $\ldots$ & $\ldots$ & L84520 & 56311.490 & 20 & \hphantom{0}15 & $\ldots$ & \hphantom{38, 3}1 \\ 
J2017+0603 & 2.896 & 23.918 & He WD & $\ldots$ & $\ldots$ & L85591 & 56318.520 & 20 & \hphantom{00}5 & $\ldots$ & \hphantom{38, }19 \\ 
J2229+2643 & 2.978 & 23.019 & He WD & $\ldots$ & y & L84517 & 56311.587 & 20 & \hphantom{00}3 & \hphantom{0}$60\pm30$\hphantom{0} & \hphantom{38, }28 \\ 
\hline
\end{tabular}
\tablefoot{
Column descriptions are the same as in Table~\ref{msps_summary_detected}.
$S{}_\mathrm{lim}$ is the upper limit on the mean flux density and taken as three times the nominal error 
of the mean flux density measurements. Flux densities by \citet[KL01]{kl01} are at 102 or 111\,MHz.
Some of the pulsars were attempted twice.
\tablefoottext{\S}{Observations were done both at 102 and 111\,MHz, and it is not clear for which frequency the 
flux densities were reported.}
\tablefoottext{$\ast$}{The catalogued position of PSR J1400$-$1438 was about $6\arcmin$ off from the position derived from over three years
of timing with the GBT (J.~Swiggum, private communication). We detected this MSP in a recent LOFAR observation using the updated
coordinates. The average profile, effective pulse width and flux density measurement for this MSP will be presented elsewhere.}
}
\tablebib{
(1)~\citet{desvignes2015}; (9)~J.~Hessels (private communication); (19)~\citet{cgj+11}; (25)~\citet{dpr+10}; (26)~\citet{jbo+07}; 
(27)~\citet{dfc+12}; (28)~\citet{wdk+00}; (29)~\citet{boyles2013}; (30)~\citet{hrm+11}; (31)~\citet{rap+12}; (32)~\citet{nbb+14};
(33)~\citet{bbb+11a}; (34)~\citet{eb01b}; (35)~\citet{jsb+10}; (36)~\citet{bgc+13}; (37)~\citet{sfl+05}; (38)~\citet{hlk+04};
(39)~\citet{llb+96}; (40)~\citet{bbb+13}; (41)~\citet{lfl+06}; (42)~\citet{rsm+13}.
}
\end{sidewaystable*}

\subsubsection{Folding and $v_\mathrm{orb}/c$ caution}\label{ranseff}

For binary MSPs, the low observing frequencies and large fractional bandwidth of the LOFAR data mean that special care 
is needed when folding the data. This is because,
in standard analyses, phase-folding is done based on a timing
model evaluated at the central frequency of the band.
If the dispersion delay across the band is large enough for the orbital phase to change
significantly during this time, then data can be assigned an incorrect orbital phase.
In such a case, the monochromatic timing model will not be able to
correctly predict the pulse period across the entire bandwidth, but
will have a frequency-dependent bias that changes with orbital phase \citep{ransom_effect}.
Given that the error introduced in the pulse
times-of-arrival (TOAs) is of the order $v_\mathrm{orb}/c \times \Delta t_\mathrm{DM}$, where $\Delta t_\mathrm{DM}$ is
the dispersion delay between the central frequency and the lowest
frequency in the band, it is clear that this effect can most easily be
countered by splitting up the band into narrower sub-bands, and then applying the timing model separately to each of these.
Alternatively, two-dimensional phase predictors can be employed.

Without using two-dimensional phase predictors the actual error in TOAs will depend
on the orbital phase of the pulsar and the projection of the orbit on the plane of the sky, and is largest
for edge-on orbits when the pulsar is at periastron -- i.e. when the pulsar's radial acceleration
is strongest. In general, the effect is larger for higher-DM pulsars in compact (short orbital period) binary systems. 
In our MSP sample, the double pulsar J0737$-$3039A has the largest value of $v_\mathrm{orb}/c\sim 10^{-3}$. For pulsars
J0621+1002, J2145$-$0750, J0751+1807, J1023+0038, J1022+1001, J2215+5135, and J1816+4510, it is about (1--2)$\times 10^{-4}$, and 
for other binaries it is of the order of $10^{-5}$ or less. 
Using a single `polyco' file (a file generated by the pulsar timing program {\tt Tempo}\footnote{\tt http://tempo.sourceforge.net/}
which contains a polynomial description of the apparent pulse phase as a function of time)
applicable to the central frequency
of the total band, would result in TOA errors of 8.4\,ms or $0.37$ of the spin phase at the lowest observed frequency of 110\,MHz for the pulsar
J0737$-$3039A. In our observations we normally split the total bandwidth into 20 parts and then
process them separately, each with their own polyco files. This helps to mitigate the effect, giving a maximum TOA error
for the lowest-frequency part on the order of $\sim 10\,\upmu$s for most MSPs, with the most problematic pulsars being J0737$-$3039A (0.64\,ms 
residual smearing),
J2215+5135 (0.16\,ms), J1816+4510 ($55\,\upmu$s), J0218+4232 ($52\,\upmu$s), J0621+1002 ($46\,\upmu$s), and J0751+1807 ($39$\,$\upmu$s). 
The largest error in terms of spin phase occurs for \object{PSR J2215+5135}, where the effect is equivalent to 0.06 rotational cycles.
For most of our observed MSPs, this error is about 0.01 cycles or less.

\subsubsection{Double pulsar J0737$-$3039A}

Despite its low declination and maximum elevation of $6.4\degr$ (as seen from the LOFAR Core), we detected the 22-ms A-pulsar in the 
double pulsar system J0737$-$3039 in a single 1-h observation (see Fig.~\ref{mspprof}).
To exclude any broadening due to the effect described in Sect.~\ref{ranseff} we have also reprocessed
the data, folding each channel separately. It did not make any noticeable difference to the profile, because at the epoch of observation
the pulsar had a relatively low orbital velocity (orbital phase $\sim$0--0.4, longitude of periastron $\sim 300.5\degr$).
From high-frequency observations it is known that the B-pulsar
is not visible anymore due to large orbital \citep[17\,$\degr$/yr;][]{ksmm+06} and geodetic \citep[5\,$\degr$/yr;][]{bkkm+08} 
precession, but is expected to reappear again in $\sim2035$ \citep{pmks+10}.
However, we might be able to catch its emission at our low frequencies with LOFAR, if a wider low-frequency beam \citep{cordes1978} 
is still grazing our line-of-sight. Thus, we folded our data with the B-pulsar
ephemeris as well, but we did not detect it after also searching in period and DM. 
Radio emission from the B-pulsar was only ever visible at two narrow orbital 
phase windows ($\sim 0.58$ and $\sim 0.78$) when emission from the A-pulsar affected the magnetosphere of the B-pulsar \citep{lbkp+04},
and we missed these windows in our observation.
Therefore, more observations are needed to rule out emission from the B-pulsar at low frequencies.

\subsubsection{PSR J1023+0038}

We detected the 1.69-ms pulsar J1023+0038, which is in
a 4.7-h orbit with a bloated stellar companion and has been called
the `missing link' between low-mass X-ray binaries (LMXBs) and radio MSPs \citep{asr+09}. We observed it
five times, with 20-min integration times per observation, before February 2013 
and detected it twice in the orbital phase ranges
0.64--0.71 and 0.84--0.91.  The pulsar was very likely eclipsed in the other three observations 
\citep[at the observed frequencies, it is expected to be eclipsed for at least orbital phases 0.0--0.5, see][]{akh+13}.  
The profile from the best
observation on January 23, 2013 is shown in Fig.~\ref{mspprof}. The profile has a two-component shape,
as seen in WSRT observations at 155--157\,MHz by \citet{asr+09}. Two observations on August 31, 2013 
and October 5, 2013 at non-eclipse orbital phases did not show any emission from PSR~J1023+0038. This was confirmed
earlier with high-frequency observations using other radio telescopes. This is evidence that
it switched again into its accreting LMXB phase \citep{patruno2013,sah+14}.

\begin{figure}[tbh]
\centering
\includegraphics[scale=0.8,trim=0 0 0 0,clip]{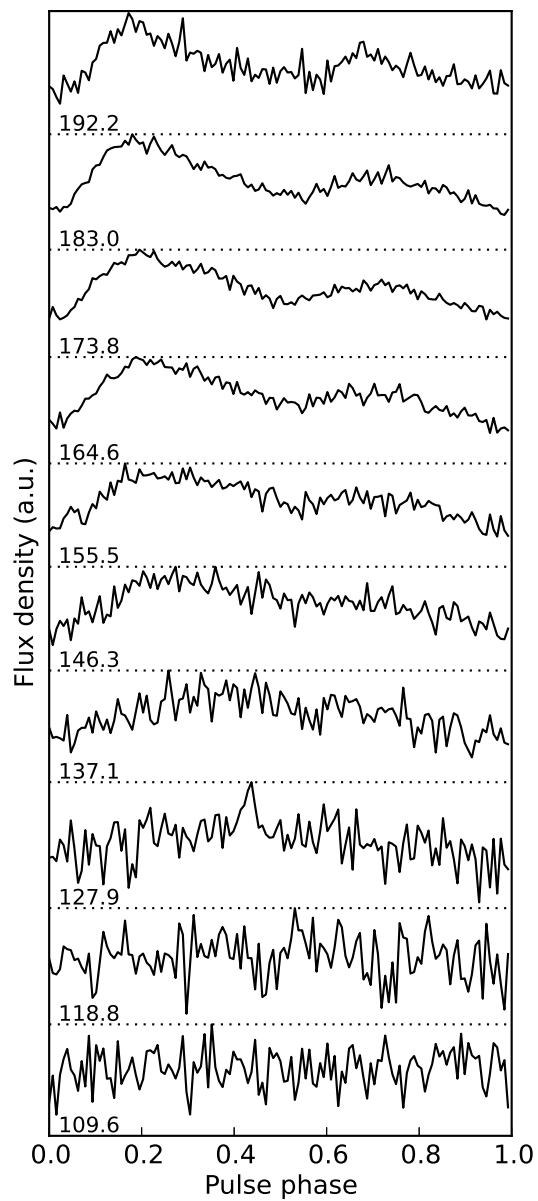}
 \caption{Profile evolution of pulsar B1937+21 within the LOFAR HBA band 
 in the range 105--197\,MHz. Each profile
 has 128 bins and is folded in $\sim9.2$-MHz wide sub-bands centred at the labelled frequency in MHz.}
 \label{b1937prof}
\end{figure}

\subsubsection{\object{PSR B1937+21}}\label{1937-prof-evol}

We successfully detected the first millisecond pulsar, B1937+21, in the LOFAR HBA band. This is the first
ever published detection of the pulse profile of this pulsar at these low frequencies (see Fig.~\ref{b1937prof}). The main pulse (MP) and
interpulse (IP) components are clearly distinguished at the highest frequencies.
Figure~\ref{b1937prof} shows the profile evolution in the HBA band from 105 to 197\,MHz. 
The rapid change in the profile shape due to scattering is evident.
Below 150\,MHz the IP is no longer visible
and below 120\,MHz the whole profile is completely washed out by scattering. 
As mentioned before, scattering time, $\tau_\mathrm{scat}$, scales as $f^{-4.4}$ with frequency, $f$. Following
the empirical relationship between $\tau_\mathrm{scat}$, DM and $f$ from \citet{bhat2004}, the $\tau_\mathrm{scat}$ 
should increase from $\sim 2$ to $\sim 19$\,ms when going from 197 to 105\,MHz,
i.e. from $\sim 1$ to almost $\sim 12$ spin 
periods ($P = 1.56$\,ms). \citet{jk09} reported on the IP giant-pulse component in the average profile from their observations
at a frequency of 238\,MHz with the Giant Metrewave Radio Telescope (GMRT). We do not see this giant-pulse-related
hump in the average profile of either the MP or the IP components, 
most likely because of much larger profile broadening. 
\citet{kuzmin-1937} reported the detection of four individual giant pulses at 112\,MHz,
though the pulse trains they present are very noisy.
We have also  performed a search for strong giant pulses in the full band, but did not find anything significant. Only searching 
the high-frequency half of the band or using the cyclic spectroscopy technique \citep{demorest2011,wds13,akhs14} may be needed.
We defer more detailed single-pulse study of PSR~B1937+21 for a subsequent paper.

\subsection{Flux densities}\label{fluxes}

\subsubsection{Contributing factors}\label{flux_definition}

For a number of reasons, flux calibration of beamformed LOFAR data is non-trivial.
The main uncertainty comes from the fact that the characteristic beam shape and sensitivity strongly depend on the elevation
and azimuth of the source. LOFAR also has a very large fractional bandwidth, thus the frequency dependence of the beam size and 
differences in the sky temperature across the bandpass should be considered. 
Overall, the following aspects have to be taken into account:

\begin{description}
 
 \item{\bf Effective area.} The total effective collecting area of the LOFAR array, $A_\mathrm{eff}$, or gain, $G$, 
 is frequency dependent. For a 48-tile HBA station the upper limit on the effective area, $a_\mathrm{eff}^\mathrm{max}$($\lambda$),
 is given by \citet{haarlem2013}\footnote{\tt http://www.astron.nl/radio-observatory/astronomers/ lofar-imaging-capabilities-sensitivity/sensitivity- lofar-array/sensiti}:
 \begin{equation}
  a_\mathrm{eff}^\mathrm{max}(\lambda) = \mathrm{min}\left\{256\lambda^2, 1200\right\} \mathrm{m}^2,
 \end{equation}
 where $\lambda$ is the observing wavelength in m.

 \item{\bf Beam model.} An accurate beam model of the LOFAR HBAs at 110--190\,MHz provides the dependence of the antenna gain, $G$, 
 on the zenith angle and azimuth for a single tile or station, and preferably takes into account the shadowing of individual
 tiles within a station at low elevations. 

 \item{\bf System temperature.} The system temperature, $T_\mathrm{sys} = T_\mathrm{sky} + T_\mathrm{A}$, 
 is also frequency dependent. The main contribution to $T_\mathrm{sys}$ comes
 from the sky temperature, $T_\mathrm{sky}$, which strongly changes with frequency as $f^{-2.55}$ \citep{lmop87}.
 We should note, that, although the spectral index is remarkably uniform for most of the Galactic emission, it does vary
 across the sky in the range from $-2.40$ to $-2.55$ \citep{rcls94}.
 The antenna temperature, $T_\mathrm{A}$, is not constant either. At the centre of the band its
 average value is about 400\,K \citep{wijnholds2013}, although it increases slightly towards higher frequencies \citep{haarlem2013}.
 Below 120\,MHz and above 180\,MHz $T_\mathrm{A}$ increases rapidly. We used a sixth-order polynomial
 function of frequency based on $T_\mathrm{sys}$ measurements from \citet{wc11} to fit for $T_\mathrm{A}$ and calculate its value 
 for each frequency channel. 

 \item{\bf Broken tiles.} In practice, not all the HBA tiles are operational at all times, 
 and on average about 5--10\% of them are unused while observing, making the effective area smaller.
 For some of the earlier observations this fraction was as high as 30\%.
  
  \item{\bf Coherent station summation.} Coherent summation of the stations should increase the sensitivity 
  linearly with the number of stations added. 
In practise, however, we found the dependence of the S/N to scale as $N_\mathrm{s}^{\gamma}$, where
$N_\mathrm{s}$ is the number of 48-tile stations, and $\gamma = 0.85$ is the coherence factor. 
The possible reasons for this non-linear dependence are correlated noise between stations and uncalibrated non-zero
phase offsets in the phase-frequency dependence for each of the polarisations/stations, which can be caused by the station's
electronics themselves or the influence of the ionosphere. This measurement was based on the BG/P data, and we used BG/P
for most of our observations. Similar tests are currently being done for the \emph{Cobalt} correlator, where
these non-zero phase offsets will be taken into account.

 \item{\bf Radio frequency interference.} RFI clipping reduces the total on-source time and bandwidth. 
 On average, 25--30\% of our data were clipped, the smallest and largest fractions being 10\% and 46\%, respectively.
 This large zapping fraction is an unfortunate consequence of keeping a frequency resolution of only 195\,kHz (this is needed 
 in order to achieve the desired time resolution).
 The typical spectral width of the RFI seen by LOFAR is $\sim 1$\,kHz, and thus higher-frequency-resolution data
 can be excised of RFI without sacrificing such a large fraction of the data.

\end{description}

\subsubsection{Calibration}\label{calib_sect}

The Stokes vector, $\vec{S}_0 = (I_0, Q_0, U_0, V_0)$, of the incident radiation is related to the measured
Stokes vector, $\vec{S} = (I, Q, U, V)$, via a $4\times4$ Mueller matrix, $M$, as $\vec{S} = M\cdot\vec{S}_0$.  The measured Stokes $I$
component will depend on the initial Stokes $\vec{S}_0$ as $I = m_{II}I_0 + m_{IQ}Q_0 + m_{IU}U_0 + m_{IV}V_0$, where 
$m_{ij}$ are corresponding elements of $M$. In our observations $m_{II}$ is the dominant factor and we can assume
that all observed power comes in Stokes $I$. Pulsars are predominantly very weakly circularly polarised, thus $m_{IV}$ is small
compared to $m_{II}$ ($\lesssim 1$\%). For linear polarisation, $m_{IQ}$ and $m_{IU}$ are of the order 5--10\%
and strongly depend on the hour angle \citep{nsk+15}. \citet{nsk+15} also report significant leakage to
circular polarisation at large hour angles, however all our observations were carried out near transit where the leakage
is minimal. Moreover, Faraday rotation from the ISM removes most of the contribution from linear polarisation
in our large observing band. Therefore, we can simply consider $I\approx m_{II}I_0$.

In our flux density measurements we used the Hamaker beam model \citep[and references therein]{hamaker06},
which is also used in the LOFAR software for initial phase calibration of interferometric imaging data \citep{haarlem2013}.
Based on this model we calculated the Jones matrices of antenna response for a given HBA station, frequency and 
sky direction using the {\tt mscorpol} package\footnote{\tt https://github.com/2baOrNot2ba/mscorpol}. Then, Jones
matrices \citep[e.g.][]{jones1941, fymat1971} can be transformed to a corresponding Mueller matrix, and in particular for the $m_{II}$:
\begin{equation}
m_{II}(f,z,A) = \frac{1}{2}\left(J_{xx}J_{xx}^{*} + J_{xy}J_{xy}^{*} + J_{yx}J_{yx}^{*} + J_{yy}J_{yy}^{*}\right),
\end{equation}
where $J_{ij}$ are components of the Jones matrices, $f$ is the frequency, $z$ is the zenith angle, and $A$ is
the azimuth \citep[e.g.][]{hbs96}. Finally, to link with the absolute flux scale 
we normalised our $m_{II}(f,z,A)$ by the corresponding values for a reference observation of Cassiopeia A 
by \citet{wc11}. In calculations of Jones matrices we always used the station CS001 as our template. The difference in
flux densities when other stations are used to derive Jones matrices is much smaller than the nominal flux uncertainty.
Using $m_{II}$, the $a_\mathrm{eff}$ for a given frequency and direction can be simply calculated as:
\begin{equation}
 a_\mathrm{eff}(f,z,A) = a_\mathrm{eff}^\mathrm{max}\times m_{II}(f,z,A).
\end{equation}

To determine the pulse-phase-averaged flux density of the pulsar (hereafter referred to as `mean
flux density'), we adapt Equation (7.10) of \citet{handbook} to obtain the flux density, $S_i$, of the
\emph{i}-th profile bin:

\begin{equation}
S_i = (\mathrm{S/N})_i\frac{\beta T_\mathrm{sys}}{G\sqrt{n_\mathrm{p(\mathit{T}_\mathrm{obs}/\mathit{N}_\mathrm{b})\mathit{B}}}}~,
\end{equation}
where $\beta$ is a digitisation correction factor, $(\mathrm{S/N})_i$  is the signal-to-noise ratio of the \emph{i}-th profile bin,
$n_\mathrm{p} = 2$ is the number of polarisations summed, $T_\mathrm{obs}$ is the observation length, $B$ is the observation bandwidth, 
$N_\mathrm{b}$ is the number of profile bins, $G = A_\mathrm{eff}/2k$ is the telescope gain, $A_\mathrm{eff}$ is the total
effective area, $k$ is the Boltzmann constant, and $T_\mathrm{sys}$ is the system
temperature. For systems with a large number of bits per sample (8 in our case) $\beta\approx 1$.

Expanding this expression to apply to LOFAR, for the flux density $S_i(f,z,A)$ in Jy of the \emph{i}-th profile bin of a given
frequency channel and sub-integration, we derive:
\begin{align}
S_i(f,z,A)=\frac{2\beta k(\mathrm{S/N})_i^\mathrm{f,z,A}[T_\mathrm{A}(f) + T_\mathrm{sky}(f,l,b)]}{N_\mathrm{s}^{\gamma} a_\mathrm{eff}(f,z,A)m_{II}(f,z,A)[1-\xi]} \notag \\
\times\frac{1}{\sqrt{n_\mathrm{p}[1-\zeta(f)](\frac{T_\mathrm{int}}{N_\mathrm{b}})\Delta f}}~,
\end{align}
where $l$ and $b$ are Galactic longitude and latitude, 
$\xi$ is the overall fraction of bad tiles, 
$\zeta(f)$ is the RFI fraction, $T_\mathrm{int}$ is the length of a sub-integration,
$\Delta f$ is the width of the frequency channel, 
and $(\mathrm{S/N})_i^\mathrm{f,z,A} = (X_i^\mathrm{f,z,A} - \langle X\rangle)/\sigma_\mathrm{X}$ 
is the signal-to-noise ratio of the \emph{i}-th 
profile bin of the individual frequency channel and sub-integration, with $X_i^\mathrm{f,z,A}$ the value of the \emph{i}-th
profile bin, $\langle X\rangle$ the mean and $\sigma_\mathrm{X}$ the rms value (and both $\langle X\rangle$ and
$\sigma_\mathrm{X}$ are calculated in the off-pulse window).
To calculate $T_\mathrm{sky}$ we use
the reference sky temperature at 408\,MHz from the skymap determined by \citet{haslam1982} and scale it to the frequency
of the channel using $f^{-2.55}$ \citep{lmop87}. At present there are few sophisticated sky models available \citep{gsm2008,lfmap}, which are based
on published reference sky maps, and that we can use in the future especially for the calibration in the LBA
frequency range 10--90\,MHz \citep[see also][]{nelles2015}.

Finally, to measure the mean flux density $S_\mathrm{mean}$ we find the average value of $S_i(f,z,A)$ as:
\begin{eqnarray}
S_\mathrm{mean} = \frac{1}{M N_\mathrm{b}}\sum_{i=1}^{N_\mathrm{b}}\sum_{j=1}^{M}S_i(f(j),z(j),A(j))~,
\end{eqnarray}
where $M = N_\mathrm{ch} N_\mathrm{int} - N_\mathrm{zap}$ with $N_\mathrm{ch}$ the total number of frequency
channels, $N_\mathrm{int}$ the total number of sub-integrations, and $N_\mathrm{zap}$ the number of completely
zapped frequency channels and sub-integrations.

We estimated the nominal uncertainty, $\sigma_i(f,z,A)$, on $S_i(f,z,A)$ by taking $(\mathrm{S/N})_i^\mathrm{f,z,A} = 1$ for every
individual channel and sub-integration, i.e. $\sigma_i(f,z,A) = S_i(f,z,A)|_{(\mathrm{S/N})_i^\mathrm{f,z,A} = 1}$.
Then, we find the nominal uncertainty, $\sigma_\mathrm{mean}$, on the mean flux density, $S_\mathrm{mean}$, as:
\begin{eqnarray}
 \sigma_\mathrm{mean}=\frac{1}{M\sqrt{N_\mathrm{b}}}\left.\sqrt{\sum_{j=1}^{M}\sigma^2_i(f(j),z(j),A(j))}\right.~.
 \label{sigmamean}
\end{eqnarray}

However, there are also other factors that can affect our flux measurements, but which are very difficult (or not possible) to take
account of, and were not addressed here. 
For completeness we note that these are:

\begin{description}

 \item{\bf Scattering.} Our flux density measurements can be seriously affected if $\tau_\mathrm{scat}$ is close to or
 exceeds a pulse period.
 In these cases, the measured S/N is reduced and we underestimate $T_\mathrm{sys}$ due to 
 the contribution from the scattered pulse signal.
 
 \item{\bf Refractive scintillation.} Diffractive scintillation should not affect our measurements because 
 the decorrelation bandwidths are lower than the width of a channel (195\,kHz) for all of our sources, so many
 scintles are averaged out. On the other hand, refractive interstellar scintillation (RISS) can change the 
 pulsar flux density by a factor of $\sim 1.5$ at low observing frequencies. 
 \citet{grc93} found
 the characteristic RISS time scale to vary from a few days to several weeks at a frequency of 74\,MHz for a sample
 of nearby pulsars. The measured modulation indices (ratio of the standard deviation of the observed
 flux densities to their mean) ranged from 0.15--0.45. The RISS time scale and modulation
 index scale with the observing frequency as $\propto f^{-2.2}$ and $\propto f^{0.57}$, respectively \citep{rcb84,crcf87}.
 Observations spanning dozens of months are required to measure the average flux densities more accurately.
 Here we provide only the flux density measurements for the best individual observation for each pulsar.
 
 \item{\bf Beam jitter.} The jitter of the tied-array beam is caused by the ionosphere and 
 can be as large as $\sim\!\!2$--$3\arcmin$, equal to the
half-width at half-maximum of the HBA full-core tied-array 
beam\footnote{For a demonstration, see: \tt http://www.astron.nl/dailyimage/ index.html?main.php?date=20140123},
but is usually much smaller than this.  However,
when the ionosphere is excited, e.g. around sunset/sunrise, it could move the source to the edge of the beam leading to inferred
flux densities lower by a factor of $\sim 2$. In future analyses forming a ring of tied-array beams around the pointing position 
could eliminate this
problem by choosing the beam with highest S/N in the post-processing. This, however, is expensive with respect to both 
data volume and processing time.

 \item{\bf Galactic plane.} Variation of the system temperature with time due to rise/set of the Galactic plane 
 can increase $T_\mathrm{sys}$ by 30--40\% if it is in the field-of-view (FoV) of the primary station beam or grating lobes.
 This is also possible for other strong background sources in the FoV.
 Also, $T_\mathrm{sys}$ can vary with pointing direction due to noise coupling effects.
 
\end{description}

To get realistic values for the systematic uncertainty on our flux density measurements, we also estimated
the flux density using images from the Multifrequency Snapshot Sky Survey \citep[MSSS,][]{heald_msss} for a 
sample of bright MSPs and slow pulsars that we could unambiguously identify. On average our mean flux density
measurements agree with those from MSSS data within $\sim40$\%. 
The same calibration technique was used
to get several monthly flux density measurements for ten bright non-millisecond pulsars in the LOFAR HBA 
range (Bilous et al. 2015, in prep.). Comparing measured fluxes to spectra fit through the literature
values, the authors estimate the uncertainty of a single LOFAR flux measurement as 50\%.
Thus, MSP fluxes reported here should also be used with this uncertainty.
  
\begin{sidewaystable*}
\caption{
Mean flux densities for the detected MSPs in the frequency range 110--188\,MHz. 
}\label{msps_fluxes}
\centering
\begin{tabular}{ld{2.3}ccccccclc}
\hline\hline
PSR & \multicolumn{1}{c}{DM} & $W_\mathrm{eff}$\hphantom{\tablefootmark{$\star$}}\hphantom{0} & $\delta$ & $S_\mathrm{mean}$ & KL01 flux density      & Predicted $S_\mathrm{mean}$          & $S_\mathrm{ref}$ & $\nu_\mathrm{ref}$ & \hphantom{$\sim$~\!00}Spectral & Reference \\
    & \multicolumn{1}{c}{(pc\,cm${}^{-3}$)}  & (ms)\hphantom{\tablefootmark{$\star$}}\hphantom{0}             & (\%)     & (mJy)               & (mJy @ 102\,MHz)       & (mJy) & (mJy)    & (MHz)          & \hphantom{$\sim$~\!000}index    & \\
\hline
J0030+0451 & 4.333 & 0.61(1)\tablefootmark{$\star$} & 12.6(2)\tablefootmark{$\star$} & \hphantom{0}99(1)\hphantom{.0} & \hphantom{\tablefootmark{$\dagger$}}$380\pm200$\tablefootmark{$\dagger$} & $80\pm40$ & $7.9\pm0.2$ & \hphantom{0}430 & \hphantom{$\sim$~\!}$-2.15^{+0.25}_{-0.52}$ & \hphantom{0}1, 38 \\
J0034$-$0534 & 13.765 & 0.54(1)\tablefootmark{$\star$} & 28.8(4)\tablefootmark{$\star$} & 491(2)\hphantom{.0} & $250\pm120$ & $400\pm100$ & $17\pm5$\hphantom{0} & \hphantom{0}436 & \hphantom{$\sim$~\!}$-2.64\pm0.05$ & \hphantom{0}2, 38 \\
J0214+5222 & 22.037 & 1.5(2)\tablefootmark{$\star$}\hphantom{0} & \hphantom{0}6.1(8)\tablefootmark{$\star$} & \hphantom{0}21.0(7) & $\ldots$ & \hphantom{0}4.1 &  \hphantom{0}0.90 & \hphantom{0}350 & \hphantom{$\sim$~\!}$-1.6$\hphantom{$0\pm0.00$} &  \hphantom{00, }32 \\
J0218+4232 & 61.252 & 0.77(2)\hphantom{\tablefootmark{$\star$}} & 33.1(8)\hphantom{\tablefootmark{$\star$}} & \hphantom{0}37.6(8) & \hphantom{\tablefootmark{$\ddag$}}$270\pm150$\tablefootmark{$\ddag$} & $530\pm70$\hphantom{0} & 30--40 & \hphantom{0}410 & \hphantom{$\sim$~\!}$-2.41\pm0.04$ & \hphantom{0}3, 38 \\
J0337+1715 & 21.316 & 0.85(4)\tablefootmark{$\star$} & 16(2)\hphantom{\tablefootmark{$\star$}}\hphantom{.0} & \hphantom{00}5.0(7) & $\ldots$ & \hphantom{0}$50\pm100$ & \hphantom{0}$\sim$~\!2~\!\hphantom{$\sim$}\hphantom{.00} & 1400 & \hphantom{$\sim$~\!}$-1.4\hphantom{0}\pm1$\hphantom{.00} & 30, 22 \\
J0407+1607 & 35.65 & 4.2(1)\tablefootmark{$\star$}\hphantom{0} & 16.4(4)\tablefootmark{$\star$} & \hphantom{0}56.5(8) & $\ldots$ & $50\pm50$ & 10.2\hphantom{0} & \hphantom{0}430 & \hphantom{$\sim$~\!}$-1.4\hphantom{0}\pm1$\hphantom{.00} & 34, 22 \\
J0621+1002 & 36.601 & 6.1(2)\hphantom{\tablefootmark{$\star$}}\hphantom{0} & 21.3(8)\hphantom{\tablefootmark{$\star$}} & \hphantom{0}20(1)\hphantom{.0} & \hphantom{\tablefootmark{$\dagger$}}$50\pm25$\tablefootmark{$\dagger$} & $170\pm380$ & $1.9\pm0.3$ & \hphantom{0}800 & \hphantom{$\sim$~\!}$-2.5\hphantom{0}\pm1.4\hphantom{0}$ & \hphantom{00, 0}4 \\
J0636+5129 & 11.107 & 0.14(2)\hphantom{\tablefootmark{$\star$}} & \hphantom{0}4.9(8)\hphantom{\tablefootmark{$\star$}} & \hphantom{00}4.1(8) & $\ldots$ & \hphantom{0}5.5  & \hphantom{0}1.8\hphantom{0} & \hphantom{0}350 & \hphantom{$\sim$~\!}$-1.2$\hphantom{$0\pm0.00$} & \hphantom{00, }32 \\
J0645+5158 & 18.247 & 0.16(2)\tablefootmark{$\star$} & \hphantom{0}1.8(2)\tablefootmark{$\star$} & \hphantom{00}4.7(7) & $\ldots$ & 20\hphantom{.0} & \hphantom{0}2.4\hphantom{0} & \hphantom{0}350 & \hphantom{$\sim$~\!}$-2.2$\hphantom{$0\pm0.00$} & \hphantom{00, }32 \\
J0737$-$3039A & 48.920 & 2.5(2)\hphantom{\tablefootmark{$\star$}}\hphantom{0} & 11.2(8)\hphantom{\tablefootmark{$\star$}} & \hphantom{0}64(10)\hphantom{.} & $\ldots$ & $40\pm80$ & $1.6\pm0.3$ & 1400 & \hphantom{$\sim$~\!}$-1.4\hphantom{0}\pm1$\hphantom{.00} & \hphantom{0}5, 22 \\
J0751+1807 & 30.249 & 0.65(5)\hphantom{\tablefootmark{$\star$}} & 19(2)\hphantom{\tablefootmark{$\star$}}\hphantom{.0} & \hphantom{00}5.5(8) & $70\pm30$ & $30\pm10$ & 10\hphantom{.00} & \hphantom{0}430 & \hphantom{$\sim$~\!}$-0.9\hphantom{0}\pm0.3$\hphantom{0} & \hphantom{0}6, \hphantom{0}4 \\
J1012+5307 & 9.023 & 0.69(2)\tablefootmark{$\star$} & 13.1(4)\tablefootmark{$\star$} & \hphantom{0}34.5(5) & $30\pm15$ & $220\pm90$\hphantom{0} & 30\hphantom{.00} & \hphantom{0}410 & \hphantom{$\sim$~\!}$-1.8\hphantom{0}\pm0.4$\hphantom{0} & \hphantom{0}7, \hphantom{0}4 \\
J1022+1001 & 10.252 & 1.00(3)\tablefootmark{$\star$} & \hphantom{0}6.1(2)\tablefootmark{$\star$} & \hphantom{0}40.7(9) & $90\pm40$ & $110\pm40$\hphantom{0} & $20\pm9$\hphantom{0} & \hphantom{0}370 & \hphantom{$\sim$~\!}$-1.7\hphantom{0}\pm0.1$\hphantom{0} & \hphantom{0}8, \hphantom{0}4 \\
J1023+0038 & 14.325 & 0.15(1)\tablefootmark{$\star$} & \hphantom{0}9.1(8)\tablefootmark{$\star$} & \hphantom{0}48(2)\hphantom{.0} & $\ldots$ & \hphantom{0}$440$--$3400$ & \hphantom{\tablefootmark{\textpilcrow}}$0.4$--$3.1$\tablefootmark{\textpilcrow} & 1600 & $\sim$~\!$-2.8$ & 31, \hphantom{0}9 \\
J1024$-$0719 & 6.485 & 0.54(4)\tablefootmark{$\star$} & 10.4(8)\tablefootmark{$\star$} & \hphantom{0}21(2)\hphantom{.0} & $200\pm100$ & $30\pm10$ & $4.6\pm1.2$ & 410/430 & \hphantom{$\sim$~\!}$-1.5\hphantom{0}\pm0.2$\hphantom{0} & 10, \hphantom{0}4 \\
J1038+0032 & 26.59 & 5.4(2)\hphantom{\tablefootmark{$\star$}}\hphantom{0} & 18.7(8)\hphantom{\tablefootmark{$\star$}} & \hphantom{0}16(1)\hphantom{.0} & $\ldots$ & \hphantom{0}$5\pm10$ & $0.20\pm0.04$ & 1400 & \hphantom{$\sim$~\!}$-1.4\hphantom{0}\pm1$\hphantom{.00} & \hphantom{0}5, 22 \\
J1231$-$1411 & 8.090 & 0.41(3)\hphantom{\tablefootmark{$\star$}} & 11.1(8)\hphantom{\tablefootmark{$\star$}} & \hphantom{0}22(3)\hphantom{.0} & $\ldots$ & $5\pm7$ & \hphantom{0}0.4\hphantom{0} & \hphantom{0}820 & \hphantom{$\sim$~\!}$-1.4\hphantom{0}\pm1$\hphantom{.00} & 27, 22 \\
B1257+12 & 10.166 & 0.63(1)\tablefootmark{$\star$} & 10.2(2)\tablefootmark{$\star$} & \hphantom{0}80.9(9) & $150\pm50$ & $180\pm90$\hphantom{0} & 20\hphantom{.00} & \hphantom{0}430 & \hphantom{$\sim$~\!}$-1.9\hphantom{0}\pm0.5$\hphantom{0} & 11, \hphantom{0}4 \\
J1453+1902 & 14.049 & 0.96(9)\hphantom{\tablefootmark{$\star$}} & 17(2)\hphantom{\tablefootmark{$\star$}}\hphantom{.0} & \hphantom{00}6(1)\hphantom{.0} & $\ldots$ & $10\pm10$ & $2.2\pm0.1$ & \hphantom{0}430 & \hphantom{$\sim$~\!}$-1.4\hphantom{0}\pm1$\hphantom{.00} & 36, 22 \\
J1544+4937 & 23.226 & 0.45(3)\hphantom{\tablefootmark{$\star$}} & 21(2)\hphantom{\tablefootmark{$\star$}}\hphantom{.0} & \hphantom{0}11.5(5) & $\ldots$ & 40\hphantom{.0} & \hphantom{0}5.4\hphantom{0} & \hphantom{0}322 & \hphantom{$\sim$~\!}$-2.3$\hphantom{$0\pm0.00$} & \hphantom{00, }37 \\
J1640+2224 & 18.426 & 0.21(1)\tablefootmark{$\star$} & \hphantom{0}6.7(4)\tablefootmark{$\star$} & \hphantom{0}15.1(9) & \hphantom{\tablefootmark{$\dagger$}}$450\pm200$\tablefootmark{$\dagger$} & $370\pm180$ & $2\pm1$ & 1472 & \hphantom{$\sim$~\!}$-2.18\pm0.12$ & \hphantom{0}4, 38 \\
J1709+2313 & 25.347 & 0.80(7)\hphantom{\tablefootmark{$\star$}} & 17(2)\hphantom{\tablefootmark{$\star$}}\hphantom{.0} & \hphantom{00}3.4(6) & $\ldots$ & $30\pm5$\hphantom{0} & $2.52\pm0.07$ & \hphantom{0}430 & \hphantom{$\sim$~\!}$-2.1\hphantom{0}\pm0.1$\hphantom{0} & \hphantom{00, }35 \\
J1713+0747 & 15.992 & 0.44(4)\hphantom{\tablefootmark{$\star$}} & \hphantom{0}9.5(8)\hphantom{\tablefootmark{$\star$}} & \hphantom{0}11(1)\hphantom{.0} & $250\pm100$ & $200\pm20$\hphantom{0} & 36\hphantom{.00} & \hphantom{0}430 & \hphantom{$\sim$~\!}$-1.5\hphantom{0}\pm0.1$\hphantom{0} & 12, \hphantom{0}4 \\
J1730$-$2304 & 9.617 & 0.78(6)\hphantom{\tablefootmark{$\star$}} & \hphantom{0}9.6(8)\hphantom{\tablefootmark{$\star$}} & 130(20)\hphantom{.} & $310\pm120$ & $350\pm120$ & $43\pm6$\hphantom{0} & \hphantom{0}436 & \hphantom{$\sim$~\!}$-1.8\hphantom{0}\pm0.3$\hphantom{0} & 13, \hphantom{0}4 \\
\hline
\end{tabular}
\tablefoot{
$S_\mathrm{mean}$ errors in parentheses are only nominal uncertainties. The realistic uncertainties are estimated to be within 50\% of the actual 
flux density values (see Sect.~\ref{calib_sect}).
Flux densities by \citet[KL01]{kl01} are at 102 or 111\,MHz.
\tablefoottext{$\dagger$}{Observations were done at 111\,MHz.}
\tablefoottext{$\ddag$}{Observations were done both at 102 and 111\,MHz, and it is not clear for which frequency the 
  flux densities were reported.}
\tablefoottext{$\star$}{Profile does not show an evident scattering tail.}
\tablefoottext{\textpilcrow}{Subject to substantial intrinsic variability, as pointed out by both \citet{asr+09} and \citet{dcrh12}.}
}
\tablebib{
(1)~\citet{lzb+00}; (2)~\citet{tbms98}; (3)~\citet{nbf+95}; (4)~\citet{kramer1998}; (5)~\citet{bjd+06};
(6)~\citet{lzc95}; (7)~\citet{nll+95}; (8)~\citet{snt97}; (9)~\citet{asr+09}; (10)~\citet{bjb+97}; (11)~\citet{wf92};
(12)~\citet{cam95a}; (13)~\citet{lnl+95}; (14)~\citet{hrm+11}; (15)~\citet{sfl+05}; (16)~\citet{ffb91};
(17)~\citet{llb+96}; (18)~\citet{jsb+10}; (19)~\citet{clm+05}; (20)~\citet{bbf+84}; (21)~\citet{fbb+90};
(22)~\citet{blv13}; (23)~\citet{sbl+96}; (24)~\citet{cgj+11}; (25)~\citet{cnt96}; (26)~\citet{ntf93};
(27)~\citet{rrc+11}; (28)~\citet{lynch2013}; (29)~\citet{jacoby_thesis2005}; (30)~\citet{ransom2014}; 
(31)~\citet{dcrh12}; (32)~\citet{slr+14}; (33)~\citet{hfs+04}; (34)~\citet{lxf+05}; (35)~\citet{lwf+04};
(36)~\citet{lmcs07}; (37)~\citet{brr+13}; (38)~\citet{kvl+15}.
}
\end{sidewaystable*}

\begin{sidewaystable*}
\setcounter{table}{3}
\caption{continued.
}\label{msps_fluxes1}
\centering
\begin{tabular}{ld{2.3}ccccccclc}
\hline\hline
PSR & \multicolumn{1}{c}{DM} & $W_\mathrm{eff}$\hphantom{\tablefootmark{$\star$}}\hphantom{0} & $\delta$ & $S_\mathrm{mean}$ & KL01 flux density      & Predicted $S_\mathrm{mean}$          & $S_\mathrm{ref}$ & $\nu_\mathrm{ref}$ & \hphantom{00}Spectral & Reference \\
    & \multicolumn{1}{c}{(pc\,cm${}^{-3}$)}  & (ms)\hphantom{\tablefootmark{$\star$}}\hphantom{0}             & (\%)     & (mJy)               & (mJy @ 102\,MHz)       & (mJy) & (mJy)   & (MHz)           & \hphantom{000}index    & \\
\hline
J1738+0333 & 33.778 & 0.58(5)\hphantom{\tablefootmark{$\star$}}\hphantom{0} & \hphantom{0}9.9(8)\hphantom{\tablefootmark{$\star$}} & \hphantom{0}17(3)\hphantom{.0} & $\ldots$ & \hphantom{0}$50\pm100$ & \hphantom{0}2\hphantom{.0} & 1400 & $-1.4\hphantom{0}\pm1$\hphantom{.00} & 29, 22 \\
J1744$-$1134 & 3.139 & 0.23(3)\tablefootmark{$\star$}\hphantom{0} & \hphantom{0}5.6(8)\tablefootmark{$\star$} & \hphantom{0}38(5)\hphantom{.0} & \hphantom{\tablefootmark{$\dagger$}}$220\pm100$\tablefootmark{$\dagger$} & $160\pm20$\hphantom{0} & $18\pm2$\hphantom{0} & \hphantom{0}436 & $-1.85\pm0.08$ & \hphantom{0}2, 38 \\
J1810+1744 & 39.659 & 0.566(3)\hphantom{\tablefootmark{$\star$}} & 34.1(2)\hphantom{\tablefootmark{$\star$}} & 563(1)\hphantom{.0} & $\ldots$ & $240\pm30$\hphantom{0} & 20\hphantom{.0} & \hphantom{0}350 & $-2.57\pm0.16$ & 14, 38 \\
J1816+4510 & 38.887 & 0.69(2)\hphantom{\tablefootmark{$\star$}}\hphantom{0} & 21.7(8)\hphantom{\tablefootmark{$\star$}} & \hphantom{0}83.5(5) & $\ldots$ & 10\hphantom{.0} & \hphantom{0}1.5 & \hphantom{0}350 & $-2$\hphantom{$.00\pm0.00$} & \hphantom{00, }32 \\
J1853+1303 & 30.570 & 0.75(6)\hphantom{\tablefootmark{$\star$}}\hphantom{0} & 18(2)\hphantom{\tablefootmark{$\star$}}\hphantom{.0} & \hphantom{0}18(2)\hphantom{.0} & $\ldots$ & $10\pm20$ & $0.4\pm0.2$ & 1400 & $-1.4\hphantom{0}\pm1$\hphantom{.00} & 15, 22 \\
B1855+09 & 13.300 & 0.85(4)\hphantom{\tablefootmark{$\star$}}\hphantom{0} & 15.9(8)\hphantom{\tablefootmark{$\star$}} & \hphantom{0}32(3)\hphantom{.0} & $450\pm250$ & $140\pm30$\hphantom{0} & $32\pm6$\hphantom{0} & \hphantom{0}430 & $-1.3\hphantom{0}\pm0.2$\hphantom{0} & \hphantom{00, }16 \\
J1905+0400 & 25.692 & 0.7(1)\hphantom{\tablefootmark{$\star$}}\hphantom{00} & 18(3)\hphantom{\tablefootmark{$\star$}}\hphantom{.0} & \hphantom{0}21(4)\hphantom{.0} & $\ldots$ & $1\pm2$ & $0.05\pm0.01$ & 1400 & $-1.4\hphantom{0}\pm1$\hphantom{.00} & 33, 22 \\
J1911$-$1114 & 30.975 & 0.64(6)\hphantom{\tablefootmark{$\star$}}\hphantom{0} & 18(2)\hphantom{\tablefootmark{$\star$}}\hphantom{.0} & \hphantom{0}38(4)\hphantom{.0} & $260\pm130$ & $470\pm110$ & $31\pm9$\hphantom{0} & \hphantom{0}400 & $-2.45\pm0.06$ & 17, 38 \\
J1918$-$0642 & 26.554 & 1.9(1)\hphantom{\tablefootmark{$\star$}}\hphantom{00} & 24(2)\hphantom{\tablefootmark{$\star$}}\hphantom{.0} & \hphantom{0}38(3)\hphantom{.0} & $\ldots$ & $30\pm5$\hphantom{0} & $5.9\pm0.6$ & \hphantom{0}350 & $-1.67\pm0.06$ & \hphantom{00, }18 \\
J1923+2515 & 18.858 & 0.37(1)\tablefootmark{$\star$}\hphantom{0} & \hphantom{0}9.7(4)\tablefootmark{$\star$} & \hphantom{0}17(1)\hphantom{.0} & $\ldots$ & $15\pm5$\hphantom{0} & $0.6\pm0.2$ & \hphantom{0}820 & $-1.7$\hphantom{$0\pm0.00$} & \hphantom{00, }28 \\
B1937+21 & 71.040 & 0.697(6)\hphantom{\tablefootmark{$\star$}} & 44.7(4)\hphantom{\tablefootmark{$\star$}} & 370(2)\hphantom{.0} & $\ldots$ & $4950\pm490$\hphantom{0} & $232\pm25$\hphantom{0} & \hphantom{0}430 & $-2.59\pm0.04$ & 16, 38\\
J1944+0907 & 24.34 & 1.48(2)\hphantom{\tablefootmark{$\star$}}\hphantom{0} & 28.6(4)\hphantom{\tablefootmark{$\star$}} & \hphantom{0}81(2)\hphantom{.0} & $\ldots$ & $20\pm20$ & $3.9\pm0.3$ & \hphantom{0}430 & $-1.4\hphantom{0}\pm1$\hphantom{.00} & 19, 22 \\
B1953+29 & 104.501 & 1.98(5)\hphantom{\tablefootmark{$\star$}}\hphantom{0} & 32.3(8)\hphantom{\tablefootmark{$\star$}} & \hphantom{0}38(1)\hphantom{.0} & $\ldots$ & $200\pm80$\hphantom{0} & $15\pm5$\hphantom{0} & \hphantom{0}430 & $-2.2\hphantom{0}\pm0.3$\hphantom{0} & 20, \hphantom{0}4 \\
B1957+20 & 29.117 & 0.576(6)\hphantom{\tablefootmark{$\star$}} & 35.8(4)\hphantom{\tablefootmark{$\star$}} & 155(1)\hphantom{.0} & $\ldots$ & \hphantom{0}$400$--$1300$ & \hphantom{${}^{+25}_{-13}$}$\sim$~\!$25^{+25}_{-13}$~\!\hphantom{$\sim$}\hphantom{.0} & \hphantom{0}430 & $-2.79\pm0.11$ & 21, 38 \\
J2019+2425 & 17.203 & 0.51(6)\hphantom{\tablefootmark{$\star$}}\hphantom{0} & 13(2)\hphantom{\tablefootmark{$\star$}}\hphantom{.0} & \hphantom{00}6(1)\hphantom{.0} & $190\pm100$ & $15\pm15$ & $2.7\pm0.7$ & \hphantom{0}430 & $-1.4\hphantom{0}\pm1$\hphantom{.00} & 26, 22 \\
J2043+1711 & 20.710 & 0.415(9)\tablefootmark{$\star$} & 17.4(4)\tablefootmark{$\star$} & \hphantom{0}34(1)\hphantom{.0} & $\ldots$ & $3\pm2$ & \hphantom{0}0.8 & \hphantom{0}350 & $-1.4\hphantom{0}\pm1$\hphantom{.00} & 14, 22 \\
J2051$-$0827 & 20.745 & 0.48(2)\tablefootmark{$\star$}\hphantom{0} & 10.7(4)\tablefootmark{$\star$} & \hphantom{0}43(3)\hphantom{.0} & \hphantom{\tablefootmark{$\dagger$}}$250\pm100$\tablefootmark{$\dagger$} & $180\pm60$\hphantom{0} & 22\hphantom{.0} & \hphantom{0}436 & $-1.8\hphantom{0}\pm0.3$\hphantom{0} & 23, \hphantom{0}4 \\
J2145$-$0750 & 8.998 & 1.17(2)\tablefootmark{$\star$}\hphantom{0} & \hphantom{0}7.3(1)\tablefootmark{$\star$} & 162(2)\hphantom{.0} & $480\pm120$ & $470\pm130$ & $100\pm30$\hphantom{0} & \hphantom{0}436 & $-1.33\pm0.18$ & \hphantom{0}2, 38 \\
J2214+3000 & 22.557 & 0.47(5)\hphantom{\tablefootmark{$\star$}}\hphantom{0} & 15(2)\hphantom{\tablefootmark{$\star$}}\hphantom{.0} & \hphantom{00}5.4(7) & $\ldots$ & $30\pm40$ & \hphantom{0}2.1 & \hphantom{0}820 & $-1.4\hphantom{0}\pm1$\hphantom{.00} & 27, 22 \\
J2215+5135 & 69.2 & 0.76(1)\hphantom{\tablefootmark{$\star$}}\hphantom{0} & 29.0(4)\hphantom{\tablefootmark{$\star$}} & \hphantom{0}53(2)\hphantom{.0} & $\ldots$ & $120\pm20$\hphantom{0} & \hphantom{0}5\hphantom{.0} & \hphantom{0}350 & $-3.22^{+0.15}_{-0.23}$ & 14, 38 \\
J2235+1506 & 18.09 & 2.5(2)\tablefootmark{$\star$}\hphantom{00} & \hphantom{0}4.2(4)\tablefootmark{$\star$} & \hphantom{00}6(1)\hphantom{.0} & \hphantom{\tablefootmark{$\ddag$}}$170\pm80$\tablefootmark{$\ddag$} & $15\pm15$ & \hphantom{0}3\hphantom{.0} & \hphantom{0}430 & $-1.4\hphantom{0}\pm1$\hphantom{.00} & 12, 22 \\
J2302+4442 & 13.762 & 0.64(4)\hphantom{\tablefootmark{$\star$}}\hphantom{0} & 12.4(8)\hphantom{\tablefootmark{$\star$}} & \hphantom{00}8.8(7) & $\ldots$ & $30\pm60$ & $1.2\pm0.4$ & 1400 & $-1.4\hphantom{0}\pm1$\hphantom{.00} & 24, 22 \\
J2317+1439 & 21.907 & 0.27(1)\tablefootmark{$\star$}\hphantom{0} & \hphantom{0}7.7(4)\tablefootmark{$\star$} & \hphantom{0}37.9(8) & $90\pm50$ & $90\pm30$ & $19\pm5$\hphantom{0} & \hphantom{0}430 & $-1.4\hphantom{0}\pm0.3$\hphantom{0} & 25, \hphantom{0}4 \\
J2322+2057 & 13.372 & 0.3(2)\hphantom{\tablefootmark{$\star$}}\hphantom{00} & \hphantom{0}7(3)\hphantom{\tablefootmark{$\star$}}\hphantom{.0} & \hphantom{00}2(1)\hphantom{.0} & $75\pm40$ & $15\pm15$ & $2.8\pm1.2$ & \hphantom{0}430 & $-1.4\hphantom{0}\pm1$\hphantom{.00} & 25, 22 \\
\hline
\end{tabular}
\tablefoot{
$S_\mathrm{mean}$ errors in parentheses are only nominal uncertainties. The realistic uncertainties are estimated to be within 50\% of the actual 
flux density values (see Sect.~\ref{calib_sect}).
Flux densities by \citet[KL01]{kl01} are at 102 or 111\,MHz.
\tablefoottext{$\dagger$}{Observations were done at 111\,MHz.}
\tablefoottext{$\ddag$}{Observations were done both at 102 and 111\,MHz, and it is not clear for which frequency the 
  flux densities were reported.}
\tablefoottext{$\star$}{Profile does not show an evident scattering tail.}
\tablefoottext{\textpilcrow}{Subject to substantial intrinsic variability, as pointed out by both \citet{asr+09} and \citet{dcrh12}.}
}
\tablebib{
(1)~\citet{lzb+00}; (2)~\citet{tbms98}; (3)~\citet{nbf+95}; (4)~\citet{kramer1998}; (5)~\citet{bjd+06};
(6)~\citet{lzc95}; (7)~\citet{nll+95}; (8)~\citet{snt97}; (9)~\citet{asr+09}; (10)~\citet{bjb+97}; (11)~\citet{wf92};
(12)~\citet{cam95a}; (13)~\citet{lnl+95}; (14)~\citet{hrm+11}; (15)~\citet{sfl+05}; (16)~\citet{ffb91};
(17)~\citet{llb+96}; (18)~\citet{jsb+10}; (19)~\citet{clm+05}; (20)~\citet{bbf+84}; (21)~\citet{fbb+90};
(22)~\citet{blv13}; (23)~\citet{sbl+96}; (24)~\citet{cgj+11}; (25)~\citet{cnt96}; (26)~\citet{ntf93};
(27)~\citet{rrc+11}; (28)~\citet{lynch2013}; (29)~\citet{jacoby_thesis2005}; (30)~\citet{ransom2014}; 
(31)~\citet{dcrh12}; (32)~\citet{slr+14}; (33)~\citet{hfs+04}; (34)~\citet{lxf+05}; (35)~\citet{lwf+04};
(36)~\citet{lmcs07}; (37)~\citet{brr+13}; (38)~\citet{kvl+15}.
}
\end{sidewaystable*}

\subsubsection{Flux measurements}\label{flux_measurements}

Table~\ref{msps_fluxes} lists the measured 110--188\,MHz flux densities of our MSP sample. 
In the Table we give the catalogued value of DM for reference, followed by our measurements of effective pulse width, $W_\mathrm{eff}$ (Col.~3), 
duty cycle, $\delta = W_\mathrm{eff}/P$ (Col.~4), and mean flux density measurements, $S_\mathrm{mean}$ (Col.~5).
Given $S_\mathrm{mean}$ errors in parentheses are nominal uncertainties only from the Eq.~\ref{sigmamean}.
For each MSP the $(\mathrm{S/N})_i^\mathrm{f,z,A}$ was calculated based on $\langle X\rangle$ and $\sigma_\mathrm{X}$
in the pre-selected off-pulse window.
For MSPs with broad profiles (due to scattering) we tried to place the off-pulse window just before the
leading edge of the profile, but most likely this still resulted in underestimated S/N and flux density. 
The $W_\mathrm{eff}$ was calculated as the integrated flux density in the pulse profile over its peak flux density.
In Col.~6 we list the flux density
measurements at 102/111\,MHz reported by \citet{kl01}. In Col.~ 7 we provide the predicted mean flux
density using the reference flux density value, $S_\mathrm{ref}$ (Col.~8) at the reference observing
frequency, $\nu_\mathrm{ref}$ (Col.~9), and spectral index (Col.~10) from published data. We give references in 
Col.~11 to all the published data we used in our calculations of the predicted flux density.
The values of the predicted mean flux densities listed are the average between the values at the edges
of our frequency range of 110--188\,MHz.
If the spectral index was not known,
we used the mean value of $-1.4\pm1$ for the pulsar population \citep{blv13}. We used this recently published value 
instead of the generally accepted value of $-1.8\pm0.2$ by \citet{maron2000} as it takes into account the sample bias in the spectral
index measurements available.
Although the work of \citet{blv13}
focused on the normal pulsar population, MSP spectra seem to be similar to those of normal pulsars \citep{kramer1998}, 
and thus we follow that assumption here.
For the non-detected MSPs, we provide upper limits $S_\mathrm{lim}$ (Table~\ref{msps_summary_nondetected}). 
We calculated $S_\mathrm{lim}$ as three times the nominal error on the mean flux density measurement.

\begin{figure}[tbh]
\centering
 \includegraphics[width=0.5\textwidth]{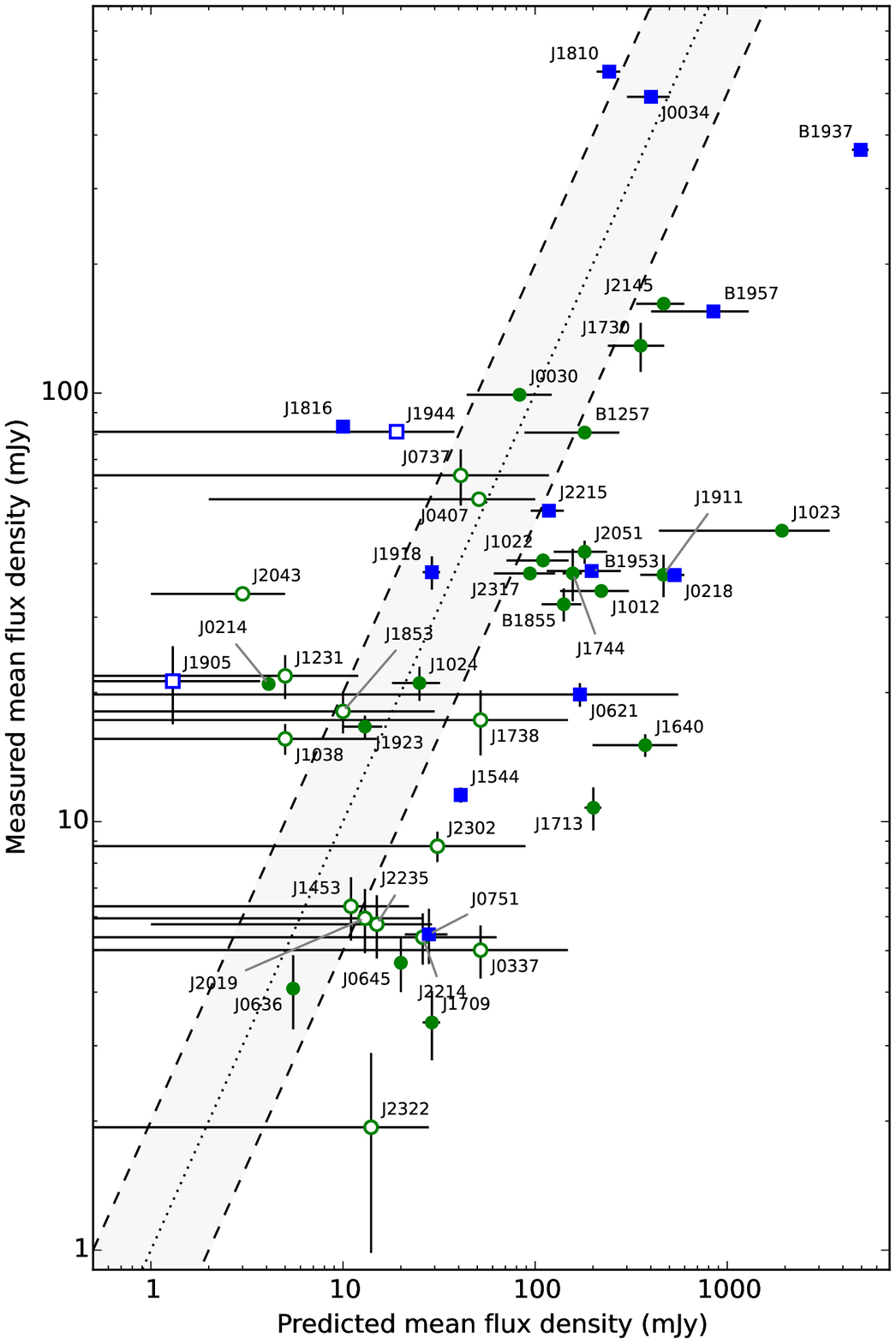}
 \caption{Measured mean flux densities at 110--188\,MHz versus predicted flux densities using high-frequency data for 48 MSPs. 
The dotted line shows the loci of measured mean fluxes equal to predicted fluxes, and dashed lines mark the region where measured mean 
flux density deviates from predicted flux density by a factor of two. Open symbols show the MSPs for which spectral indices are not known, 
and the average index from \citet{blv13} was used, while solid symbols are for those with known spectral indices (see Table~\ref{msps_fluxes}).
Blue squares and green circles indicate MSPs with measured duty cycles $>20$\% and $<20$\%, respectively. Pulsars are labelled
by their right ascension.
}
 \label{msp-flux-meas-est}
\end{figure}
\begin{figure}[tbh]
\centering
 \includegraphics[width=0.5\textwidth]{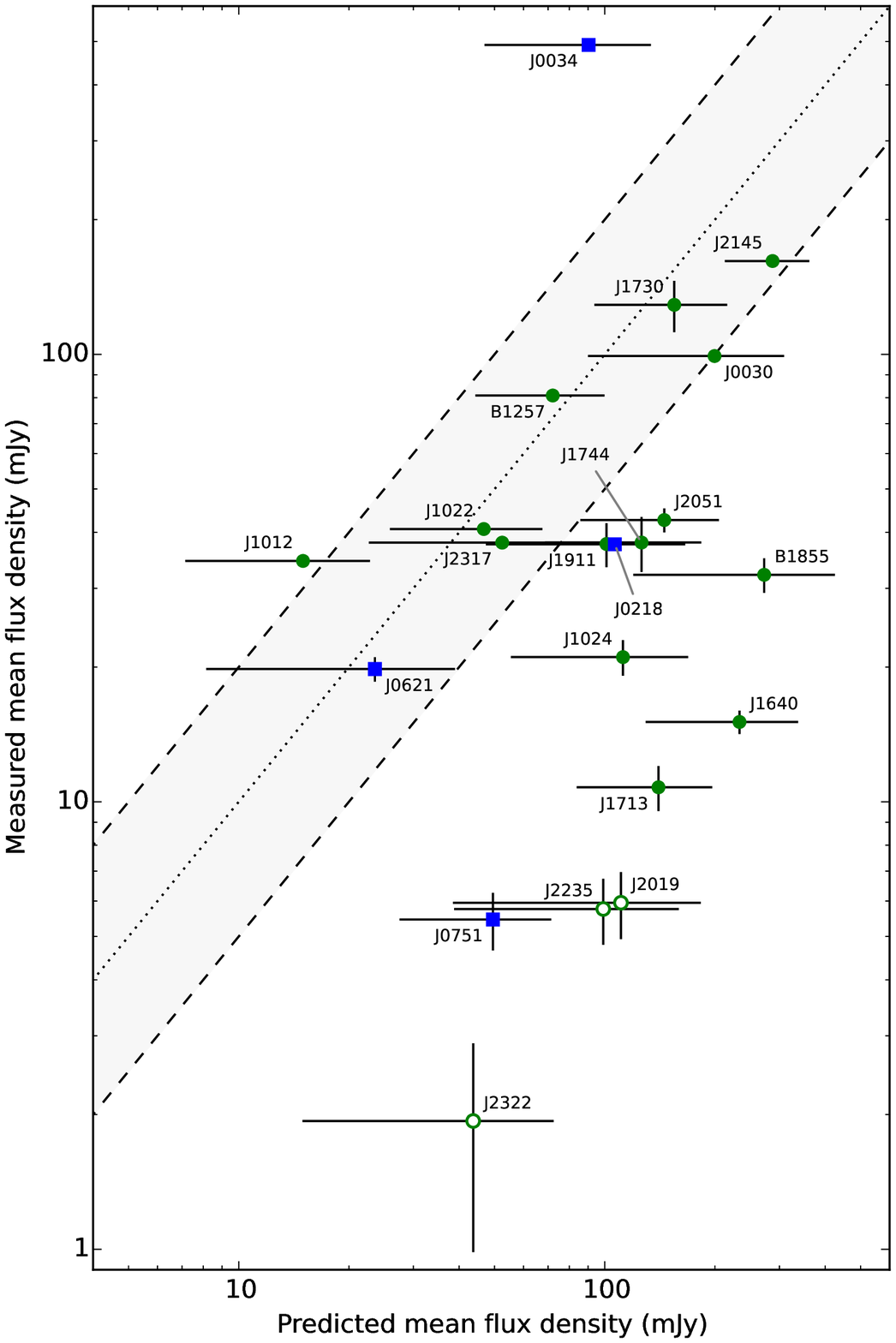}
 \caption{Same as the Fig.~\ref{msp-flux-meas-est}, but predicted mean flux densities are those from \citet{kl01} scaled 
 to the frequency range of 110--188~MHz using the spectral indices from Table~\ref{msps_fluxes}. 
}
 \label{msp-flux-meas-est-kl01}
\end{figure}

Figure~\ref{msp-flux-meas-est} shows the comparison between the measured and predicted flux densities for our 48 detected MSPs.
Predicted flux densities were calculated from the available high-frequency data and known spectral indices as given in Table~\ref{msps_fluxes}. 
For at least a third of the MSPs the main uncertainty on the predicted flux density comes from our 
poor knowledge of the spectral indices. However, for about another third of the MSPs, the measured flux densities are still lower
than the predicted ones even 
including the uncertainty on their spectral indices. This can be explained by contributions
from other factors given in Sect.~\ref{flux_definition}. For instance, the profiles of PSRs B1937+21, B1957+20, J0218+4232, B1953+29
are dominated by scattering, so our flux density measurements are underestimated for them.
PSR J1023+0038 has a very large predicted flux of up to $\sim 1.7$\,Jy. 
This could be due to significant intrinsic
variability \citep{asr+09}. For others, we note that
their profiles are quite narrow and show only a small scattering tail or none at all.
It is thus possible that the spectra of these pulsars turnover within or above our observing band.
For example, \citet{kl01} report a turnover for PSR J1012+5307 near 100\,MHz. We can not rule out this possibility, 
but more accurate measurements are needed at 
both HBA and LBA bands. Also, \citet{kvl+15} recently reported that PSRs J1640+2224
and J1911$-$1114 likely have a spectral break near 100\,MHz.

Refractive scintillation could also affect our measurements, although not by more than a factor of $\sim 1.5$ (see Sect.~\ref{flux_definition}).
Another possible explanation for the discrepancy between measured and predicted flux density values,
is that published flux densities at high frequencies
were over- or underestimated, most likely due to scintillation as was recently noted by \citet{lbb+13}.
To exclude the influence of 
overestimated flux densities at higher frequencies, we used the flux density measurements at frequencies 102/111\,MHz from \citet{kl01} 
to predict flux densities at 150\,MHz. The corresponding plot is shown in Fig.~\ref{msp-flux-meas-est-kl01}. If there is a real
turnover in the spectra we would expect our flux density measurements to be higher than the 
predicted flux density and, hence, located in the top-left
part of the plot. 
Our flux density value for PSR~J1012+5307 is indeed large which speaks in favour of
spectral turnover for this MSP. The same could also apply for PSR~J0034$-$0534, although its flux density
is consistent with what is expected from high-frequency measurements, and the flux density value reported by \citet{kl01} 
could be underestimated. For \object{PSR J1640+2224} \citet{kvl+15} report a possible turnover below 100\,MHz, however our flux
measurement is still lower than that predicted from the \citet{kl01} measurements at 102/111\,MHz. Assuming all flux measurements
are not affected by other factors (see Sect.~\ref{calib_sect}), turnover must occur sharply below 100\,MHz.
Similarly to the predictions from high-frequency data, our measured flux densities lie below
the predicted ones for about ten out of 21 pulsars with available flux densities from \citet{kl01}. 
Not ruling out other explanations including our own flux densities 
being somewhat underestimated due to scattering, RISS, beam jitter or contribution to the system temperature 
from the Galactic plane in the FoV, the 100-MHz flux density measurements from \citet{kl01} could be 
subject to a significant bias due to 
diffractive scintillation. The decorrelation bandwidth, $\Delta\nu_\mathrm{d}$, at 100\,MHz is of the order of 
$\lesssim 80$\,kHz \citep{cwb85}, and the receiver bandwidth, $B$, of \citet{kl01} was only 160\,kHz, 
comparable to the frequency scale of diffractive 
scintillation. This would modulate the pulsar flux density by a factor of 
$\sqrt{\Delta\nu_\mathrm{d}/B}$ \citep{popov_soglasnov1984}.
If a pulsar is weak enough to be only detected during the favourable ISM conditions, then the measured flux density 
will be overestimated.

LOFAR's observing bandwidth is much larger than the characteristic decorrelation bandwidth 
at 150\,MHz ($\Delta\nu_\mathrm{d}\lesssim 0.2$\,MHz), so diffractive scintillation cannot affect our measurements.
However, to make firm flux density estimates one needs to perform a long-term monitoring campaign to account for RISS.

\subsection{LBA detections}\label{lba}

We also observed nine bright MSPs with the LOFAR LBAs in the frequency range 15--93\,MHz.
We detected three MSPs, namely PSRs J0030+0451, J0034$-$0534, and J2145$-$0750. All detected
and non-detected MSPs in LBA observations are listed in the last column of Table~\ref{msps_summary_detected}.
Only \object{PSR J2317+1439} was observed for 20\,min, for the others the integration time was 1\,h.

It is currently unclear which effect plays the dominant role in the non-detections of MSPs below 100\,MHz: scattering,
increased sky temperature,
or smaller flux densities due to a possible turnover in their spectra. However, using the empirical relationship from \citet{bhat2004} 
for DM $= 10$\,pc\,cm$^{-3}$ one would expect the scattering time to be about 0.37\,ms at 57\,MHz, or 
about $\sim0.07$ cycles or less for PSRs J1012+5307, J1022+1001, J1024$-$0719, and B1257+12. 
Hence, scattering is most likely not the primary reason for the non-detections of these four MSPs.
From HBA profiles of these pulsars we determined rough upper limits on the $\tau_\mathrm{scat}$ at 57\,MHz to be
9, 6.4, 28, and 9\,ms, respectively. The limits do not contradict the prediction of $\tau_\mathrm{scat}$
from the \citet{bhat2004} relationship. However, although rough, our limits still give the possibility for scattering
to play a role in the non-detections of PSRs J1012+5307, J1024$-$0719, and B1257+12, as the uncertainty in the \citet{bhat2004}
relationship can be up to two orders of magnitude.
For \object{PSR J1022+1001} the spin period is about three times longer than our rough limit on $\tau_\mathrm{scat}$, thus most likely 
other factors contribute to the non-detection of this MSP -- more so than scattering.

These four non-detected MSPs are located at high Galactic latitudes
similar to the three MSPs detected with the LBAs, and the sky temperatures in these directions are very similar.
However, the spectral indices for two of the detected MSPs, PSRs J0030+0451 and J0034$-$0534 
($-2.2\pm 0.2$ and $-2.6\pm1$, respectively, see Table~\ref{msps_fluxes}; approximately equivalent to the spectral index of 
Galactic synchrotron radiation), are steeper than for these four non-detected MSPs.
Hence, the influence of the increased background temperature at low radio frequencies for PSRs J0030+0451
and J0034$-$0534 is much smaller than for the non-detected MSPs with shallower spectral indices.
Alternatively, these four non-detected MSPs might have a turnover in their radio spectra 
at frequencies $\lesssim 120$\,MHz (for PSR~J1012+5307
\citet{kl01} reported a possible turnover in its spectrum near 100\,MHz). 

For PSR~J2317+1439 the predicted scattering time at 57\,MHz is about 3\,ms, comparable to the pulse period. The rough upper
limit on $\tau_\mathrm{scat}$ from the HBA profile is somewhat lower, about $2.5$\,ms. Thus, similarly to the other four non-detected MSPs,
scattering is not the dominant factor in its non-detection.

For the black widow \object{PSR J1810+1744}, with a DM of 39.7\,pc\,cm$^{-3}$, the scattering time is already four times longer
than the spin period, and the profile is scattered out \citep{akhs14}. One can already see profile evolution within 
the HBA band for this MSP in Fig.~\ref{msps-prof-evolution} (left), where the profile at the low-frequency edge is notably 
more scattered than at the high-frequency end of the HBA band.

For completeness, we should also consider the influence of $v_\mathrm{orb}/c$ smearing on LBA profiles of binary MSPs as
was discussed in Sect.~\ref{ranseff}, as this effect becomes even more significant 
at lower observing frequencies. PSRs J0030+0451 and J1024$-$0719 are isolated, so this effect is not relevant for them.
For the remaining seven MSPs we calculated the extra smearing of the profile due to this effect in a 3.9-MHz frequency
band near a frequency of 58\,MHz, where we have our highest sensitivity.
For four out of seven MSPs
(PSRs J1012+5307, J1022+1001, B1257+12, and J2145-0750) this effect is minor with less than $0.01$ cycles of extra broadening
of the profiles. For PSR J2317+1439, which we did not detect, the broadening is $0.04$ cycles, still quite small, so we should have
been able to detect it regardless of this effect. In contrast, for PSR J1810+1744 the effect is quite significant producing
additional profile scattering of about $0.1$ cycles. However, as we already showed, the expected broadening due to the scattering
itself is 100 times larger.

\begin{figure*}[htb]
\centering
 \includegraphics[scale=0.5]{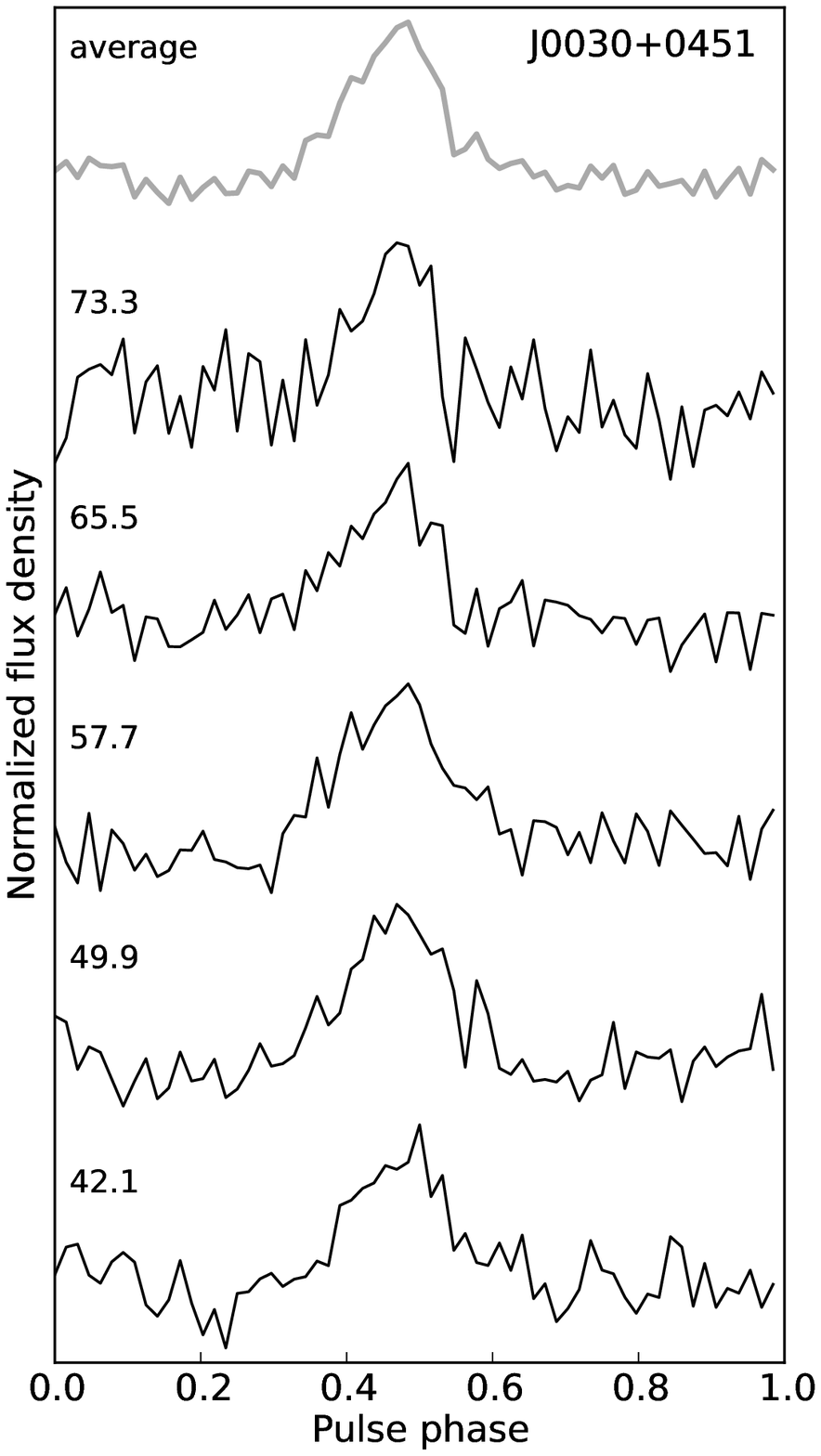}\includegraphics[scale=0.5]{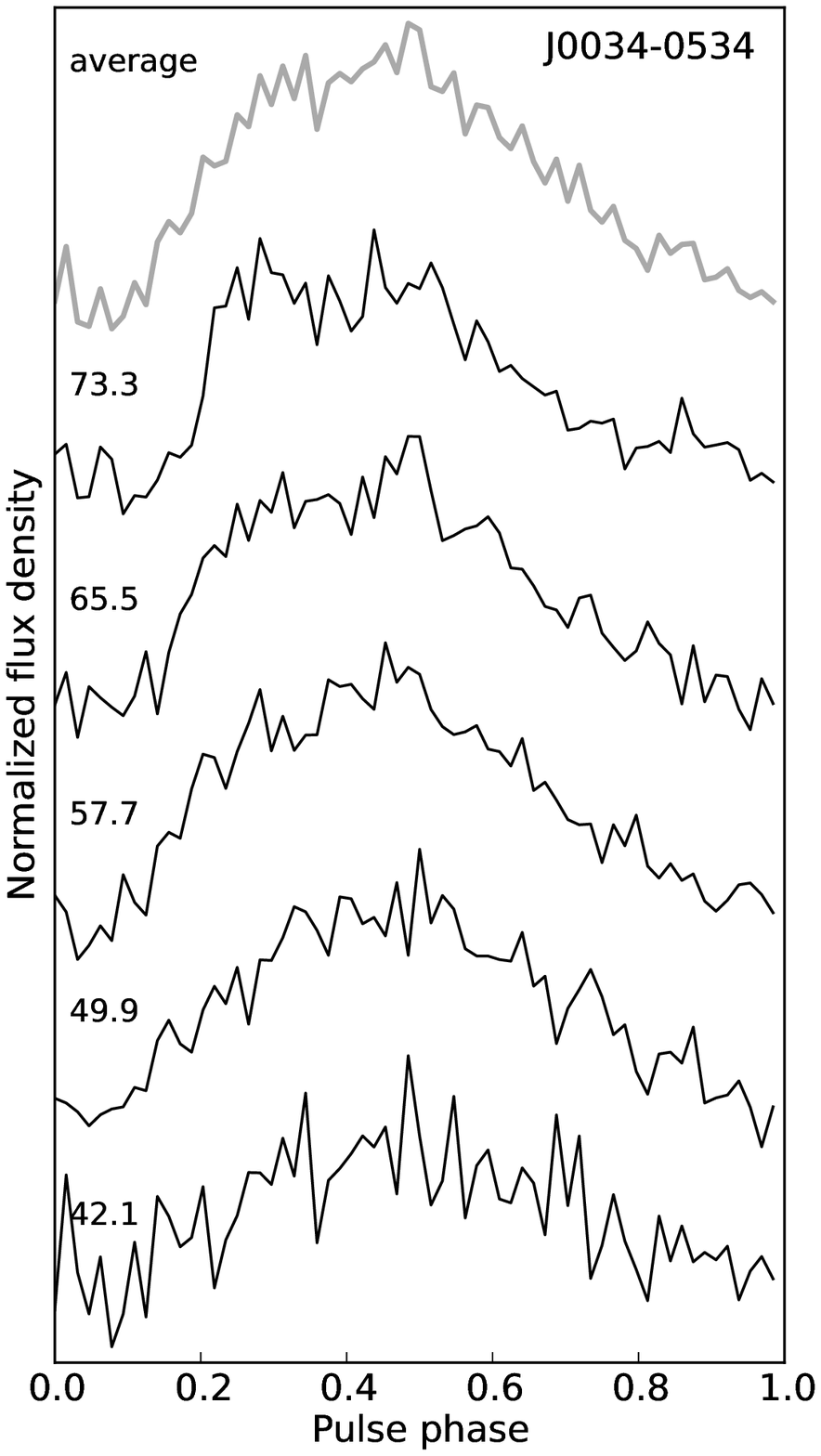}\includegraphics[scale=0.5]{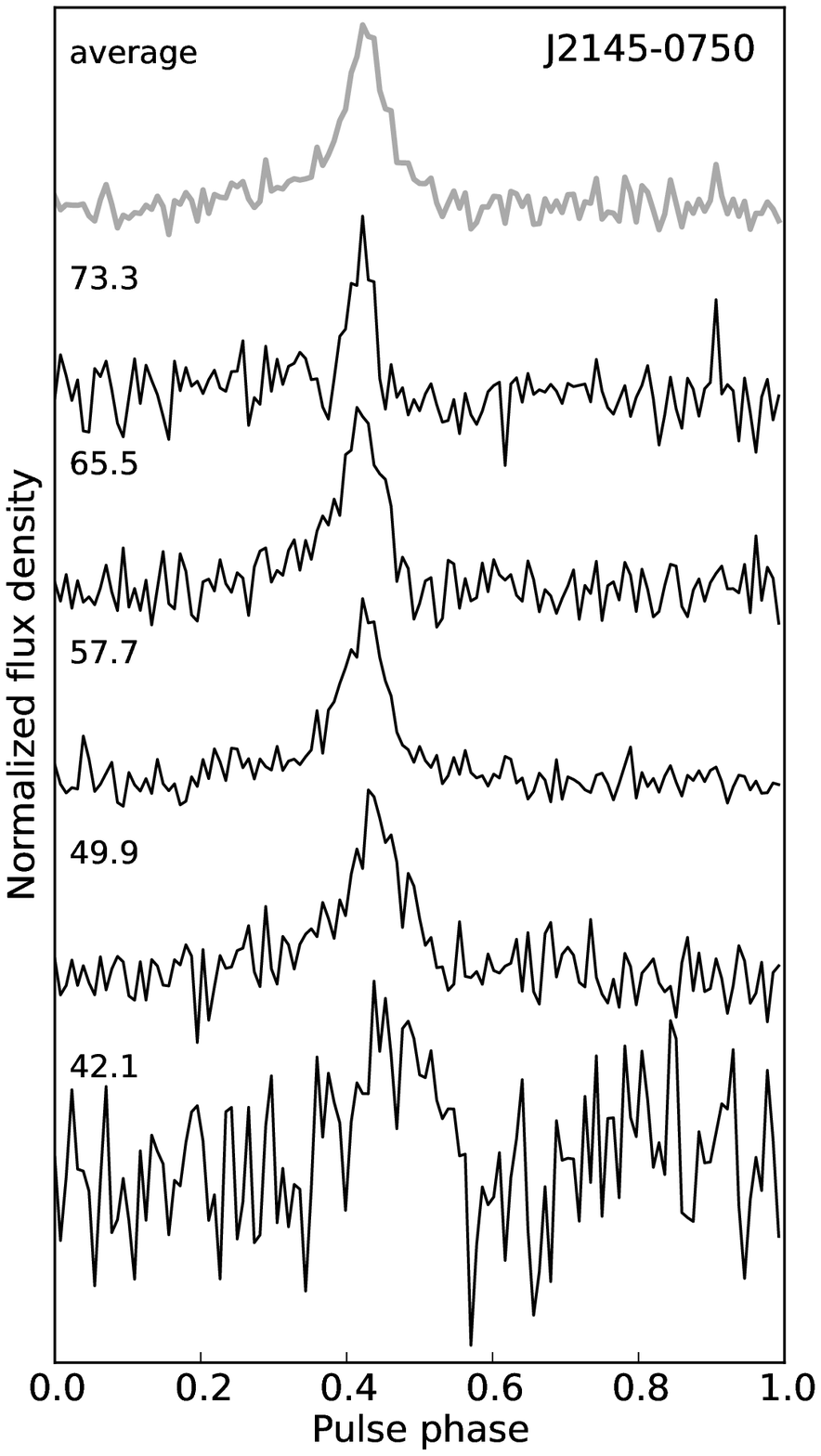}
 \caption{Profile evolution of the three MSPs detected
 in the LBA frequency range of 38.2--77.2\,MHz. The grey profile on top of each panel is the average profile within the 39-MHz wide band.
 All profiles are from 1-h LBA observations. There are 64 bins in the profiles of PSRs J0030+0451 and J0034$-$0534 
 and 128 bins in the profiles of PSR J2145$-$0750. The numbers on the left in each panel show the central frequency in MHz of the 
 corresponding 7.8-MHz sub-band.
 }
 \label{lba_profs}
\end{figure*}

Figure~\ref{lba_profs} shows the profile evolution for the three detected MSPs.
The profile on top of each panel is the average profile from 38--77\,MHz.
Among these three, PSR J0034$-$0534 has the largest DM of 13.8\,pc\,cm$^{-3}$ and shows significant profile
evolution with its profile clearly being scattered towards lower frequencies. The profile shape in the total LBA band
is largely determined by scattering at the lower part of the band,
although the profile in the higher-frequency sub-band at 73.3\,MHz 
resembles the profile in the HBA band (see Fig.~\ref{mspprof}). At this observing frequency we measured the scattering
time to be about $440\,\upmu$s by fitting the trailing part of the profile with an exponential function.
The extra profile broadening due to the $v_\mathrm{orb}/c$ effect is about 0.05 cycles for a 7.8-MHz sub-band around 57.7\,MHz. 
It does scatter the profile but not at a large enough level to scatter it out completely for the lowest sub-band shown in 
Fig.~\ref{lba_profs} (middle).
Therefore, the profile broadening must be dominated by scattering in the ISM.

The profiles of the other two MSPs, PSRs J0030+0451 and J2145$-$0750, do not show significant evolution in comparison
to PSR~J0034$-$0534. They have smaller DMs (4.3, 9.0, 13.8\,pc\,cm$^{-3}$ for PSRs J0030+0451, J2145$-$0750, and 
J0034$-$0534, respectively),
and the scattering conditions must be temperate towards them. We measured the scattering time to be about $300\,\upmu$s for
\object{PSR J0030+0451} at 42.1\,MHz and less than $700\,\upmu$s for PSR J2145$-$0750 at 57.7\,MHz.
Although there is no significant broadening of the
profile for PSR J2145$-$0750 its profile almost vanishes at 42\,MHz. 
For PSR J2145$-$0750 the sky temperature is four times higher at 42\,MHz than at 73\,MHz. The LOFAR LBA bandpass response 
is not uniform and peaks at approximately 58\,MHz \citep[see][Fig.~19]{haarlem2013}. 
However, we do not think that lower sensitivity is the reason for the much weaker profile at 42\,MHz.
For PSR J0030+0451 the change in the sky temperature and bandpass response is the same, but it has
a clearly detected profile at 42\,MHz, only slightly weaker than the profile at 58\,MHz.
Hence, we believe that this weakening is due to a turnover in the spectrum of PSR~J2145$-$0750.   
Our assumption qualitatively agrees with the recent conclusion by \citet{drt+13}, who found
that the radio spectrum of PSR J2145$-$0750 is best fit with a model with spectral curvature and
a roll-over frequency of 730\,MHz using data from the LWA. Recently, \citet{kvl+15} also reported on
a spectral turnover between 100 and 400\,MHz based on archival and LWA data.

\citet{srb+14} also report on the detection of the same three MSPs with the LWA between 30--88\,MHz from their
6-h observations of PSR J0030+0451 and 8-h observations of PSRs J0034$-$0534 and J2145$-$0750.

\begin{figure*}[htbp]
\centering
 \includegraphics[scale=0.75]{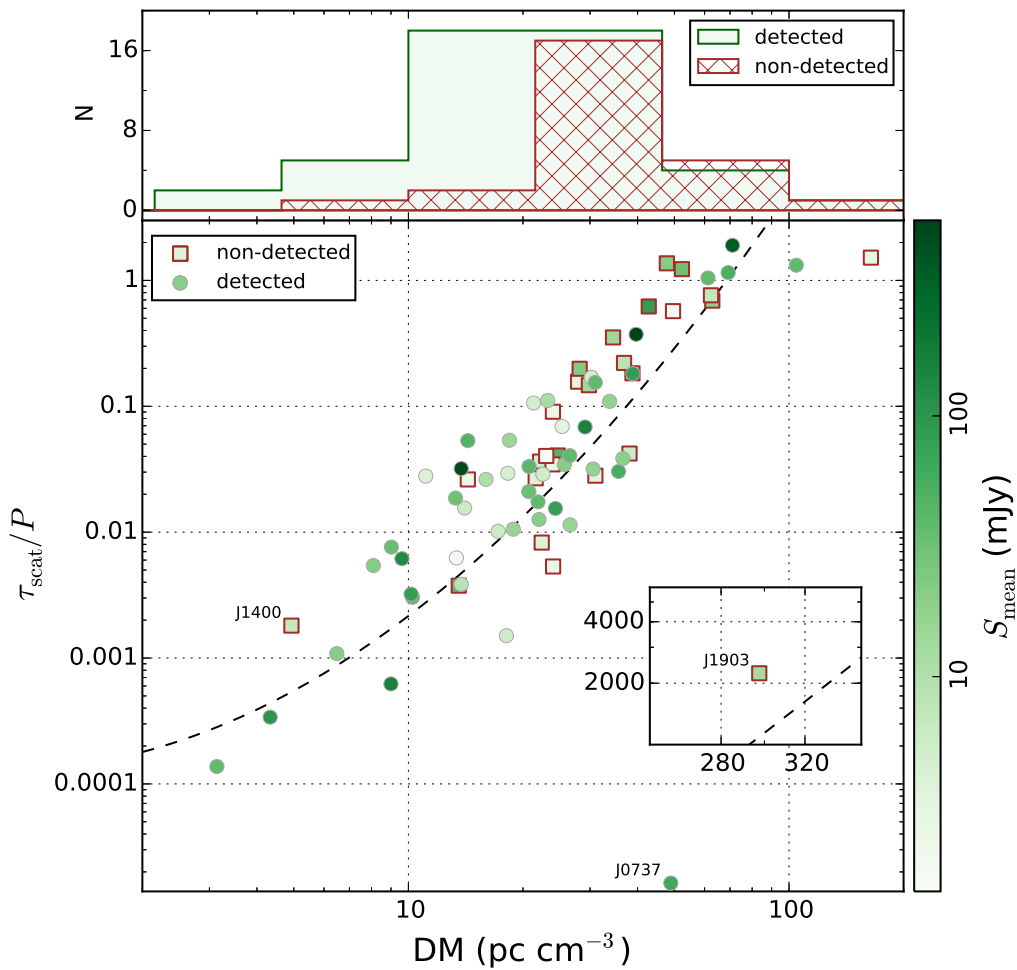}
\caption{MSP detectability, or dependence of the ratio of predicted scattering time at 150\,MHz 
over the pulsar period, $\tau_\mathrm{scat}/P$, 
on dispersion measure, DM. The green colour map corresponds to the measured mean flux density, $S_\mathrm{mean}$, 
at 110--188\,MHz. Non-detected MSPs
are shown with brown squares and their green intensity level corresponds to the measured upper limit of the mean flux
density at 110--188\,MHz. Scattering times at 150\,MHz were calculated from the NE2001 Galactic free electron density model \citep{cl+02}
assuming a Kolmogorov spectrum of electron density inhomogeneities. 
The dashed line shows the empirical dependence between $\tau_\mathrm{scat}$ and DM at 150\,MHz from \citet{bhat2004}, where $\tau_\mathrm{scat}$ 
is arbitrarily normalised by $P = 4.075$\,ms, which is the median value
of MSP periods in our sample of detected and non-detected MSPs.
The top subplot shows the histograms of DMs for detected and non-detected MSPs. 
The inset plot is the same as the main plot but only for the non-detected PSR J1903+0327, which did not fit into the range 
of the main plot. \object{PSR J1400$-$1438} was recently detected with LOFAR using the updated coordinates (J.~Swiggum, private
communication) derived from over three years of timing with the GBT. The previously used catalogued position was off by about $6\arcmin$, which is
about two times larger than the full-width at half-maximum of LOFAR's full-core tied-array beam at 150\,MHz.
} 
 \label{msps-detectability}
\end{figure*}

\section{Discussion}\label{discussion}

\subsection{Detectability of MSPs}\label{detectability}

In order to explore the degree to which propagation in the ISM affects the detectability of the MSPs presented here,
we plotted the ratio of the predicted $\tau_\mathrm{scat}$ at 150\,MHz, over the pulsar period, $P$,
on DM and $S_\mathrm{mean}$ at 110--188\,MHz.
This dependence is shown in Fig.~\ref{msps-detectability}
for both detected and non-detected MSPs, with the green colour indicating the $S_\mathrm{mean}$.
For non-detected MSPs, the green intensity shows the upper limit on $S_\mathrm{mean}$ as given
in Table~\ref{msps_summary_nondetected} (Col.~10).
For $\tau_\mathrm{scat}$ we used the NE2001 Galactic free electron density model \citep{cl+02} and assumed a
Kolmogorov spectrum of electron density inhomogeneities (see Table~\ref{msps_dms}, last column).

The histogram on top of Fig.~\ref{msps-detectability} presents the number of detected and non-detected MSPs
as a function of DM. We detected 25 out of 28 observed MSPs with DMs $\lesssim 20$\,pc\,cm$^{-3}$.
On the high-DM side, we did not detect two MSPs with DM $> 110$\,pc\,cm$^{-3}$ that we tried, 
namely \object{PSR J1949+3106} with DM $=164$\,pc\,cm$^{-3}$ and \object{PSR J1903+0327} with DM $=297$\,pc\,cm$^{-3}$.
On the other hand, one of the detected MSPs, \object{PSR B1953+29}, has a DM of 104.5\,pc\,cm$^{-3}$.
For the DM range of 20--100\,pc\,cm$^{-3}$ we have almost parity between detected (22) and non-detected MSPs (23).

There is no clear dependence of the MSP detectability on the predicted scattering measure. If it were the case, we would see
the majority of non-detected MSPs clustering close to or above the line $\tau_\mathrm{scat}/P = 1$, where a pulse
profile would be completely washed out by the scattering.
However, most of our non-detected MSPs lie below $\tau_\mathrm{scat}/P = 1$ line populating the DM range of $\sim$20--50\,pc\,cm$^{-3}$.
The dependence of $\tau_\mathrm{scat}/P$ on DM follows the empirical relationship between DM and $\tau_\mathrm{scat}$ 
provided by \citet{bhat2004}. However, there is more than two orders of magnitude of scatter in this relationship,
hence its predictive power for individual pulsars is very limited. Therefore, it is certainly possible that for some of the non-detected MSPs
the scattering conditions are much worse than we can infer from their DM (and/or their detectability depends
on other factors).

Another factor that can play an equal or even larger role than scattering in the MSP detectability is our lack of sensitivity
resulting from the increased sky temperature background at 150\,MHz. 
In Fig.~\ref{msps-detectability}, all five detected MSPs with $\mathrm{DM} > 40$\,pc\,cm$^{-3}$
have $S_\mathrm{mean}$ above 27\,mJy, larger than the average value for the rest of the detected MSPs. Also, except
for the double pulsar, PSR~J0737$-$3039A, four others pulsars -- PSRs J0218+4232, B1937+21, B1953+29, and J2215+5135 -- 
have significantly scattered profiles, and their $S_\mathrm{mean}$ most 
likely is underestimated. For PSRs J0218+4232, B1937+21, and B1953+29 spectral indices derived from high-frequency data are
$-2.8$, $-2.3$, and $-2.2$, respectively, significantly larger than the average for the pulsar population and comparable
with the spectral index of the synchrotron background radiation.
In fact, PSRs B1937+21 and J0218+4232 were first recognised as steep-spectrum, highly linearly polarised, compact radio sources
in interferometric imaging data \citep{erickson1980,nbf+95}, which is what led to their discovery.
Most of the other detected MSPs have shallower spectral indices.
Together with the possible turnover in their spectra and scattering, this could hinder MSP detection at 150\,MHz.
For instance, PSR J1713+0747 is very bright (S/N $\sim 4000$) at 1.4\,GHz \citep[see e.g.][]{dlc+14}. However, 
the LOFAR detection was very weak, with $S_\mathrm{mean}$ of only about 11\,mJy (S/N $\sim 8$), and the profile is not very
scattered ($\tau_\mathrm{scat} \sim 120\,\upmu$s). Thus, in this case the main reason for a weak detection is not due to the
scattering, but rather the spectral index of roughly $-1.5$ being smaller than for the synchrotron background radiation, or
the possible turnover in its radio spectrum between 200 and 1400\,MHz, or both.

\begin{figure*}[htbp]
\centering
 \includegraphics[scale=0.75]{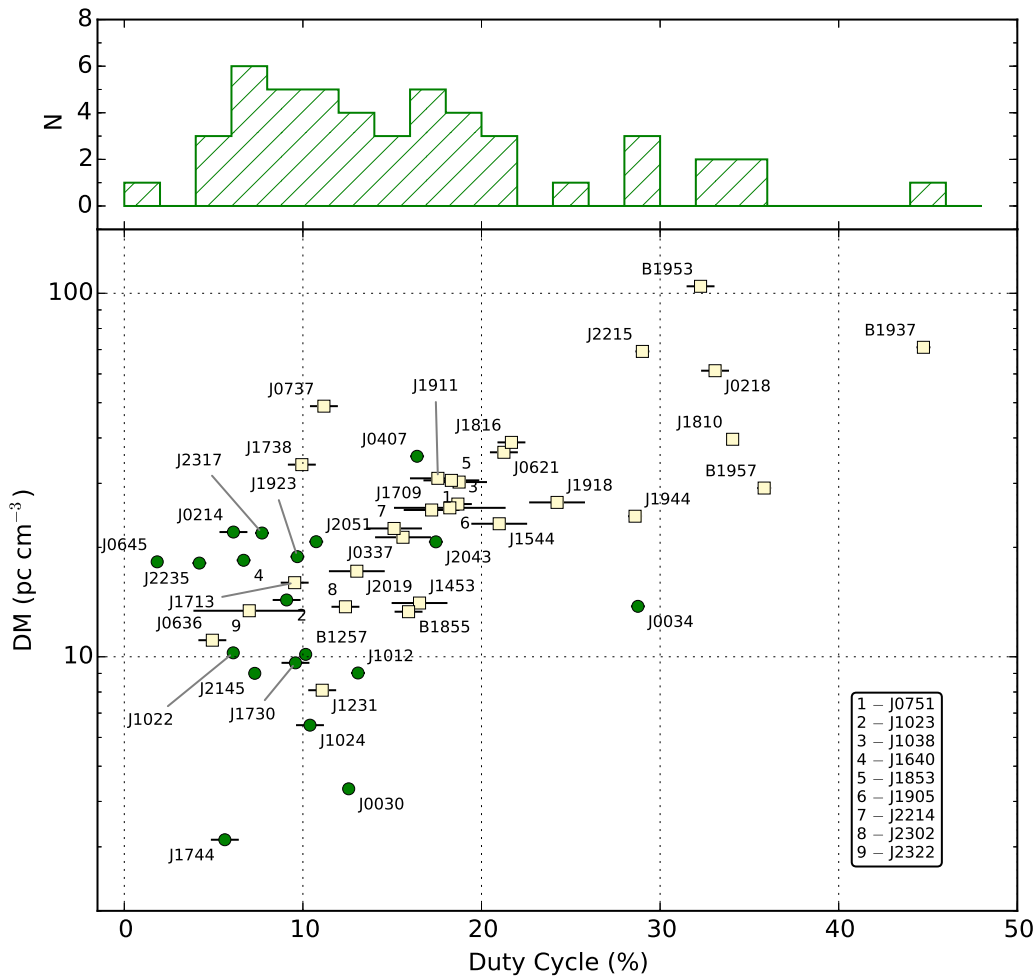}
 \caption{Dispersion measure versus the MSP duty cycle, $\delta$. Green circles show MSPs with profiles which do not
 show an evident scattering tail -- see $\delta$ values marked by an asterisk ($^\star$) in Table~\ref{msps_fluxes}. 
 Light yellow squares show all other MSPs.
 Errors in DMs including $\delta\mathrm{DM}_\mathrm{orb}$ (see Sect.~\ref{sect_dmvar}) are smaller than the size of the marker for all MSPs.
 For MSPs without error bars, the uncertainty on $\delta$ is also less than the size of the marker. The top subplot
 shows the histogram of duty cycles for all detected MSPs.
 }
 \label{msp-dc-dm}
\end{figure*}

Three of our non-detected MSPs, PSRs J0613$-$0200, B1620$-$26, and J2229+2643, were previously reported to be detected 
at 102/111\,MHz by \citet{kl01}. As we discussed in Sect.~\ref{flux_measurements}, their detection and flux measurements could be 
significantly biased due to diffractive scintillation. PSR J0613$-$0200 was observed by \citet{stappers2008} 
with the WSRT at 150\,MHz and they did not detect it. Also, for these and other non-detected MSPs
additional factors discussed in Sect.~\ref{flux_definition} could affect their detection, most notably 
the ionospheric beam jitter and RISS.

\subsection{Profile widths}

Table~\ref{msps_fluxes} lists measured values of $W_\mathrm{eff}$ and $\delta$ in Cols.~3 and 4 for the total
observing bandwidth of 110--188\,MHz.
At low observing frequencies scattering broadens the intrinsic 
pulse profiles, and our measured $W_\mathrm{eff}$ is essentially affected by scattering for many
of the observed MSP profiles. 
For MSPs with narrow, visibly unscattered profiles, $W_\mathrm{eff}$ and $\delta$ in Table~\ref{msps_fluxes}
are marked with an asterisk ($^\star$). Figure~\ref{msp-dc-dm} shows $\delta$ vs DM for detected MSPs.
Almost all MSPs with $\delta\gtrsim 20$\% show profiles widened by scattering (yellow circles).
Also, between them there is a trend of $\delta$ being larger with an increasing DM. The only exception here
is PSR J0034$-$0534 which has a very broad profile ($\delta\approx 29$\%), which mostly likely is
genuinely broad by comparing with the WSRT profile at 376\,MHz (see Fig.~\ref{mspfreq}).

For the majority of the detected MSPs, $\delta<20$\%. Among them, MSPs with low DMs $\lesssim20$\,pc\,cm$^{-3}$
have sharp narrow profiles and are weakly (if at all) affected by scattering (green circles).
Others, at moderately high DMs, have profiles that are strongly affected by scattering 
and/or intrinsically weak at 110--188\,MHz.

Table~\ref{msps_fluxes} gives nominal 1-$\sigma$ errors on $W_\mathrm{eff}$ and $\delta$ values. 
The actual error for $W_\mathrm{eff}$ can be up to two times larger for the majority of binary MSPs depending on the ratio
$v_\mathrm{orb}/c$ (see Sect.~\ref{ranseff}). The actual error on $\delta$ could be larger by 0.1\%, and
does not affect our conclusions on MSP widths. 

\subsection{DM variations}\label{sect_dmvar}

It is known that, on time scales of months to years,  
DM and $\tau_\mathrm{scat}$ towards some pulsars change due to the varying line-of-sight through the ISM \citep[e.g.][]{hlk+04}.
For instance, \citet{hs08} showed that detectable variations of the scattering measure towards PSR~B1737+13 
occurred roughly on a time scale of a fortnight
for observations in the 1.1--1.5-GHz range. 
\citet{kcs+13} reported on long-term DM variations for MSPs on time scales of a few months.
Correction for such DM variations is crucial for high-precision timing.
 
To measure DM at the epochs of the individual observations we split the observing band into four--five sub-bands and for each
of them generated a single TOA for the whole observation by cross-correlation with a profile template.
The template was based on the average pulse profile of all available observations in the total HBA band.
We then used the pulsar timing package {\tt Tempo2}\footnote{\tt https://bitbucket.org/mkeith/tempo2} \citep{hem+06}
to fit the generated TOAs only for DM.
In Table~\ref{msps_dms} we present the weighted average, $\mathrm{DM}_\mathrm{HBA}$, of our DM measurements (Col.~5) 
for each of the individual observing
epochs during the given time span (Col.~7). The reference epoch that corresponds to the middle of the observing time span 
is given in Col.~6. The number of observations, $N_\mathrm{obs}$, performed within a given span is listed in Col.~8.
For 11 out of 48 detected MSPs we carried out only one exploratory observation, therefore 
the presented DMs are individual DM measurements for the epoch of the observation.
We calculated the DM offset, $\Delta\mathrm{DM} = \mathrm{DM}_\mathrm{ref} - \mathrm{DM}_\mathrm{HBA}$, 
between the catalogue DM value, $\mathrm{DM}_\mathrm{ref}$,
and LOFAR's DM measurement, $\mathrm{DM}_\mathrm{HBA}$ (Col.~9), with the uncertainty, $\sigma_{\Delta\mathrm{DM}}$, 
being the quadrature sum of uncertainties of both $\mathrm{DM}_\mathrm{ref}$ and $\mathrm{DM}_\mathrm{HBA}$.

Column 11 of Table~\ref{msps_dms} lists measured DMs for the best individual observations. The uncertainties
for most of the MSPs are lower than the errors on $\mathrm{DM}_\mathrm{ref}$,
owing to the low observing frequency and large fractional bandwidth of LOFAR. The uncertainty on the weighted average DM value, $\mathrm{DM}_\mathrm{HBA}$,
on the other hand is larger than the uncertainty of the individual measurements for the majority of the MSPs. 
This is most likely evidence for short-term DM variations. However, care should be taken to account
for artificial orbital DM variations for binary MSPs as was discussed in Sect.~\ref{ranseff}. These DM
variations, $\delta\mathrm{DM}_\mathrm{orb} = v_\mathrm{orb}/c\times\mathrm{DM}$, are larger for higher-DM
pulsars in compact binary orbits. We list $\delta\mathrm{DM}_\mathrm{orb}$ for all binary MSPs in Col.~10 of
Table~\ref{msps_dms}. For $v_\mathrm{orb}$ we used the maximum orbital velocity of the pulsar at periastron,
$v_\mathrm{p}/c=(2\pi x/P_\mathrm{b})[(1+e)/(1-e)]^{1/2}$, where $x$ is the projected semi-major axis, $e$ is the
eccentricity, and $P_\mathrm{b}$ is the orbital period of the pulsar. 
In our binary MSP sample $|\delta\mathrm{DM}_\mathrm{orb}|$ is highest for the double pulsar J0737$-$3039A
(0.054\,pc\,cm$^{-3}$), and lowest for \object{PSR B1257+12} ($<10^{-7}$\,pc\,cm$^{-3}$).
The influence of $\delta\mathrm{DM}_\mathrm{orb}$ on the weighted average $\mathrm{DM}_\mathrm{HBA}$ and 
$\Delta\mathrm{DM}$ should be even smaller, as the actual $\delta\mathrm{DM}_\mathrm{orb}$ 
depends on $\vec{v_\mathrm{orb}}$ at the epoch of observation. 

\begin{figure*}[htbp]
\centering
 \includegraphics[scale=0.75]{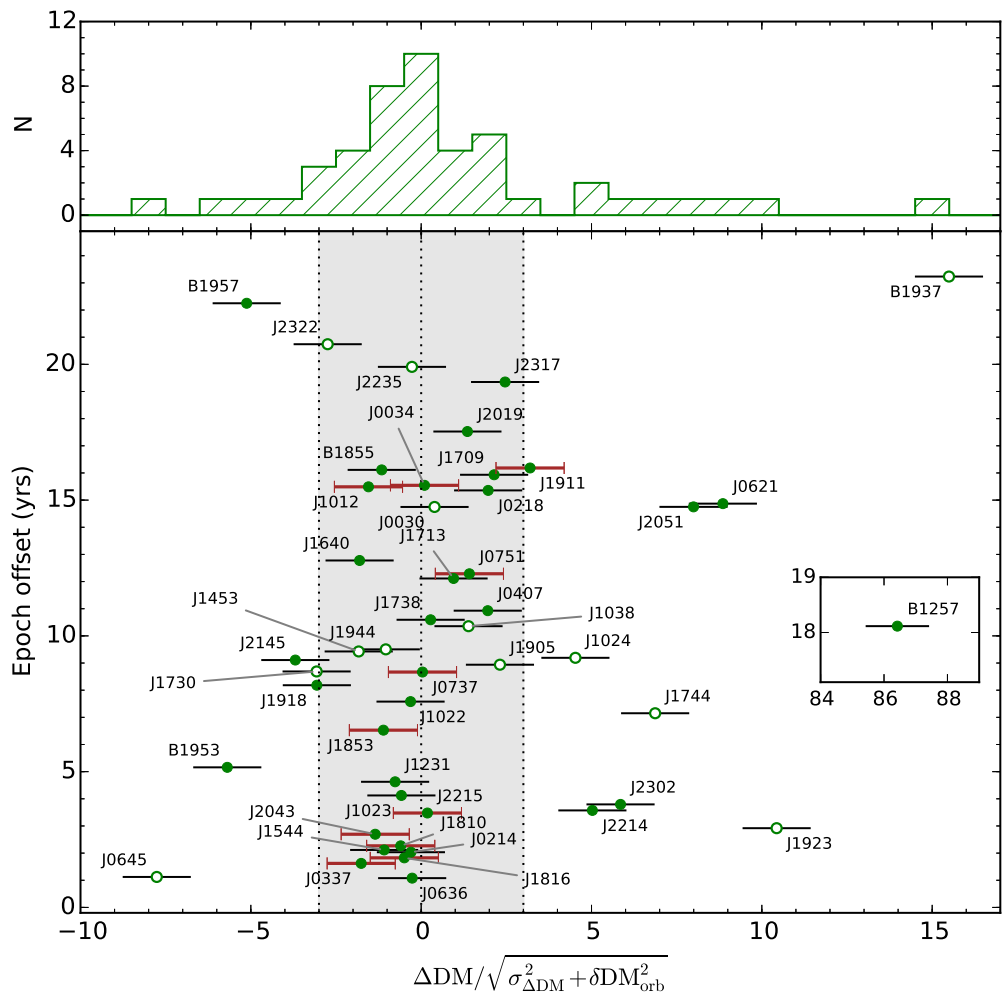}
\caption{{\bf Top.} Histogram of normalised DM offsets, $\Delta\mathrm{DM} = \mathrm{DM}_\mathrm{ref}-\mathrm{DM}_\mathrm{HBA}$.
Values of $\Delta\mathrm{DM}$ are normalised by the quadrature sum of DM offset uncertainty, $\sigma_{\Delta\mathrm{DM}}$, and
$\delta\mathrm{DM}_\mathrm{orb}$ (see text).
{\bf Bottom.} Dependence of normalised $\Delta\mathrm{DM}$ on the epoch offset between reference catalogue epoch and LOFAR epoch 
of DM measurements. Isolated MSPs are shown with open circles, and binary MSPs with filled circles. 
Brown bold error bars with caps are for MSPs, where $\delta\mathrm{DM}_\mathrm{orb}$ is larger than $\sigma_{\Delta\mathrm{DM}}$.  The grey area
shows the region where $|\Delta\mathrm{DM}| < 3\sqrt{\sigma^2_{\Delta\mathrm{DM}}+\delta\mathrm{DM}^2_\mathrm{orb}}$. 
} 
 \label{dmvar}
\end{figure*}

\citet{ars95} have shown that the structure of the ISM has a steep red power spectrum of electron density that results in DM
variations becoming more prominent at longer time scales. Hence, one would expect to see larger DM offsets 
for larger epoch offsets. In Fig.~\ref{dmvar} we plotted DM offset, $\Delta\mathrm{DM}$, against the epoch
offset between the reference catalogue epoch and LOFAR epoch of DM measurements. To account for different uncertainties
of DM measurements for different MSPs and, more importantly, to exclude the influence from $\delta\mathrm{DM}_\mathrm{orb}$
for binary MSPs, we normalised $\Delta\mathrm{DM}$ by the quadrature sum of its uncertainty, $\sigma_{\Delta\mathrm{DM}}$,
and $\delta\mathrm{DM}_\mathrm{orb}$. Note, that for the average DM value over a given time span 
$\delta\mathrm{DM}_\mathrm{orb}$ should be also taken as the average contribution from all ($N_\mathrm{obs}$)
observations. We will, however, use the most conservative estimate on $\delta\mathrm{DM}_\mathrm{orb}$ to avoid
any influence on our conclusions.

For about 25\% of the MSPs, $\Delta\mathrm{DM}$ is more than three times
its DM uncertainty including $\delta\mathrm{DM}_\mathrm{orb}$ as well. The rest of the DM offsets lie within 
$\pm3\sigma_{\Delta\mathrm{DM}}$ area including all MSPs for which the $\delta\mathrm{DM}_\mathrm{orb}$ is larger
than $\sigma_{\Delta\mathrm{DM}}$ (brown error bars). 
We do not see any obvious trends for $\Delta\mathrm{DM}$ between binary and
isolated MSPs. Finally, we also do not see any trends for $\Delta\mathrm{DM}$ to become larger for larger epoch
offsets. There are six MSPs with large $\Delta\mathrm{DM}$ within five years from the reference DM epoch.
This discrepancy in DM may be due to large secular DM drift in the corresponding line of sights.
Another possibility for large DM differences at such short-term scales
may reflect the fact that at low frequencies the radio emission 
probes somewhat different regions of interstellar space, thus providing a slightly different DM \citep{css15}.
Multifrequency simultaneous observations and DM measurements are needed to confirm this point conclusively or rule it out.

\citet{kcs+13} presented DM variations for 20 MSPs from the Parkes Pulsar Timing Array (PPTA); eight sources
overlap with our sample. Unfortunately, observing time spans do not overlap, with the last 
measurement in their paper being for MJD~55800. Hence,
we cannot make a direct comparison of DM offsets. However, a few of the overlapping MSPs show large regular DM drifts, 
in particular PSRs J1730$-$2304 and B1937+21, so we can try to extrapolate the DM offsets at our observing epochs
for at least these two MSPs. For PSR~J1730$-$2304 the extrapolation to the reference epoch of MJD~56474 gives
$\Delta\mathrm{DM}$ of about $4.5\times 10^{-3}$\,pc\,cm$^{-3}$, comparable with our measured offset of $0.006$\,pc\,cm$^{-3}$.
We see similar agreement for PSR~B1937+21, where DM shows a secular decrease with 
an expected value at the reference epoch of MJD~56385 of $\sim 5\times 10^{-3}$\,pc\,cm$^{-3}$ smaller than the $\mathrm{DM}_\mathrm{ref}$. 
This is somewhat smaller than our offset of 0.016\,pc\,cm$^{-3}$, but the amount of DM drift in \citet{kcs+13} for this MSP 
does show significant variations \citep[see also][]{ktr94}. One of the three LOFAR observations of PSR B1937+21 at MJD~56258.75
also overlaps with the intense multifrequency observing campaign of this pulsar presented in \citet{yss+14}. Our DM measurement
for this observing epoch is about $5\times 10^{-4}$\,pc\,cm$^{-3}$ larger than theirs. In their timing analysis they accounted 
for the broadening of the profile due to scattering in each frequency channel across the 310--390\,MHz band, which likely explains
the observed DM offset between our measurements.

Precise LOFAR DM measurements on time scales of two--four weeks
could provide valuable data for DM corrections of higher-frequency pulsar timing data, especially for the MSPs that do
not show variations within the uncertainties of their measurements at higher frequencies, 
e.g. PSRs J1022+1001, J1713+0747, and J2145$-$0750.
We have already started regular monthly observations of MSPs with LOFAR and we also perform
observations with shorter cadence using international LOFAR stations.  With these regular timing observations
we will measure DM values and provide them to the broader pulsar community for use in ISM studies and high-precision timing.
A detailed analysis of DM variations will be presented elsewhere.

\begin{sidewaystable*}
\vskip 10mm
\caption{Dispersion measures of detected MSPs in the frequency range 110--188\,MHz.
}\label{msps_dms}
\centering
\begin{tabular}{lccccccccllld{3.1}}
\hline\hline
PSR & \multicolumn{3}{c}{Catalogue}& \multicolumn{4}{c}{LOFAR} & $\Delta\mathrm{DM}$ & $\delta\mathrm{DM}_\mathrm{orb}$ & $\mathrm{DM}_\mathtt{Tempo2}$ & $\mathrm{DM}_\mathtt{pdmp}$ & \multicolumn{1}{c}{$\tau_\mathrm{scat}$} \\
\cline{2-4}\cline{5-8}
    & $\mathrm{DM}_\mathrm{ref}$ & Epoch & Ref & $\mathrm{DM}_\mathrm{HBA}$ & Epoch & Span & $N_\mathrm{obs}$ & (pc\,cm${}^{-3}$) & ($10^{-3}$\,pc & (pc\,cm${}^{-3}$) & (pc\,cm${}^{-3}$) & \multicolumn{1}{c}{($\upmu$s)} \\
    & (pc\,cm${}^{-3}$) & (MJD) &  & (pc cm${}^{-3}$) & (MJD) & (days) & & & cm${}^{-3}$) &  &  & \\
\hline
J0030+0451    & \hphantom{0}$4.333(1)$\hphantom{000}     & 50984.40   & \hphantom{0}1 & \hphantom{0}$4.3326(2)$\hphantom{00} & 56371.00 & 154.6 & \hphantom{0}6 & $+0.0004(10)$\hphantom{0} & \hphantom{00}--- & \hphantom{0}4.332613(14) & \hphantom{0}4.3326(1) &  1.7 \\
J0034$-$0534  & $13.76517(4)$\hphantom{0}  & 50690.00   & \hphantom{0}2 & $13.76508(17)$ & 56367.52 & 161.6 & \hphantom{0}6 &   $+0.00009(17)$ & \hphantom{0}0.906 & 13.765034(7) & 13.76498(8) & 60 \\
J0214+5222    & $22.0354(34)$\hphantom{0}  & 55974.00   & \hphantom{0}6 & $22.03644(24)$ & 56716.60 & 139.6 & \hphantom{0}6 & $-0.0010(34)$\hphantom{0} & \hphantom{0}0.549 & 22.0361(1) & 22.035(2) & 310 \\
J0218+4232    & $61.252(5)$\hphantom{000}    & 50864.00   & \hphantom{0}3 & $61.23889(32)$ & 56473.30 & --- & \hphantom{0}1 & $+0.013(5)$\hphantom{000} & \hphantom{0}4.357 & 61.2389(3) & 61.2395(3) &  2430 \\
J0337+1715    & $21.3162(3)$\hphantom{00}   & 55920.00   & \hphantom{0}4 & \hphantom{\tablefootmark{$\ddag$}}$21.319(1)$\tablefootmark{$\ddag$}\hphantom{000} & 56512.23 & --- & \hphantom{0}1 & $-0.0028(10)$\hphantom{0} & \hphantom{0}1.159 & \hphantom{0000}--- & 21.319(1) &  290 \\
J0407+1607    & $35.65(2)$\hphantom{0000}     & 52799.00   & 20 & $35.61091(17)$ & 56790.52 & --- & \hphantom{0}1 & $+0.04(2)$\hphantom{0000} & \hphantom{0}0.413 & 35.61091(17) & 35.6105(9) & 780 \\
J0621+1002    & $36.6010(6)$\hphantom{00}   & 50944.00   & \hphantom{0}5 & $36.522(8)$\hphantom{000} & 56375.26 & 172.5 & \hphantom{0}6 & $+0.079(8)$\hphantom{000} & \hphantom{0}3.859 & 36.5349(81) & 36.5377(63) & 1110 \\
J0636+5129    & $11.10598(6)$\hphantom{0}  & 56307.00   & \hphantom{0}6 & $11.1068(31)$\hphantom{0} & 56700.88 & 105.7 & \hphantom{0}2 & $-0.0008(31)$\hphantom{0} & \hphantom{0}0.109 & 11.10647(21) & 11.1064(3) & 80 \\
J0645+5158    & $18.247536(9)$ & 56143.00   & \hphantom{0}6 & $18.24847(12)$ & 56552.87 & 461.7 & \hphantom{0}9 & $-0.00093(12)$ & \hphantom{00}--- & 18.24846(15) & 18.2485(6) &  260 \\
J0737$-$3039A & $48.920(5)$\hphantom{000}    & 53156.00   & \hphantom{0}7 & $48.9179(55)$\hphantom{0} & 56321.94 & --- & \hphantom{0}1 & $+0.002(7)$\hphantom{000} & 53.73\hphantom{0} & 48.9179(55) & 48.915(12) &  0.4 \\
J0751+1807    & $30.2489(3)$\hphantom{00}   & 51800.00   & \hphantom{0}8 & $30.244(1)$\hphantom{000} & 56288.03 & \hphantom{0}16.0 & \hphantom{0}2 & $+0.005(1)$\hphantom{000} & \hphantom{0}3.316 & 30.2434(5) & 30.2425(15) &  590 \\
J1012+5307    & \hphantom{0}$9.0233(2)$\hphantom{00}    & 50700.00   & \hphantom{0}9 & \hphantom{0}$9.02433(8)$\hphantom{0} & 56357.49 & 154.6 & \hphantom{0}6 & $-0.00103(22)$ & \hphantom{0}0.631 & \hphantom{0}9.024355(76) & \hphantom{0}9.0245(2) & 40 \\
J1022+1001    & $10.2521(1)$\hphantom{00}   & 53589.00   & 10 & $10.2530(24)$\hphantom{0} & 56357.46 & 154.6 & \hphantom{0}5 & $-0.0009(24)$\hphantom{0} & \hphantom{0}1.601 & 10.25327(16) & 10.2533(1) &  50 \\
J1023+0038    & $14.325(10)$\hphantom{00}   & 54802.00   & 11 & $14.3309(2)$\hphantom{00} & 56308.15 & \hphantom{0}14.0 & \hphantom{0}2 & $-0.006(10)$\hphantom{00} & \hphantom{0}1.805 & 14.3311(2) & 14.3315(1) &  90 \\
J1024$-$0719  & \hphantom{0}$6.48520(8)$\hphantom{0}   & 53000.00   & 10 & \hphantom{0}$6.48447(14)$ & 56357.46 & 154.6 & \hphantom{0}4 & $+0.00073(16)$ & \hphantom{00}--- & \hphantom{0}6.48446(25) & \hphantom{0}6.4846(4) & 5.6 \\
J1038+0032    & $26.59(20)$\hphantom{000}    & 53000.00   & 30 & $26.312(2)$\hphantom{000} & 56782.80 & --- & \hphantom{0}1 & $+0.28(20)$\hphantom{000} & \hphantom{00}--- & 26.3121(22)\hphantom{00} & 26.3069(55) & 330 \\
J1231$-$1411  & \hphantom{0}$8.090(1)$\hphantom{000}     & 55100.00   & 27 & \hphantom{0}$8.09100(54)$ & 56789.86 & --- & \hphantom{0}1 & $-0.001(1)$\hphantom{000} & \hphantom{0}0.646 & \hphantom{0}8.09100(54) & \hphantom{0}8.0908(11) & 20 \\
B1257+12      & $10.16550(3)$\hphantom{0}  & 49750.00   & 12 & $10.15397(13)$ & 56366.55 & 172.5 & \hphantom{0}6 & $+0.0115(1)$\hphantom{00} & $<10^{-7}$ & 10.15405(11) & 10.1542(1) &  20 \\
J1453+1902    & $14.049(4)$\hphantom{000}    & 53337.00   & 31 & \hphantom{\tablefootmark{$\ddag$}}$14.0569(16)$\tablefootmark{$\ddag$}\hphantom{0} & 56779.99 & --- & \hphantom{0}1 & $-0.0079(43)$\hphantom{0} & \hphantom{00}--- & \hphantom{0000}--- & 14.0569(16) & 90 \\
J1544+4937    & $23.2258(1)$\hphantom{00}   & 56007.00   & 32 & $23.227(1)$\hphantom{000} & 56780.00 & --- & \hphantom{0}1 & $-0.0012(10)$\hphantom{0} & \hphantom{0}0.460 & 23.227(1) & 23.2277(5) & 240 \\
J1640+2224    & $18.4260(8)$\hphantom{00}   & 51700.00   & 13 & $18.42765(12)$ & 56366.68 & 172.6 & \hphantom{0}7 & $-0.00165(81)$ & \hphantom{0}0.423 & 18.427656(5) & 18.4280(2) & 170 \\
J1709+2313    & $25.3474(2)$\hphantom{00}   & 51145.00   & 33 & \hphantom{\tablefootmark{$\ddag$}}$25.3416(24)$\tablefootmark{$\ddag$}\hphantom{0} & 56964.53 & --- & \hphantom{0}1 & $+0.0058(24)$\hphantom{0} & \hphantom{0}1.241 & \hphantom{0000}--- & 25.3416(24) & 320 \\
J1713+0747    & $15.9915(2)$\hphantom{00}   & 52000.00   & 14 & $15.9907(6)$\hphantom{00} & 56424.60 & 286.3 & 10 & $+0.0008(6)$\hphantom{00} & \hphantom{0}0.555 & 15.99044(35) & 15.9909(5) & 120 \\
J1730$-$2304  & \hphantom{0}$9.617(2)$\hphantom{000}     & 53300.00   & 14 & \hphantom{0}$9.62314(4)$\hphantom{0} & 56473.94 & \hphantom{0}33.9 & \hphantom{0}2 & $-0.006(2)$\hphantom{000} & \hphantom{00}--- & \hphantom{0}9.6232(4) & \hphantom{0}9.6233(6) & 50 \\
\hline
\end{tabular}
\tablefoot{
$N_\mathrm{obs}$ is the number of observations within the given span (Col.~7) used for calculation of $\mathrm{DM}_\mathrm{HBA}$;
$\Delta\mathrm{DM}$ is the DM offset between the reference catalogue DM value, $\mathrm{DM}_\mathrm{ref}$, 
and LOFAR's DM measurement, $\mathrm{DM}_\mathrm{HBA}$, i.e. $\Delta\mathrm{DM} = \mathrm{DM}_\mathrm{ref} - \mathrm{DM}_\mathrm{HBA}$;
$\delta\mathrm{DM}_\mathrm{orb} = v_\mathrm{orb}/c\times\mathrm{DM}$ is conservative upper limit on 
``artificial'' DM orbital vatiations for binary MSPs (see text);
$\mathrm{DM}_\mathrm{Tempo2}$ and $\mathrm{DM}_\mathrm{pdmp}$ are DM measurements for the best individual observation 
using {\tt Tempo2} and {\tt pdmp}, respectively;
$\tau_\mathrm{scat}$ is the predicted scattering time at 150\,MHz from the NE2001 model~\citep{cl+02} scaled from 1\,GHz using 
a Kolmogorov spectrum of electron density inhomogeneities. 
\tablefoottext{$\ddag$}{Calculated using {\tt pdmp}.}
}
\tablebib{
(1)~\citet{aaa+09e}; (2)~\citet{aaa+10b}; (3)~\citet{hlk+04}; (4)~\citet{ransom2014}; (5)~\citet{sna+02}; (6)~\citet{slr+14}; 
(7)~\citet{ksm+06}; (8)~\citet{nss+05}; (9)~\citet{lcw+01}; (10)~\citet{hbo06}; (11)~\citet{asr+09}; (12)~\citet{kw03}; (13)~\citet{llww05};
(14)~\citet{vbc+09}; (15)~\citet{jacoby_thesis2005}; (16)~\citet{hrm+11}; (17)~\citet{gsf+11}; (18)~\citet{tsb+99}; (19)~\citet{jsb+10};
(20)~\citet{lynch2013}; (21)~\citet{cbl+95}; (22)~\citet{clm+05}; (23)~\citet{aft94}; (24)~\citet{ntf93}; (25)~\citet{gfc+12};
(26)~\citet{dlk+01}; (27)~\citet{rrc+11}; (28)~\citet{cnt93}; (29)~\citet{cgj+11}; (30)~\citet{bjd+06}; (31)~\citet{lmcs07};
(32)~\citet{brr+13}; (33)~\citet{lwf+04}; (34)~J.~Hessels (private communication).
}
\end{sidewaystable*}

\begin{sidewaystable*}
\setcounter{table}{4}
\vskip 10mm
\caption{continued.
}\label{msps_dms1}
\centering
\begin{tabular}{lccccccccllld{3.0}}
\hline\hline
PSR & \multicolumn{3}{c}{Catalogue}& \multicolumn{4}{c}{LOFAR} & $\Delta\mathrm{DM}$ & $\delta\mathrm{DM}_\mathrm{orb}$ & $\mathrm{DM}_\mathtt{Tempo2}$ & $\mathrm{DM}_\mathtt{pdmp}$ & \multicolumn{1}{c}{$\tau_\mathrm{scat}$} \\
\cline{2-4}\cline{5-8}
    & $\mathrm{DM}_\mathrm{ref}$ & Epoch & Ref & $\mathrm{DM}_\mathrm{HBA}$ & Epoch & Span & $N_\mathrm{obs}$ & (pc\,cm${}^{-3}$) & ($10^{-3}$\,pc & (pc\,cm${}^{-3}$) & (pc\,cm${}^{-3}$) & \multicolumn{1}{c}{($\upmu$s)} \\
    & (pc\,cm${}^{-3}$) & (MJD) &  & (pc cm${}^{-3}$) & (MJD) & (days) & & & cm${}^{-3}$) & & & \\
\hline
J1738+0333    & \hphantom{0}$33.778(9)$\hphantom{0000}    & 52500.00   & 15 & \hphantom{0}$33.7754(11)$\hphantom{0} & 56369.21 & 175.5 & 7 & $+0.0026(91)$\hphantom{0} & \hphantom{0}2.378 & \hphantom{0}33.77481(44) & \hphantom{0}33.7747(13) & 640 \\
J1744$-$1134  & \hphantom{00}$3.13908(4)$\hphantom{00}   & 53742.00   & 10 & \hphantom{00}$3.13808(14)$ & 56353.78 & 144.6 & 6 & $+0.0010(1)$\hphantom{00} & \hphantom{00}--- & \hphantom{00}3.13815(8) & \hphantom{00}3.1383(2) & $<1$ \\
J1810+1744    & \hphantom{0}$39.659298(52)$ & 55520.78  & 34 & \hphantom{0}$39.66041(28)$ & 56349.31 & 135.6 & 6 & $-0.0011(3)$\hphantom{00} & \hphantom{0}1.828 & \hphantom{0}39.6600(2) & \hphantom{0}39.66050(3) & 620 \\
J1816+4510    & \hphantom{0}$38.8874(4)$\hphantom{000}   & 56047.00   & \hphantom{0}6  & \hphantom{0}$38.88972(13)$ & 56713.83 & 136.7 & 5 & $-0.0023(4)$\hphantom{00} & \hphantom{0}4.666 & \hphantom{0}38.8897(1) & \hphantom{0}38.8905(2) & 580 \\
J1853+1303    & \hphantom{0}$30.5701(6)$\hphantom{000}   & 54000.00   & 17 & \hphantom{0}$30.57121(17)$ & 56384.24 & 145.6 & 2 & $-0.0011(6)$\hphantom{00} & \hphantom{0}0.784 & \hphantom{0}30.57140(46) & \hphantom{0}30.5691(9) &  130 \\
B1855+09      & \hphantom{0}$13.300(4)$\hphantom{0000}    & 50481.00   & 14 & \hphantom{0}$13.30482(89)$ & 56366.29 & 169.5 & 7 & $-0.0048(41)$\hphantom{0} & \hphantom{0}0.724 & \hphantom{0}13.3047(4) & \hphantom{0}13.3042(15) &  100 \\
J1905+0400    & \hphantom{0}$25.6923(12)$\hphantom{00}  & 53700.00   & 17 & \hphantom{0}$25.68947(24)$ & 56964.66 & --- & 1 & $+0.0028(12)$\hphantom{0} & \hphantom{00}--- & \hphantom{0}25.68947(24) & \hphantom{0}25.6885(38) & 130 \\
J1911$-$1114  & \hphantom{0}$30.9750(9)$\hphantom{000}   & 50458.00   & 18 & \hphantom{0}$30.96909(69)$ & 56368.80 & 164.5 & 5 & $+0.0059(11)$\hphantom{0} & \hphantom{0}1.462 & \hphantom{0}30.96904(43) & \hphantom{0}30.9690(6) &  560 \\
J1918$-$0642  & \hphantom{0}$26.554(10)$\hphantom{000}   & 53400.00   & 19 & \hphantom{0}$26.590(6)$\hphantom{000} & 56388.75 & 204.4 & 3 & $-0.036(12)$\hphantom{00} & \hphantom{0}1.477 & \hphantom{0}26.5913(11) & \hphantom{0}26.5952(27) &  310 \\
J1923+2515    & \hphantom{0}$18.85766(19)$\hphantom{0} & 55322.00   & 20 & \hphantom{0}$18.85561(5)$\hphantom{0} & 56387.77 & 138.6 & 2 & $+0.00205(20)$ & \hphantom{00}--- & \hphantom{0}18.85567(12) & \hphantom{0}18.8557(2) &  40 \\
B1937+21      & \hphantom{0}$71.0398(2)$\hphantom{000}   & 47899.50   & 21 & \hphantom{0}$71.024(1)$\hphantom{000} & 56385.37 & 253.2 & 3 & $+0.016(1)$\hphantom{000} & \hphantom{00}--- & \hphantom{0}71.02373(55) & \hphantom{0}71.0252(1) &  2960 \\
J1944+0907    & \hphantom{0}$24.34(2)$\hphantom{00000}     & 52913.00   & 22 & \hphantom{0}$24.36079(38)$ & 56384.27 & 145.6 & 2 &$-0.02(2)$\hphantom{0000} & \hphantom{00}--- & \hphantom{0}24.3611(6) & \hphantom{0}24.3611(7) &  80 \\
B1953+29      & $104.501(3)$\hphantom{0000}   & 54500.00   & 17 & $104.5263(26)$\hphantom{0} & 56384.27 & 145.6 & 2 & $-0.025(4)$\hphantom{000} & \hphantom{0}2.035 & 104.523(1) & 104.5202(25) &  8100 \\
B1957+20      & \hphantom{0}$29.1168(7)$\hphantom{000}   & 48196.00   & 23 & \hphantom{0}$29.12151(34)$ & 56322.45 & \hphantom{0}71.8 & 4 & $-0.0047(8)$\hphantom{00} & \hphantom{0}0.494 & \hphantom{0}29.12143(13) & \hphantom{0}29.12130(12) &  110 \\
J2019+2425    & \hphantom{0}$17.203(5)$\hphantom{0000}    & 50000.00   & 24 & \hphantom{0}$17.1959(14)$\hphantom{0} & 56401.26 & 179.5 & 3 & $+0.007(5)$\hphantom{000} & \hphantom{0}0.634 & \hphantom{0}17.19544(84) & \hphantom{0}17.1952(7) &  40 \\
J2043+1711    & \hphantom{0}$20.70987(3)$\hphantom{00}  & 55400.00   & 25 & \hphantom{0}$20.7121(1)$\hphantom{00} & 56384.33 & 145.6 & 3 & $-0.0022(1)$\hphantom{00} & \hphantom{0}1.650 & \hphantom{0}20.71232(19) & \hphantom{0}20.7124(3) &  50 \\
J2051$-$0827  & \hphantom{0}$20.7449(4)$\hphantom{000}   & 51000.00   & 26 & \hphantom{0}$20.7299(17)$\hphantom{0} & 56387.81 & 138.6 & 2 & $+0.015(2)$\hphantom{000} & \hphantom{0}0.686 & \hphantom{0}20.73016(17) & \hphantom{0}20.73120(35) &  150 \\
J2145$-$0750  & \hphantom{00}$8.9977(14)$\hphantom{00}   & 53040.00   & 14 & \hphantom{00}$9.00401(13)$ & 56367.40 & 161.6 & 7 & $-0.0063(14)$\hphantom{0} & \hphantom{0}0.973 & \hphantom{00}9.00401(4) & \hphantom{00}9.00380(13) &  10 \\
J2214+3000    & \hphantom{0}$22.557(1)$\hphantom{0000}    & 55100.00   & 27 & \hphantom{0}$22.5448(22)$\hphantom{0} & 56404.82 & 172.5 & 3 & $+0.0122(24)$\hphantom{0} & \hphantom{0}0.233 & \hphantom{0}22.5431(42) & \hphantom{0}22.5507(9) &  90 \\
J2215+5135    & \hphantom{0}$69.2$\hphantom{000000(0)}           & 55135(30) & 16 & \hphantom{0}$69.1976(15)$\hphantom{0} & 56404.84 & 172.5 & 2 & $+0.0024(15)$\hphantom{0} & 13.14\hphantom{0} & \hphantom{0}69.1959(7) & \hphantom{0}69.1972(2) &  3010 \\
J2235+1506    & \hphantom{0}$18.09(5)$\hphantom{00000}     & 49250.00   & 28 & \hphantom{0}$18.10352(56)$ & 56521.02 & --- & 1 & $-0.014(50)$\hphantom{00} & \hphantom{00}--- & \hphantom{0}18.10352(56) & \hphantom{0}18.1037(20) &  90 \\
J2302+4442    & \hphantom{0}$13.762(6)$\hphantom{0000}    & 55000.00   & 29 & \hphantom{0}$13.7268(1)$\hphantom{00} & 56385.90 & 148.6 & 2 & $+0.035(6)$\hphantom{000} & \hphantom{0}0.409 & \hphantom{0}13.72662(34) & \hphantom{0}13.7274(5) &  20 \\
J2317+1439    & \hphantom{0}$21.907(3)$\hphantom{0000}    & 49300.00   & 28 & \hphantom{0}$21.89873(14)$ & 56367.48 & 161.5 & 6 & $+0.008(3)$\hphantom{000} & \hphantom{0}1.499 & \hphantom{0}21.89876(26) & \hphantom{0}21.8991(1) &  60 \\
J2322+2057    & \hphantom{0}$13.372(5)$\hphantom{0000}    & 48900.00   & 24 & \hphantom{0}$13.38574(25)$ & 56475.68 & \hphantom{0}30.9 & 2 & $-0.014(5)$\hphantom{000} & \hphantom{00}--- & \hphantom{0}13.3853(8) & \hphantom{0}13.3859(9) &  30 \\
\hline
\end{tabular}
\tablefoot{
$N_\mathrm{obs}$ is the number of observations within the given span (Col.~7) used for calculation of $\mathrm{DM}_\mathrm{HBA}$;
$\Delta\mathrm{DM}$ is the DM offset between the reference catalogue DM value, $\mathrm{DM}_\mathrm{ref}$, 
and LOFAR's DM measurement, $\mathrm{DM}_\mathrm{HBA}$, i.e. $\Delta\mathrm{DM} = \mathrm{DM}_\mathrm{ref} - \mathrm{DM}_\mathrm{HBA}$;
$\delta\mathrm{DM}_\mathrm{orb} = v_\mathrm{orb}/c\times\mathrm{DM}$ is conservative upper limit on 
``artificial'' DM orbital vatiations for binary MSPs (see text);
$\mathrm{DM}_\mathrm{Tempo2}$ and $\mathrm{DM}_\mathrm{pdmp}$ are DM measurements for the best individual observation 
using {\tt Tempo2} and {\tt pdmp}, respectively;
$\tau_\mathrm{scat}$ is the predicted scattering time at 150\,MHz from the NE2001 model~\citep{cl+02} scaled from 1\,GHz using 
a Kolmogorov spectrum of electron density inhomogeneities. 
\tablefoottext{$\ddag$}{Calculated using {\tt pdmp}.}
}
\tablebib{
(1)~\citet{aaa+09e}; (2)~\citet{aaa+10b}; (3)~\citet{hlk+04}; (4)~\citet{ransom2014}; (5)~\citet{sna+02}; (6)~\citet{slr+14}; 
(7)~\citet{ksm+06}; (8)~\citet{nss+05}; (9)~\citet{lcw+01}; (10)~\citet{hbo06}; (11)~\citet{asr+09}; (12)~\citet{kw03}; (13)~\citet{llww05};
(14)~\citet{vbc+09}; (15)~\citet{jacoby_thesis2005}; (16)~\citet{hrm+11}; (17)~\citet{gsf+11}; (18)~\citet{tsb+99}; (19)~\citet{jsb+10};
(20)~\citet{lynch2013}; (21)~\citet{cbl+95}; (22)~\citet{clm+05}; (23)~\citet{aft94}; (24)~\citet{ntf93}; (25)~\citet{gfc+12};
(26)~\citet{dlk+01}; (27)~\citet{rrc+11}; (28)~\citet{cnt93}; (29)~\citet{cgj+11}; (30)~\citet{bjd+06}; (31)~\citet{lmcs07};
(32)~\citet{brr+13}; (33)~\citet{lwf+04}; (34)~J.~Hessels (private communication).
}
\end{sidewaystable*}

\begin{figure*}[htbp]
\centering
 \includegraphics[scale=0.5]{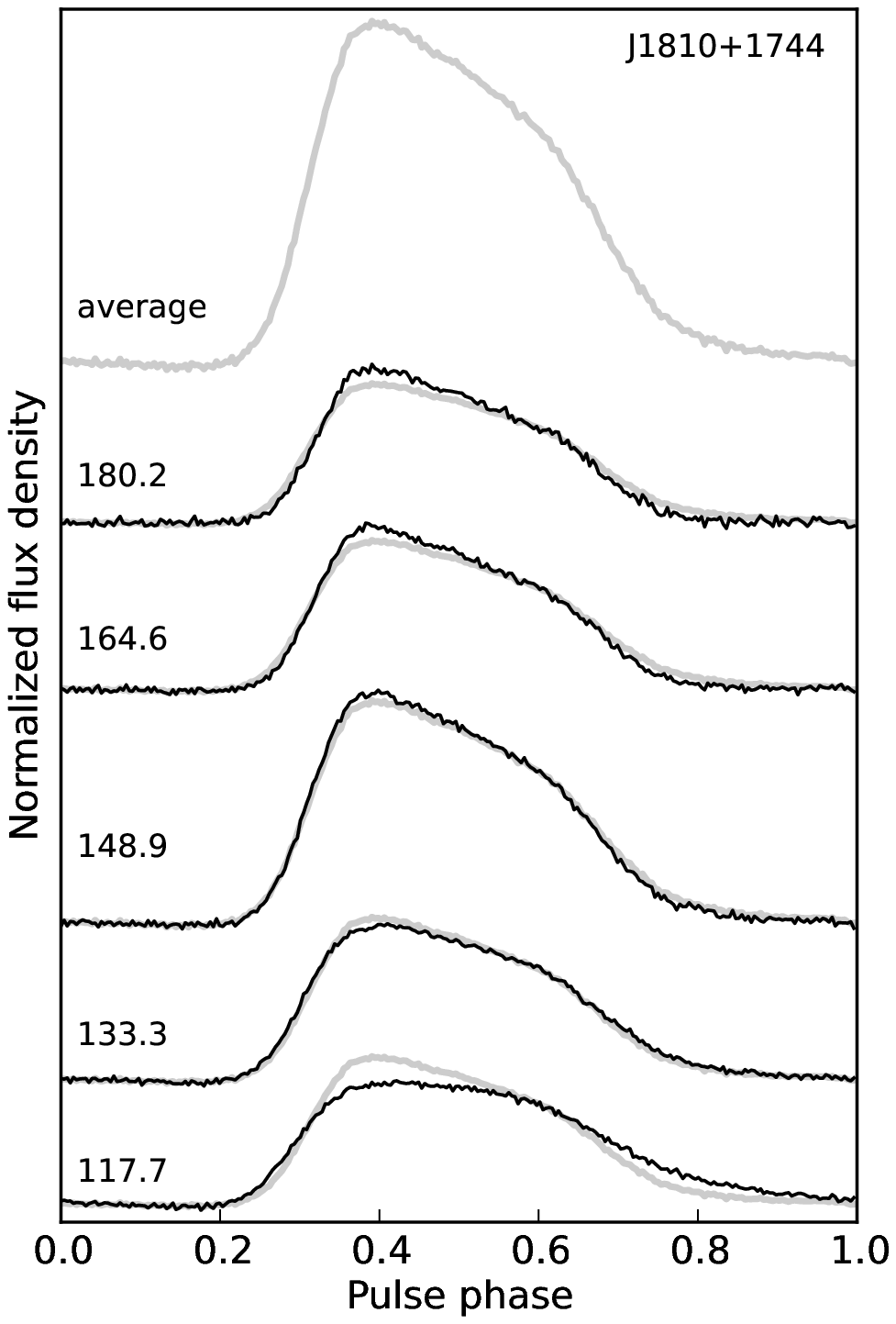}\includegraphics[scale=0.5]{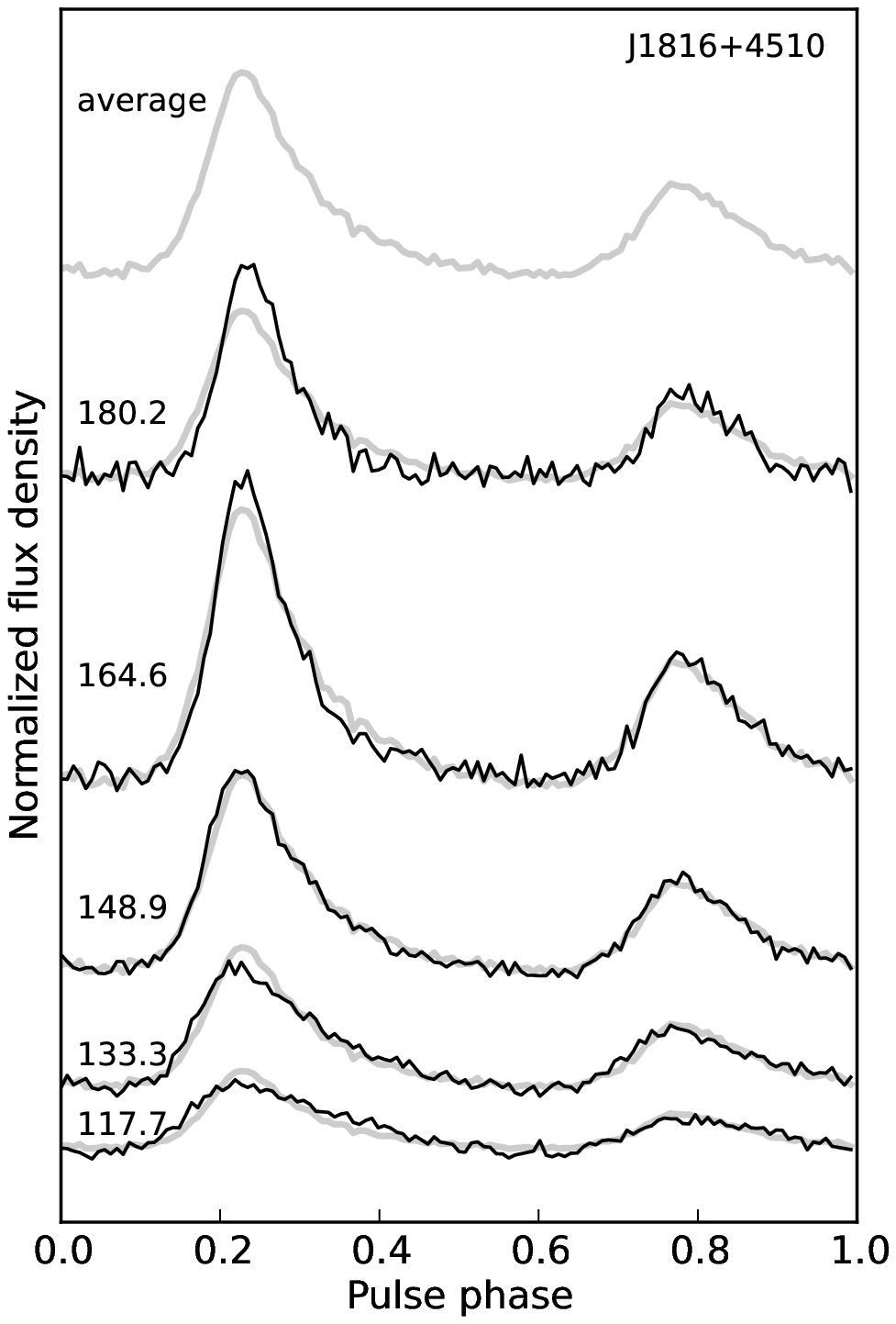}\includegraphics[scale=0.5]{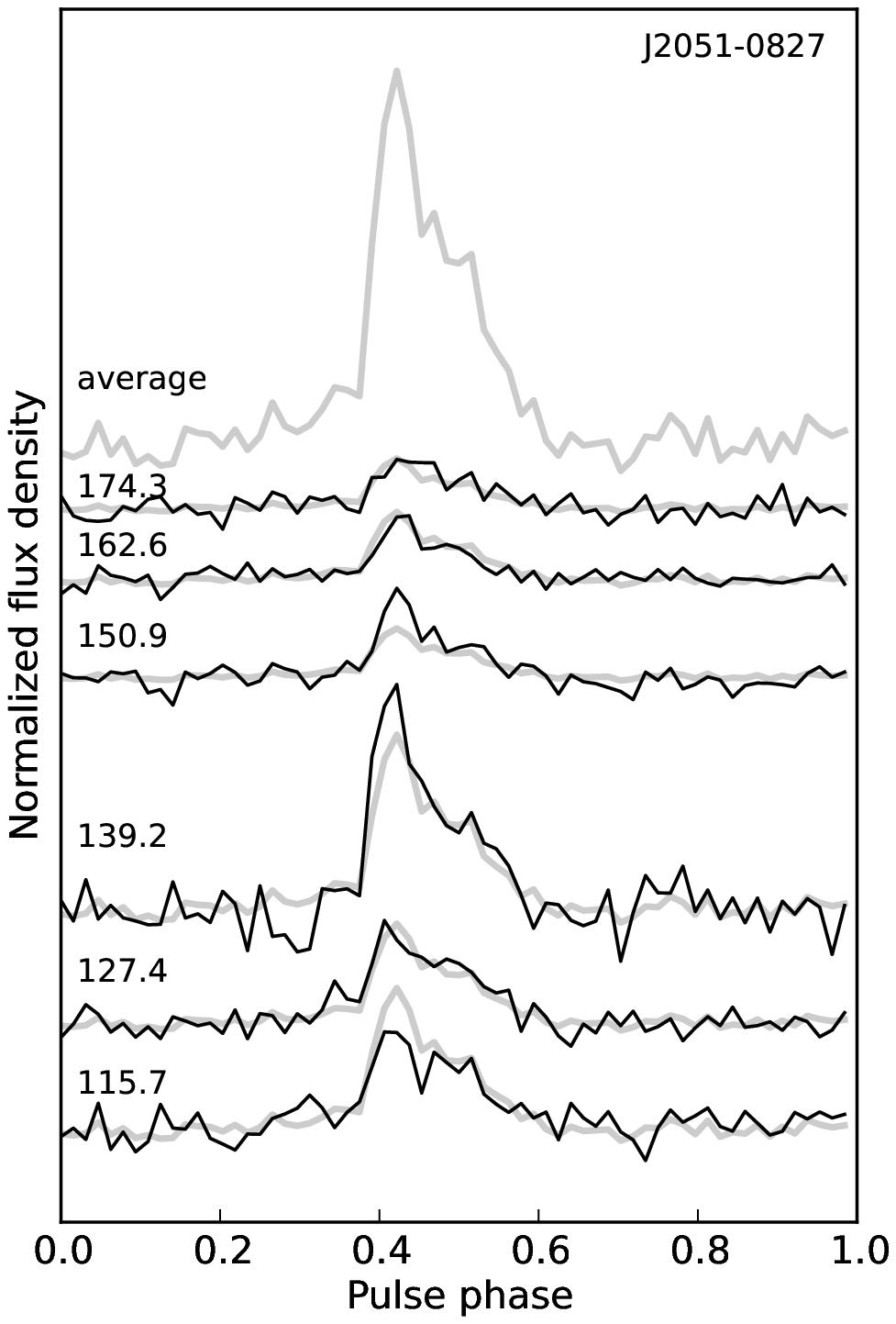}
 \caption{Profile evolution for three MSPs within the LOFAR HBA band.
 The number on the left in each panel gives the central frequency of the corresponding sub-band in MHz.
 The profile in grey on top of each panel is the average 
 profile within the 78-MHz wide band. 
 The profiles for each sub-band are normalised by its noise rms deviation.
 Each sub-band profile is overlaid with the average profile of the total band, in grey. 
 To keep the integrated flux densities the same, 
 the overlaid total-band average profile (in grey) is normalised by the ratio of areas under the total-band profile and the sub-band profile.
 }
 \label{msps-prof-evolution}
\end{figure*}

\subsection{Profile variations}\label{profvar}

In the previous Section we assumed there is no profile variation within the observing
band. However, at LOFAR frequencies this is usually not the case as there are several factors that can affect 
the observed profile shape within the band, such as the intrinsic profile evolution and profile smearing due to scattering.
For pulsar timing, normally
a single profile template is used for cross-correlation with the observed profile to get a TOA. If the profile at the low-frequency 
edge of the band is significantly scattered, it will artificially increase the uncertainty at the lower end of the band and
produce a later TOA, which can be wrongly accounted for by fitting for DM, resulting in an inaccurate
DM with larger DM uncertainty than without profile variation.
The importance of taking pulse profile evolution into account was already demonstrated in \citet{hshk+12}, where they tested
the accuracy of the dispersion law through wide-band simultaneous observations with LOFAR
at 40--190\,MHz and the Lovell and Effelsberg radio telescopes at 1.4 and 8\,GHz, respectively.

To see if this was indeed the case for our MSPs, we calculated DM values using 
both the {\tt Tempo2} and {\tt pdmp} \citep{pdmp} programs. 
For {\tt pdmp}, DM optimisation is based on the signal-to-noise ratio of the entire profile, while for {\tt Tempo2}
it is based on cross-correlation between template and observed profile.
In the ideal case of a simple, unscattered Gaussian profile, identical across the whole observing band, both {\tt Tempo2} and {\tt pdmp}
should provide the same DM measurements. For scattered profiles, one would expect {\tt pdmp} to give larger DM values
due to overcompensation for scatter-broadening, although a DM measurement by {\tt Tempo2} would be biased
as well. If there is a variation of the profile in amplitude due to changes in the bandpass gain or intrinsic to the pulsar itself, 
then this should not matter for {\tt pdmp}, but for {\tt Tempo2} it will depend on how well the match is between the template
and profiles across the band. Therefore, generally in the case of any profile variation one would expect to see
a discrepancy between DM measurements provided by {\tt Tempo2} and {\tt pdmp}. 

Columns 11 and 12 of Table~\ref{msps_dms}
list the {\tt Tempo2} and {\tt pdmp} DM measurements for the best individual observations. To characterise the difference
between these values we also calculated the parameter $\epsilon_\mathrm{DM}$ as:
$$
\epsilon_\mathrm{DM} = \frac{|\mathrm{DM}_\mathrm{Tempo2}-\mathrm{DM}_\mathrm{pdmp}|}{\sqrt{\delta\mathrm{DM}^2_\mathrm{Tempo2} + \delta\mathrm{DM}^2_\mathrm{pdmp}}}~,
$$
where $\delta\mathrm{DM}_\mathrm{Tempo2}$ and $\delta\mathrm{DM}_\mathrm{pdmp}$ are DM uncertainties from {\tt Tempo2} and {\tt pdmp}.
For 30 MSPs in our sample the value of $\epsilon_\mathrm{DM}$ is less than one, and for all of them we do not see
any profile variation. For the other 15 pulsars $\epsilon_\mathrm{DM} > 1$, and \emph{all} of them manifest a visible change
of profile shape and/or amplitude across the band. Among these 15 pulsars, six MSPs have $\epsilon_\mathrm{DM} > 2$, 
namely PSRs J1640+2224 ($\epsilon_\mathrm{DM}=2.1$), 
J1853+1303 (2.2), B1937+21 (2.65), J2051$-$0827 (2.65), J1810+1744 (3.1), and J1816+4510 (3.4) 
in ascending value of $\epsilon_\mathrm{DM}$. Figure~\ref{b1937prof} already shows the profile evolution of PSR
B1937+21, and that profile becomes completely scattered out below 150\,MHz. Profile evolution for the other three MSPs
with the largest $\epsilon_\mathrm{DM}$ values is shown on Fig.~\ref{msps-prof-evolution}.
One can see how the profiles of PSRs J1810+1744 and J1816+4510 are scattered towards lower frequencies. For PSR~J2051$-$0827
scattering does not seem to be relevant, but the pulse amplitude differs significantly at the centre and edges of the band.
Therefore, the $\epsilon_\mathrm{DM}$ parameter could be used as indirect indication for noticeable
profile variation within the observing band, and whether it is needed to be taken into account for accurate DM 
measurements.

It is clear that for accurate measurements of DM, especially at low frequencies, profile evolution
must be taken into account. Even very small profile variations at higher frequencies can also systematically 
bias the derived arrival times
and must be taken into account for high-precision pulsar timing. Frequency-dependent profile templates
that take profile broadening due to scattering into account must be used for this purpose, as implemented
in the software package PulsePortraiture\footnote{\tt https://github.com/pennucci/PulsePortraiture} \citep{pdr14},
or using a Generative pulsar timing analysis framework \citep{lah15}. Recently, \citet{lpm+15} considered
the double-lens model that reproduces the parameters of the observed diffractive scintillation with high accuracy,
and discussed its prospects for removing scattering to improve pulsar timing.

\begin{figure*}[phtb]
\centering
 \includegraphics[scale=0.25]{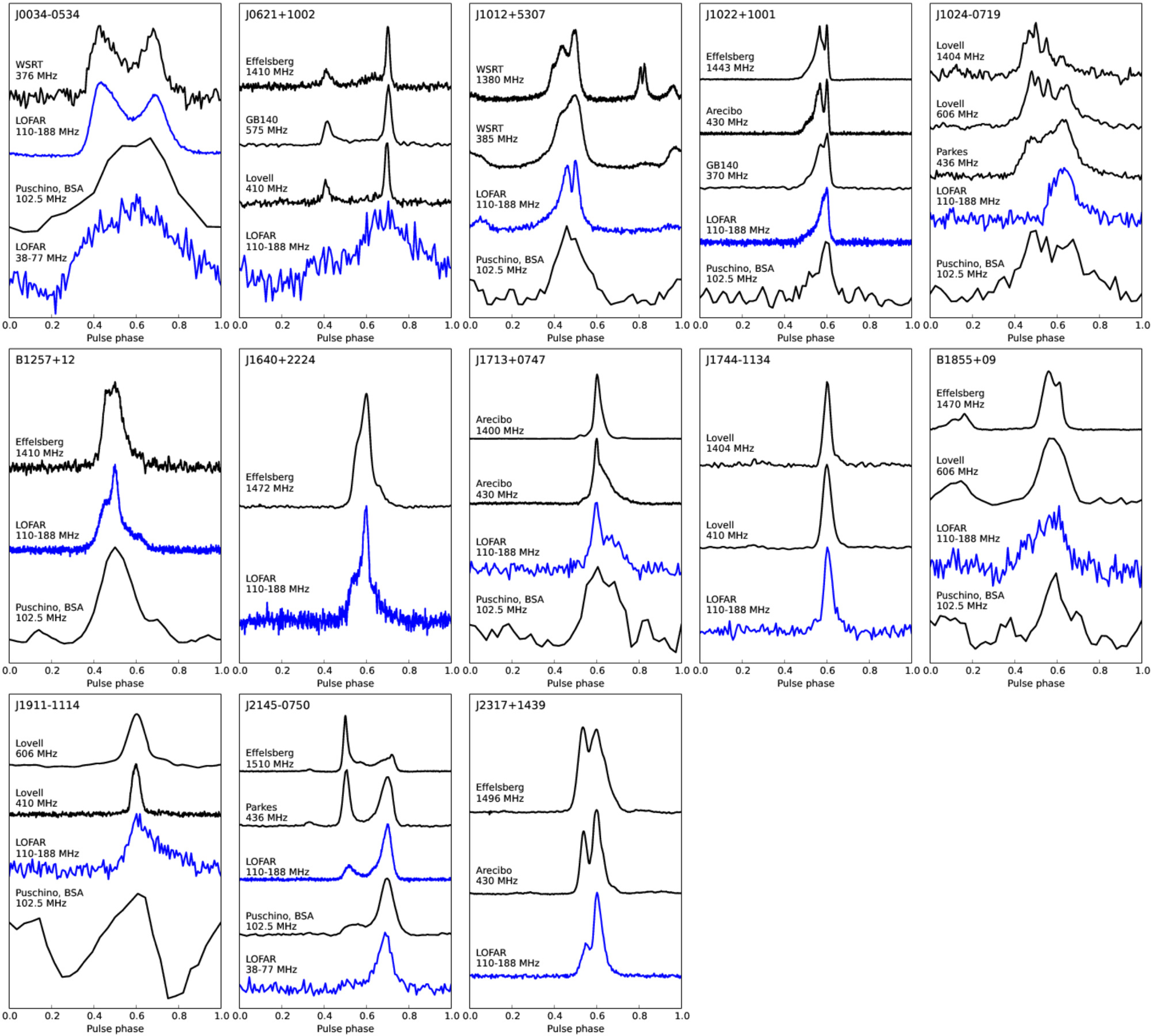}
 \caption{LOFAR MSP average profiles (blue) in comparison with profiles at other frequencies (black) mostly
 from the EPN Database of Pulsar Profiles~\citep{ljs+98}, where available. Each profile is normalised to peak amplitude. 
 These profiles are taken
 from: Pushchino~-- \citet{kl99} and \citet{kl96}, WSRT~-- Stappers (private communication), Effelsberg~-- \citet{kramer1998}, 
 GB140~-- \citet{cnst96} and \citet{snt97}, Parkes~-- \citet{bjb+97} and \citet{lor94}, Arecibo~-- \citet{cnst96}, \citet{fwc93}
 and \citet{cnt93}, and Lovell~-- \citet{stc99}, \citet{bjb+97}, \citet{gl98}, and \citet{llb+96}.
}
 \label{mspfreq}
\end{figure*}

\subsection{Multifrequency MSP profiles}\label{comparison}

Here we discuss qualitatively the wide-band (100--1510\,MHz) profile evolution and compare our findings
to other available low-frequency data. Quantitative comparison including high-energy light
curves will be presented in subsequent work.

For qualitative comparison we used radio profiles from the 
EPN Database of Pulsar Profiles\footnote{\tt http://www.epta.eu.org/epndb/} \citep{ljs+98}.
For these EPN profiles timing information is not available, therefore we visually aligned them
by the phase of the profile peaks. In case of two or more components in a profile we followed
alignment from previously published data \citep{kramer1999} or our own reasoning. These profiles
are presented in Fig.~\ref{mspfreq}, where
we also plot (where available) the previous low-frequency Pushchino profiles at 102/111\,MHz
from \citet{kl99}. LOFAR profiles (blue) are generally of better quality, with higher signal-to-noise ratio and
time resolution. More importantly, they do not suffer from intra-channel dispersion smearing, as
the Pushchino profiles do. Hence, they provide a clearer view of profile evolution at low frequencies.
\citet{stappers2008} detected eight MSPs at 115--175\,MHz with the WSRT. 
Profiles of two pulsars, PSRs J0034$-$0534 and J2145$-$0750, are presented in their paper \citep{stappers2008}, 
and their shapes fully resemble the shape
of the LOFAR profiles. Contrary to our LOFAR observations in 2013 and 
BSA observations by \citet{kl99} in 1999, \citet{stappers2008} did not detect PSR J0218+4232 in 2008. 
This could simply be due to a lack of sensitivity of WSRT,
but it might also hint in favour of regular variations in the scattering
conditions in the ISM on time scales of several years towards this pulsar. Regular monitoring with LOFAR
will answer this question for this and other pulsars in our MSP timing campaign.

Based on how the MSP profiles in Fig.~\ref{mspfreq} evolved in the LOFAR HBA band, we can clearly
divide the presented MSPs into two distinct groups: a) with LOFAR profiles strongly affected by scattering,
and b) LOFAR profiles that do not show profile broadening due to scattering, or do so very weakly.
Only two MSPs from Fig.~\ref{mspfreq}, PSRs J0621+1002 and J1911$-$1114, fall into the first group.
From Fig.~\ref{mspprof} we can certainly
conclude that there are more MSPs with significantly scattered profiles in the LOFAR HBA band, e.g.
PSRs J0218+4232, B1937+21, B1953+29, but we will leave the discussion of their profile evolution
for a future paper. 
Both MSPs with scattered profiles in Fig.~\ref{mspfreq} have relatively
large DMs, 37 and 31\,pc\,cm$^{-3}$ for PSRs J0621+1002 and J1911$-$1114, respectively.
They do not seem to show the development of other profile components. In the case of PSR J0621+1002
the visual alignment was arbitrary as the scattering moves the peak of the
profile to later pulse phases (see e.g. the LBA profile of PSR J0034$-$0534 in Fig.~\ref{lba_profs}).

The other 11 MSPs from Fig.~\ref{mspfreq} form the second group,
all with unscattered or weakly scattered LOFAR profiles.
In fact, for PSR J2145$-$0750 \citet{kl99}
claimed that the separation between components decreased at 102\,MHz, which they
interpreted as an effect caused by the quadrupole magnetic field. \citet{kramer1999} did not see this
abnormal frequency dependence at higher frequencies neither for this nor the other MSPs they studied.
This conclusion by \citet{kl99} might be due to the fact that they did not use coherently dedispersed
profiles and they were affected by intrachannel dispersion that made it more difficult to measure the
separation. \citet{kramer1999} also argued against this conjecture by \citet{kl99} because the trailing
component itself consists of two overlapping components \citep[see also][]{sallmen1998}. Similarly
to \citet{kl99} we do not see a decrease in the component separation for PSR J2145$-$0750 at LOFAR frequencies 
in comparison with higher radio frequencies. Also, there is no broadening of the profile due to scattering in
the HBA band, and we could only measure the scattering time to be less than 0.7\,ms in the LBA band at 57.7\,MHz.

It is quite interesting that for those MSPs with overlapping
profile components (except maybe for PSR~J0034$-$0534), which constitute half of the second group,
the leading component weakens while the trailing component becomes dominant. In particular, this is
clearly evident for PSRs~J1022+1001, B1257+12, J1640+2224, J2145$-$0750, and J2317+1439. This might
explain the apparent narrowing of MSP profiles at low frequencies. \citet{dwd+10}
demonstrated that such an enhancement of the trailing component in MSP profiles is caused
by aberration and retardation. There is also an opposing co-rotation effect, namely the weakening
of the trailing side of the profile due to asymmetry of curvature radiation about the dipole axis.
Therefore, whether an MSP profile shows such a rotation asymmetry is determined by the dominant effect.
\citet{dwd+10} showed that a sharp edge of the trailing component should be accompanied by a maximum in the gradient
of the polarisation angle curve at the same spin phase. We indeed confirm this for 
PSR J1022+1001~\citep[see][]{nsk+15}.

The profile evolution of PSR~J1024$-$0719 is a very interesting case. At present
we aligned the LOFAR profile with the peak of the trailing component of the 436-MHz Parkes profile, because
i) the tail is reminiscent of the tail in the high-frequency profiles, and ii) the apparent tendency
for a weakening leading component from 1.4\,GHz to 436\,MHz. However, the Pushchino profile at
102\,MHz is inconclusive and looks somewhat similar to the two-component profile at 436\,MHz.  Proper 
profile alignment is needed to make a firm conclusion.

In general, without considering the broadening of profiles due to scattering, the MSP profiles
continue to show constant separation between profile components and their widths at frequencies
below 200\,MHz. The same lack of profile evolution of MSPs was reported by \citet{kramer1999}
between frequencies of several GHz down to about 400\,MHz. This is different from what is observed 
for normal pulsars and indicates that the emission must originate from a very compact region in the magnetosphere.

\section{Summary}\label{summary}

We have carried out a LOFAR census of 75 Galactic MSPs in the frequency range 110--188\,MHz, and detected
48 of them (25 of which were detected at these frequencies for the first time).
We have also detected three MSPs out of nine observed with the LOFAR LBAs
in the frequency range 38--77\,MHz, namely PSRs J0030+0451, J0034$-$0534, and J2145$-$0750. 
For the detected MSPs:

\begin{itemize}

 \item We provide average profiles for all detected MSPs, with about half of them being the best high-quality profiles so far at these
 frequencies. For 25 MSPs the presented profiles are the first at 110--188\,MHz. About 35\% of the MSPs show strong 
 narrow profiles, another 25\% exhibit undoubtedly scattered profiles, and the rest have low signal-to-noise profiles.
 
 \item We measure the mean flux density at 110--188\,MHz and compare it with the predicted values derived
 from high-frequency observations. For at least a third of the MSPs, the main uncertainty on the predicted flux density 
 originates from poor knowledge of the spectral index. For about another third, the measured flux densities are still lower
than the predicted ones even within the uncertainty on their spectral indices. In Sect.~\ref{flux_definition} and 
Sect.~\ref{flux_measurements} we consider different factors that could affect our flux measurements.
 
 \item We also measure the effective pulse width at 150\,MHz. For the majority of the detected MSPs, their
 pulse duty cycle, $\delta$, is less than 20\%. Almost all MSPs with $\delta\gtrsim20$\% show profiles
widened by scattering.

 \item There is no clear dependence of the MSP detectability on the predicted scattering. We discuss this and
 other possible factors that affect the MSP detectability in Sect.~\ref{detectability}.

 \item We present average values of DM and their offsets from the catalogue values. For 14 MSPs in our
 sample the absolute value of their DM offsets is more than three times larger than their errors, and we do not
 see any trend of DM offsets becoming larger for larger epoch offsets. 

 \item Finally, we qualitatively compare LOFAR profiles for a subset of MSPs with their profiles
 at higher radio frequencies. If not broadened by scattering, they show apparent consistency
 in the width of their profile components and their separation within the frequency range from 150\,MHz
 to a few GHz. This is very different from what is observed for normal pulsars and suggests a compact
 emission region in the MSP magnetosphere. 

\end{itemize}

\begin{acknowledgements}
The LOFAR facilities in the Netherlands and other countries, under different ownership,
are operated through the International LOFAR Telescope foundation (ILT) as an international 
observatory open to the global astronomical community under a joint scientific
policy. In the Netherlands, LOFAR is funded through the BSIK program for interdisciplinary 
research and improvement of the knowledge infrastructure. Additional funding is
provided through the European Regional Development Fund (EFRO) and the innovation
program EZ/KOMPAS of the Collaboration of the Northern Provinces (SNN). ASTRON
is part of the Netherlands Organisation for Scientific Research (NWO).
VK thanks Michiel Brentjens, Stefan Wijnholds, Ger de Bruyn and George Heald for very helpful 
discussions regarding the LOFAR beam model and flux calibration, and ASTRON's Radio
Observatory Science Support for the help with the observations. 
We are endlessly thankful to Willem van Straten for implementing code for the {\tt dspsr} program suite
to read raw LOFAR pulsar data.
VK is grateful to Anne Archibald who provided the best available ephemerides for PSR J0337+1715, 
to Patrick Lazarus for providing the RFI zapping package {\tt CoastGuard},
and Tobia Carozzi for development of the {\tt mscorpol} package.
We acknowledge the use of pulsar profiles at high radio frequencies from the 
European Pulsar Network (EPN) database at the Max-Planck-Institut f\"ur Radioastronomie,
and its successor, the new EPN Database of Pulsar Profiles, developed by Michael Keith and
maintained by the University of Manchester.
The research leading to these results has received funding from the
European Research Council under the European Union's Seventh Framework
Programme (FP7/2007-2013) / ERC grant agreement nr. 337062 (DRAGNET; PI JWTH).
SO is supported by the Alexander von Humboldt Foundation.
SC acknowledges financial support from the UnivEarthS Labex program of
Sorbonne Paris Cit\'{e} (ANR-10-LABX-0023 and ANR-11-IDEX-0005-02), and
JvL acknowledges funding from the European Research Council under the European Union's 
Seventh Framework Programme (FP7/2007-2013) / ERC Grant Agreement n. 617199.
\end{acknowledgements}

\bibliographystyle{aa}
\bibliography{msps.bib}

\end{document}